\documentclass[a4paper,11pt]{article}
\pdfoutput=1 % if your are submitting a pdflatex

\usepackage{jheppub}
\usepackage[T1]{fontenc}
\usepackage{amsmath}
\usepackage{multirow}
\usepackage{bm}

\usepackage[dvipsnames]{xcolor}
\usepackage{tikz}
\usepackage{framed}
\usetikzlibrary{patterns}
\usetikzlibrary{snakes}

\usepackage{subcaption}
\usepackage{amsfonts}
\usepackage{tikz-cd,tikz}
\usepackage{url}
\usepackage{amsmath}
\usepackage{calrsfs}
\DeclareMathAlphabet{\pazocal}{OMS}{zplm}{m}{n}
\newcommand{\bP}{\mathbf{P}}
\newcommand{\bQ}{\mathbf{Q}}

\newcommand{\bR}{\pazocal{P}}

\newcommand{\adsst}{AdS$_3\times$S$^3\times$T$^4$}
\newcommand{\eps}{\epsilon}

\newcommand{\psu}{PSU$(1,1|2)$}
\newcommand{\bT}{\mathbf{T}}
\newcommand{\wT}{\mathbb{T}}
\newcommand{\mT}{\mathcal{T}}
\newcommand{\beq}{\begin{equation}}
\newcommand{\eeq}{\end{equation}}
\newcommand{\beqa}{\begin{eqnarray}}
\newcommand{\eeqa}{\end{eqnarray}}
\newcommand{\la}{\label}

\newcommand{\bal}{\begin{equation}\begin{aligned}}
\newcommand{\eal}{\end{aligned}\end{equation}}

\def\a {\alpha}

\def\AA {{\sigma}}
\def\ss {\mathsf{s}}

\title{Deriving the \adsst \,Quantum Spectral Curve II: T-system and QSC}

\author[a]{Andrea Cavagli\`a,}
\author[b]{Davide Polvara,}
\author[a]{Nicol\`o Primi,}
\author[c,d,1]{Alessandro Sfondrini,}
\author[a]{Roberto Tateo}

\affiliation[a]{
Dipartimento di Fisica, Universit\`a di Torino and INFN, Sezione di Torino,
Via Pietro Giuria 1, I-10125 Torino, Italy
}

\affiliation[b]{
Institut f\"ur Theoretische Physik, Universit\"at Hamburg,
Luruper Chaussee 149, 22761 Hamburg, Germany
}

\affiliation[c]{
School of Mathematics, University of Birmingham,
Watson Building, Edgbaston, Birmingham B15 2TT, UK
}

\affiliation[d]{
INFN, Sezione di Padova,
via Marzolo 8, 35131 Padova, Italy
}

\note{On leave from the University of Padova, Italy.}

\emailAdd{andrea.cavaglia@unito.it}
\emailAdd{davide.polvara@gmail.com}
\emailAdd{nicolo.primi@unito.it}
\emailAdd{a.sfondrini@bham.ac.uk}
\emailAdd{roberto.tateo@unito.it}

\abstract{
We complete the systematic derivation of the Quantum Spectral Curve (QSC) for strings on \adsst\ with pure Ramond--Ramond (RR) flux, building on the extended Y-system obtained in our preceding paper. We establish the precise relation between the two formalisms of the Thermodynamic Bethe Ansatz and the QSC, derive the previously conjectured QSC, and, along the way, uncover additional relations satisfied by its solutions. Our analysis also clarifies how massless excitations, although not explicit among the basic QSC building blocks, are encoded in their analytic structure. The derivation culminates in the so-called $\bP\bm{\mu}$-system, a compact formulation of the QSC in terms of a small set of functions and their analytic continuations.
}

\begin{document}
\begin{flushright}\small{ZMP-HH/26-30}\end{flushright}
\maketitle
\section{Introduction}
One of the most important examples of a holographic duality is AdS$_3$/CFT$_2$~\cite {Maldacena:1997re}, which can be realised on a number of spacetimes. Two well-studied such backgrounds, which preserve maximal supersymmetry, have the form AdS$_3\times $S$^3\times $M$^4$, where M$^4$ can be either T$^4$ or S$^3\times $S$^1$. Both can be supported by a mixture of Ramond-Ramond (RR) and Neveu-Schwarz-Neveu-Schwarz (NSNS) fluxes and are classically integrable~\cite{Babichenko:2009dk,Sundin:2012gc, Cagnazzo:2012se}.

While the string worldsheet theory can be solved exactly in the presence of pure NSNS flux as a Wess--Zumino--Witten model~\cite{Maldacena:2000hw, Maldacena:2000kv}, which yields a simple and highly degenerate spectrum, turning on RR flux makes the worldsheet-CFT description non-local and quite unwieldy. At best, this allows us to introduce RR flux perturbatively~\cite{Cho:2018nfn,Alday:2026jkn}, whereby the spectrum becomes non-degenerate and very non-trivial --- similar to that of AdS$_5\times $S$^5$. Similar results can be found by the amplitude bootstrap program~\cite{Alday:2024rjs,Chester:2024wnb,Alday:2026jkn}.

Integrability provides the natural tool to tackle the spectral problem of AdS$_3$/CFT$_2$ theories {in the large central charge limit (analogous to the planar limit in Lagrangian gauge theories)}. In fact, each of the three backgrounds remains classically integrable for any flux values~\cite{Cagnazzo:2012se}, providing in principle a way to determine their finite-volume spectrum exactly. While recent progress has been made in the integrability description of the spectrum of the AdS$_3\times $S$^3\times $S$^3\times$S$^1$ theory~\cite{Cavaglia:2025icd, Chernikov:2025jko, Frolov:2025syj}, the best-studied background is certainly \adsst. For the pure RR case, which is the focus of this paper, two alternative formulations of the spectral problem have been independently proposed: the mirror Thermodynamic Bethe Ansatz (TBA)~\cite{Frolov:2021bwp}, and the Quantum Spectral Curve (QSC)~\cite{Cavaglia:2021eqr, Ekhammar:2021pys}.

The mirror TBA follows from the exact worldsheet S-matrix of~\cite{Borsato:2012ud,Borsato:2014hja,Frolov:2021fmj} (and it is also known in the mixed-flux case~\cite{Frolov:2025uwz} see also~\cite{Hoare:2013lja,Lloyd:2014bsa,OhlssonSax:2023qrk,Frolov:2025tda}) and provides a direct
description in terms of the pseudo-energies of massive, massless, and auxiliary excitations. The QSC gives a much more compact formulation in terms of a finite set of
Q-functions and their analytic continuation properties, and has been used to study the spectrum of short operators in \cite{Cavaglia:2022xld} and, more recently,~\cite{Ekhammar:2026ykk}.

While in AdS$_5\times $S$^5$ and AdS$_4\times $CP$^3$ the QSC was directly derived from the TBA~\cite{Gromov:2014caa,Bombardelli:2017vhk}, the \adsst\, Quantum Spectral Curve has been conjectured by directly gluing two \psu \, Q-systems through analytic continuation across branch cuts in the
spectral-parameter plane, and the precise relation with the mirror Thermodynamic Bethe Ansatz has remained unclear. 
 In particular, massless modes are manifest in the TBA equations, whereas they are not directly visible in the QSC. In the latter framework, there is no Q-function associated to the massless modes themselves. However, it was suggested that the presence of massless excitations is linked to an unusual feature of the Q-functions of the model, i.e. the fact that they have non-quadratic branch cuts~\cite{Cavaglia:2021eqr, Ekhammar:2021pys}. In this paper, we clarify how massless excitations emerge through the monodromy equations relating two \psu\, T-hooks, and how they consequently affect the branch-cut structure of the Q-functions.

The main goal of this paper is to bridge the two formalisms by developing  a systematic derivation of the Quantum Spectral Curve (in particular, its concise formulation known as the ``$\bP \bm{\mu}$-system'') from the Thermodynamic Bethe Ansatz, as was done in the AdS$_5$ and AdS$_4$ cases \cite{Gromov:2014caa, Bombardelli:2017vhk}. In our previous work~\cite{Cavaglia:2026scj}, 
we initiated this program by deriving the theory's extended Y-system directly from the mirror TBA. The result consists of two copies of a \psu \, Y-system, supplemented by a complete set of local 
{``discontinuity relations'', which describe the monodromy of Y functions around branch points located at coupling-dependent positions in the rapidity plane.} We also showed, through an explicit inversion procedure, that this extended Y-system is equivalent to
the ground-state TBA. This provides a local, state-independent description of the spectral problem and is the natural starting point for the derivation of the Quantum Spectral Curve.

The present paper completes this derivation, going through several steps: from the Y-system to a T-system and finally to the QSC. In the first step, the analytic Y-system of~\cite{Cavaglia:2026scj} is reformulated  as a T-system living on two \psu\,  T-hooks. The T-functions satisfy the Hirota bilinear difference equations, originally introduced in the context of $\tau$-functions in integrable hierarchies~\cite{Hirota:1971zz} and later
generalised to quantum integrable systems~\cite{Zabrodin:1996vm}. Each Y-function can be expressed as a ratio of four neighbouring T-functions, so that the Y-system equations follow automatically from the Hirota equations.
 Unlike the Y-functions, the T-functions are not uniquely defined. They enjoy a large gauge freedom under transformations that leave all Y-functions invariant. We exploit this freedom to construct distinguished gauges, inspired by \cite{Gromov:2011cx}, in
which the analytic properties of the T-functions become substantially simpler. 
Based on the Wronskian solution of the
\psu \,  T-system~\cite{Tsuboi:2011iz}, we then parametrise the T-functions in terms of a finite number of more elementary building blocks. Translating all the discontinuity equations to these building blocks, and under some natural assumptions,\footnote{In some cases, we shall assume the simplest possible solution to some of the functional equations encountered, as will be detailed in the main text.} we find a closed set of functional and monodromy relations.

These relations
take precisely the form of the $\bP\bm{\mu}$-system, one of the equivalent formulations of the Quantum Spectral Curve, which was first advanced in the AdS$_5$ context in~\cite{Gromov:2013pga} and for the AdS$_3$ model under consideration in \cite{Cavaglia:2021eqr, Ekhammar:2021pys}. We match these equations and highlight previously unnoticed constraints among the QSC quantities.

While some of the methods we use echo the AdS$_5$  construction of \cite{Gromov:2011cx,Gromov:2014caa}, the massless sector turns out to be essential throughout the present derivation. As already emphasised in~\cite{Cavaglia:2026scj}, massless excitations do not occupy a node of either T-hook and enter the extended Y-system only through discontinuity relations.  We show that these discontinuities are precisely
what is needed to connect the two wings of the T-system and to close the $\bP\bm{\mu}$-system equations. In this sense, the presence of the massless sector, although not manifest in the QSC variables, is encoded in their monodromy data. Our main results can be summarised as follows:

\begin{enumerate}
\item Focusing on a sector symmetric under `left-right' reflection in each PSU(1,1|2) T-hook, we derive the \adsst \, $\bP\bm{\mu}$-system from the extended Y-system of~\cite{Cavaglia:2026scj}, thereby completing the derivation from the mirror TBA. 

\item We show an explicit relation between the presence of the massless modes and the non-quadratic branch-cut structure of the \adsst \,  QSC. Such a relation between the non-quadratic nature of branch points and the massless modes was speculated to be true in \cite{Cavaglia:2021eqr, Ekhammar:2021pys}. Here we see explicitly how it works.

\item We prove that exact massless Bethe roots correspond to zeros of the $\mu$-functions lying on their branch cuts. 
This property was previously observed in the large-volume limit
in~\cite{Ekhammar:2024kzp}; here we derive an exact relation beyond the large-volume approximation.

\item We find an additional constraint relating the monodromy data of the two \psu \, wings. In particular, the two $\bm{\mu}$ matrices introduced for the two wings in \cite{Cavaglia:2021eqr, Ekhammar:2021pys} are not independent, and the QSC can be formulated in terms of a single independent $\bm{\mu}$ matrix together with an additional discrete freedom, i.e. a sign factor that we denote as $\AA=\pm 1$.
\end{enumerate}

As mentioned, in this paper, we will restrict our attention to a sector of the theory satisfying an additional left-right symmetry within each T-hook (which corresponds to a Hodge symmetry at the level of the Q-functions of the Quantum Spectral Curve).  At the same time, we keep fully general with respect to the chiral symmetry of the theory, corresponding to the exchange of the two PSU(1,1|2) halves of the system, which we \emph{do not} assume to be related. This simplifies our derivation, while still capturing a wide class of physically relevant states.\footnote{To the best of our knowledge,  this covers all states which have been studied so far at weak coupling or numerically with the QSC~\cite{Cavaglia:2022xld, Ekhammar:2026ykk}.
}

The rest of this paper is organised as follows. Section
\ref{sec:notations} introduces the notation and conventions used throughout. Section \ref{sec:Y-system} reviews the Y-system and the discontinuity relations derived in~\cite{Cavaglia:2026scj}, rewritten in a notation adapted to the
T-hook description. Section \ref{sec:T-system} introduces the 
T-system for \adsst, consisting of two copies of a \psu \, T-system, and explains its relation to the Y-system. In section \ref{sec:Tgauges}, we construct the two distinguished gauges denoted by $\bT$ and $\wT$, and derive their analytic
properties from the discontinuity relations of the Y-system. In section \ref{sec:Pmusystem}, we introduce a  parametrisation of T-functions in terms of more elementary building blocks, and derive the $\bP\bm{\mu}$-system 
 of monodromy equations,  comparing it with the QSC proposed in~\cite{Cavaglia:2021eqr,Ekhammar:2021pys}. Section
\ref{sec:conclusions} summarises our results and discusses possible future directions. Finally, the paper contains four appendices with technical details of the derivation.

\section{Notation and analytic conventions}
\label{sec:notations}
All the integrability building blocks we will deal with -- Y functions, T-functions and Q functions -- are multivalued functions of the spectral parameter $u$. They are defined on intricate Riemann surfaces arising from the TBA equations, the Y-system, and the discontinuity relations. These surfaces contain infinite ladders of branch cuts of two types:
\begin{enumerate}
\item ``long'' branch cuts, at $u\in(-\infty+i n/h, -2+i n/h) \cup (+2+i n/h, +\infty+i n/h)$, $n\in\mathbb{Z}$, which we refer to as \textit{mirror} branch cuts;
\item ``short'' branch cuts, at $u\in(-2+i n/h, 2 + in/h)$ , $n\in\mathbb{Z}$, which we refer to as \textit{string} branch cuts.
\end{enumerate}
{Above, $h$ is a fundamental parameter of the integrability formulation, representing a dressed coupling constant which encodes the dependence on the string tension and fluxes.\footnote{The constant $h$ is the analogue of the ``interpolating function'' of ABJM theory~\cite{Gromov:2014eha}. The dependence of $h$ on the physical parameters of the theory should eventually be fixed using some method independent of integrability. We do not consider this problem in the present paper.}}

\begin{figure}[htbp]
    \centering
    \begin{subfigure}{0.45\textwidth}
        \centering
        % First figure
        \includegraphics[width=\linewidth]{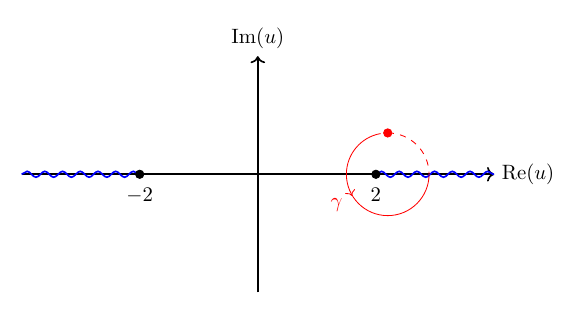}
    \end{subfigure}
    \hfill
    \begin{subfigure}{0.45\textwidth}
        \centering
        % Second figure
        \includegraphics[width=\linewidth]{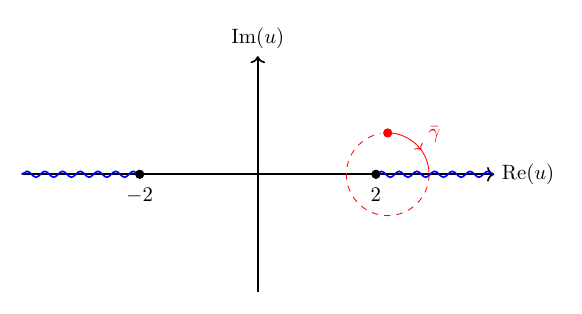}
    \end{subfigure}
\caption{The analytic continuation paths $\gamma$ and $\bar{\gamma}$ for functions with ``long'' branch cuts, as defined in~\cite{Cavaglia:2026scj}, from which this image has been taken. The path $\gamma$ crosses the long cuts from below, corresponding to a counterclockwise continuation around the branch point at $u=+2$ and a clockwise continuation around the branch point at $u=-2$. The path $\bar{\gamma}$ follows the opposite convention.}
\label{fig:gamma_path}
\end{figure}

Branch points with the same imaginary part on a given sheet will always be assumed to lead to the same hidden sheets.
At least in the initial stages of our analysis, the relevant quantities are most naturally defined on Riemann sheets with long cuts, since it is on these sheets that the Y-system and T-system relations hold. For a function $f$ defined on such a sheet, we define its discontinuity across the branch cut at $\Im(u)=N/h$ as follows:
\bal
\la{eq:def_discontin_f_N}
\bigl[ f \bigr]_N(u) \equiv f(u+\frac{i}{h}N+i0)-f(u+\frac{i}{h}N-i0) \, .\\
\eal
This definition is valid for $|\Re(u)|>2$, and can be analytically continued to define the LHS for arbitrary $u$.
An equivalent definition, valid for generic $u$ on the first Riemann sheet, is
\bal
\bigl[ f \bigr]_N(u) \equiv f\left(u+\frac{i}{h}N \right)-f^\gamma\left(u+\frac{i}{h}N \right)\,,
\eal
where $\gamma$ denotes the analytic continuation path that enters the long cuts from below, as shown in figure~\ref{fig:gamma_path}. We denote by $\bar{\gamma}$ the geometrically reversed path, obtained by winding clockwise around the long cuts in the half-plane $\Re(u)>0$, as also shown in figure~\ref{fig:gamma_path}. Although $\bar{\gamma}$ reverses the orientation of $\gamma$, the corresponding analytic continuations are in general not inverse to one another, since the branch points of the model are not quadratic.

In a similar way, we define the symmetric combination of the two determinations of $f$ across the long branch cut at $\Im(u)=N/h$ by
\bal
\{f\}_N & \equiv
f\left(u+\frac{i}{h}N\right)+f^\gamma\left(u+\frac{i}{h}N\right) \,.
\eal
Strictly speaking, this quantity is not a discontinuity; nevertheless, for brevity, we will refer to it as the ``symmetrised discontinuity'' throughout the paper.

We will also use the following convention for shifts of the spectral parameter:
\begin{equation}
f^{[\pm m]}=f\left(u \pm \frac{i}{h}m\right)\,.
\end{equation}
where the shifted function is evaluated on the Riemann sheet on which all branch cuts are long.

Finally, it will sometimes be necessary to consider a different Riemann section of a function $f$ originally defined with long branch cuts. We denote this new section by $\hat{f}$. 
The functions $\hat f$ and $f$ coincide in the widest cut-free strip containing the real axis. 
Outside this strip, $\hat f$ is obtained by analytic continuation from the neighbouring strip, choosing at each step the determination corresponding to the short-cut convention.

For example, if $f$ has branch cuts at $\Im(u)=\pm 1/h,\pm 3/h,\ldots$, then

\begin{equation}
\hat{f}(u)=f(u),
\qquad |\Im(u)|<\frac{1}{h},
\end{equation}
whereas, in the strip $1/h<\Im(u)<3/h$,
\begin{equation}
\hat{f}(u)
= \left((f^+(u))^{\gamma} \right)^-
\end{equation}
while in the strip $-3/h<\Im(u)<-1/h$,
\begin{equation}
\hat{f}(u)
= \left((f^-(u))^{\bar\gamma} \right)^+.
\end{equation}
In other words, one first shifts the argument to the adjacent strip where $f$ and $\hat f$ coincide, analytically continues $f$ across the long cut at $\Im(u)=1/h$ ($\Im(u)=-1/h$) by following the path $\gamma$ ($\bar\gamma$), and finally shifts the argument back by $-i/h$ ($+i/h$). The same construction applies in the other strips.

With a slight abuse of notation, we will also use the shift notation introduced above for functions with short cuts. Thus, for example, we will write $\hat{f}^{[+n]}$ to denote a shift by $i n/h$ performed on the Riemann sheet possessing short branch cuts.

\begin{figure}[htbp]
    \centering
    \begin{subfigure}{0.49\textwidth}
        \centering
        % First figure
        \includegraphics[width=\linewidth]{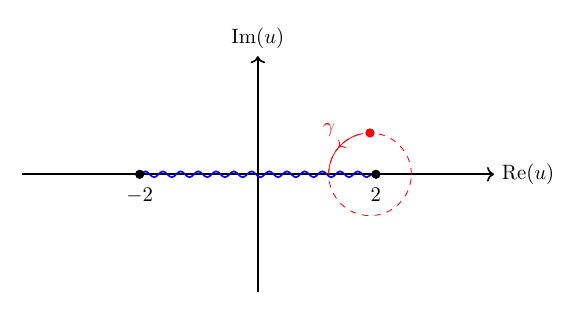}
    \end{subfigure}
    \begin{subfigure}{0.49\textwidth}
        \centering
        % Second figure
        \includegraphics[width=\linewidth]{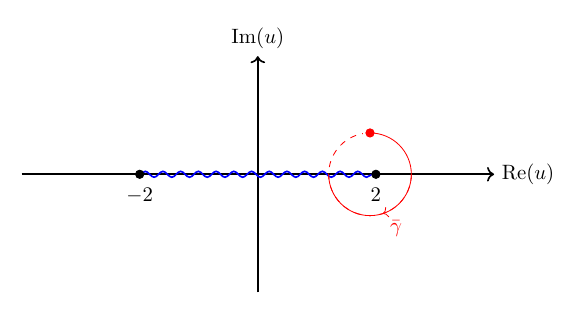}
    \end{subfigure}
    \caption{The analytic continuation paths $\gamma$ and $\bar{\gamma}$ for functions with ``short'' branch cuts. The path $\gamma$ crosses the short cuts from above, corresponding to a counterclockwise continuation around the branch point at $u=+2$ and a clockwise continuation around the branch point at $u=-2$. The path $\bar{\gamma}$ follows the opposite convention.}
    \label{fig:analytic_cont_short}
\end{figure}

The analytic continuation paths $\gamma$ and $\bar{\gamma}$ for functions with a short branch cut along $(-2,2)$ are shown in figure~\ref{fig:analytic_cont_short}.

\section{Y and T systems for \adsst}
\label{sec:YT-system}

\subsection{The extended Y-system}
\label{sec:Y-system}

In part I of our work~\cite{Cavaglia:2026scj}, we derived the analytic Y-system that describes type IIB string theory on the \adsst\, background supported by Ramond-Ramond fluxes, and proved that it is equivalent to the ground state TBA of \cite{Frolov:2021bwp}.

\begin{figure}[htbp]
    \centering

    \begin{subfigure}{0.49\linewidth}
        \centering
        \includegraphics[width=\linewidth]{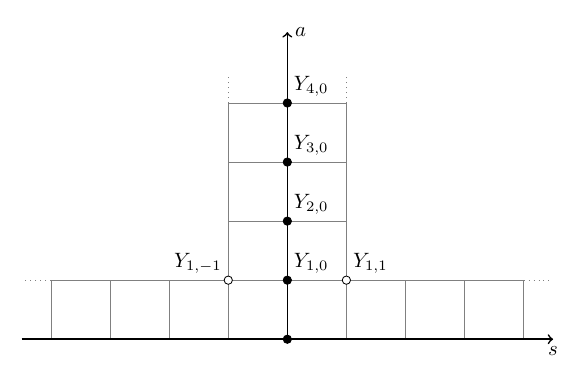}
    \end{subfigure}
    \hfill
    \begin{subfigure}{0.49\linewidth}
        \centering
        \includegraphics[width=\linewidth]{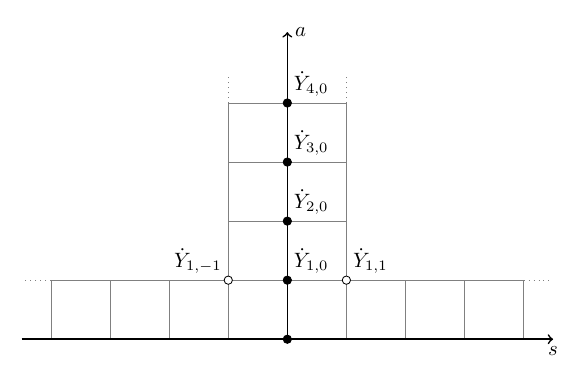}
    \end{subfigure}

    \caption{The two $\text{PSU}(1,1|2)$ T-hooks with the Y-functions drawn on top of them.}
    \label{fig:Thooks_with_Y}
\end{figure}

Starting from such a Y-system, we will provide the derivation of the Quantum Spectral Curve. 
Since this derivation hinges heavily on the T-hook diagrams, it is useful to switch to a more ``T-hook friendly'' notation with respect to the one used in \cite{Frolov:2021bwp} and in \cite{Cavaglia:2026scj}. 
Furthermore, the QSC has been proposed only in the $\mu\to0$ limit \footnote{By $\mu$ here we mean the twist parameter breaking supersymmetry introduced in~\cite{Frolov:2021bwp}. We stress that this is not the $\mu$ function appearing in the $\bP \bm{\mu}$-system.}, in which the protected vacuum serves as the reference state. Accordingly, from this point onward we set $\mu=0$ in the fermionic Y-functions. Thus, we relabel the Y-functions on the first T-hook (associated with left-massive particles and their excitations) as 
\begin{eqnarray}
&&Y_{a,0} \equiv Y_{n = a} , \\
&& Y_{1, 1} \equiv -1/Y_{+}^{(\alpha = 1)}, \;\;\; Y_{1, -1} \equiv -1/Y_{+}^{(\alpha = 2)} \, ,
\end{eqnarray}
and those on the second T-hook (associated with right-massive particles and their excitations) as
\begin{eqnarray}
&& \dot Y_{a,0} \equiv \dot{Y}_{{n} = a},\\
&& \dot{Y}_{1, 1} \equiv -Y_{-}^{(\alpha = 1)}, \;\;\; \dot{Y}_{1, -1} \equiv -Y_{{-}}^{(\alpha = 2)} \, .
\end{eqnarray}
In this notation, we list the Y functions of the Y-system, together with their branch points on the sheet with long cuts, in Table \ref{table:Yfunc}. We also depict the two $\text{PSU}(1,1|2)$ T-hooks in figure \ref{fig:Thooks_with_Y}.
\begin{table}[htbp]
    \centering
    \caption{The Y functions in the Y-system and their branch points.}
    \label{table:Yfunc}
    \renewcommand{\arraystretch}{1.5} % Adds a bit of vertical padding to rows
    \begin{tabular}{|c|l|}
        \hline
         Function &  Position of branch points \\ 
        \hline\hline
        \large $Y_{a,0}(u)$ & \multirow{2}{*}{\large \quad $u = \pm 2 + i\frac{n}{h}\,, \hspace{.5cm} n = \pm a, \pm (a+2), \pm (a+4), \dots$}  \\ 
        \cline{1-1}
        \large $\dot Y_{a,0}(u)$ & \\ 
        \cline{1-1}
        \hline
        \large $Y_0(u)$ & \large \quad $u = \pm 2 + i\frac{2n}{h}\,, \hspace{.5cm} n = 0, \pm 1, \pm 2, \dots$ \\ 
        \hline
        \large $Y_{1,\pm1}(u)$ & \large \quad $u = \pm 2 + i\frac{2n}{h}\,, \hspace{.5cm} n = 0, \pm 1, \pm 2, \dots$ \\ 
        \hline
        \large $\dot Y_{1,\pm1}(u)$ & \large \quad $u = \pm 2 + i\frac{2n}{h}\,, \hspace{.5cm} n = 0, \pm 1, \pm 2, \dots$ \\ 
        \hline
        \large $Y_{1,
        \pm1}(u)\;\dot Y_{1,\pm 1}(u)$ & \large \quad $u = \pm 2 + i\frac{2n}{h}\,, \hspace{.5cm} n = 0, \pm 1, \pm 2, \dots$
        \\ 
        \hline
        \large $Y_{1,
        \pm1}(u) / \dot Y_{1,\pm 1}(u)$ & \large \quad $u = \pm 2 + i\frac{2n}{h}\,, \hspace{.5cm} n =  \pm 1, \pm 2, \pm 3\dots$ \\\hline
        \end{tabular}
\end{table}

The functional equations that these Y functions must satisfy are given by the following $\text{PSU}(1,1|2)^{\otimes 2}$ Y-system:
\begin{eqnarray}
\label{def:Ysystem}
  Y_{a,0}^+    Y_{a,0}^- &=& \frac{\left( (1 + Y_{1,1} ) (1+ Y_{1,-1} ) \right)^{\delta_{a,1} }}{(1 + 1/Y_{a+1,0} )(1 + 1/Y_{a-1,0} )^{1-\delta_{a,1}} },\;\; a\geq 1,\\
  \label{def:Ysystem2}
  \dot{Y}_{a,0}^+    \dot{Y}_{a,0}^- &=& \frac{\left( (1 + \dot Y_{1,1} ) (1+ \dot Y_{1,-1} ) \right)^{\delta_{a,1} }}{(1 + 1/\dot Y_{a+1,0} )(1 + 1/\dot Y_{a-1,0} )^{1-\delta_{a,1}} },\;\; a\geq 1.
\end{eqnarray}
These equations can be derived from the mirror TBA equations for the ground state, as shown in~\cite{Frolov:2021bwp}. To obtain a closed set of functional relations, one must supplement the Y-system with discontinuity relations. We now recall these relations from~\cite{Cavaglia:2026scj}, grouped by the different classes of Y-functions; these expressions, derived from the ground state TBA, were conjectured to be universal.

\paragraph{Auxiliary Y functions.}

We begin with the auxiliary fermionic Y-functions.
They satisfy the following monodromy equations concerning the cut on the real axis:

\begin{eqnarray}
\label{eq:discYpmnew1}
&&\dot Y_{1,\pm1}^{\bar\gamma}= \frac{1}{Y_{1,\pm1}  \left(1 + Y^{\bar{\gamma}}_0\right)} \, ,\quad 
 Y_{1,\pm1}^{\bar\gamma}= \frac{1}{\dot Y_{1,\pm1}  \left(1 + Y^{\bar\gamma}_0\right)} \, ,\\
&&\dot Y_{1,\pm1}^\gamma= \frac{1}{Y_{1,\pm1}  \left(1 + Y_0\right)} \, ,\quad Y_{1,\pm1}^\gamma= \frac{1}{\dot Y_{1,\pm1}  \left(1 + Y_0\right)} \, .
\end{eqnarray}
These equations imply convenient relations for the product and ratio of $Y_{1,\pm1}$ and $\dot Y_{1,\pm1}$:
\bal
\label{eq:Yaux_product}
\left[\log\frac{\dot{Y}_{1,\pm1}}{{Y}_{1,\pm1}}\right]_0=0 ,\quad \left[\log{Y_{1,\pm1} \dot{Y}_{1,\pm1}}\right]_0=2\log(1 + Y_0 )+2\log \left( Y_{1,\pm1} \dot{Y}_{1,\pm1} \right)\,. 
\eal
Moreover, there are discontinuity relations describing  the other branch cuts, with imaginary part $\pm2N/h$, with $N\in \mathbb{N}^+$:
\begin{equation}
\label{eq:disc1}
\left[ \log Y_{1,\pm 1}\dot Y_{1,\pm 1}\right]_{\pm 2 N}=\sum^{N}_{a=1} \left[\log\left(1+\dot{Y}_{a,0}\right) \right]_{\pm(2N-a)}+\sum^{N}_{a=1} \left[\log\left(1+Y_{a,0}\right) \right]_{\pm(2N-a)} \,,
\end{equation}
\begin{equation}
\label{eq:disc2}
\left[ \log\frac{\dot Y_{1,\pm1}}{Y_{1,\pm1}}\right]_{\pm 2 N}=\sum^{N}_{a=1} \left[\log\left(1+\dot{Y}_{a,0}\right) \right]_{\pm(2N-a)}-\sum^{N}_{a=1} \left[\log\left(1+Y_{a,0}\right) \right]_{\pm(2N-a)} \,.
\end{equation}

\paragraph{Massless Y functions.}
For the massless Y function, we have the following symmetrised discontinuity on the real axis 
\bal
\label{Y0symmdisc}
\{\log{Y_0}\}_0=& \log{\dot Y_{1,1} \dot Y_{1,-1}}+2\log (1+Y_0)\\
&
+ \log\left(1+Y^\gamma_{1,1}\right)\left(1+Y^\gamma_{1,-1}\right)\left(1+\frac{1}{\dot Y^\gamma_{1,1}}\right)\left(1+\frac{1}{\dot Y^\gamma_{1,-1}}\right)\,.
\eal
The corresponding discontinuity across the same cut is then given by
\bal
\label{Y0disc}
[\log Y_0 ]_0 =&-\log{\dot Y_{1,1} \dot Y_{1,-1}}-2\log \left(1+\frac{1}{Y_0}\right)\\
&
- \log\left(1+Y^\gamma_{1,1}\right)\left(1+Y^\gamma_{1,-1}\right)\left(1+\frac{1}{\dot Y^\gamma_{1,1}}\right)\left(1+\frac{1}{\dot Y^\gamma_{1,-1}}\right)\,,
\eal
and  the discontinuity relations on the higher branch cuts at $\pm2N$ are
\bal
\label{eq:lnY0_disc_p2N}
[\log{Y_0}]_{+2N}= &- [\log Y_{1,\pm 1} \dot Y_{1,\pm 1}]_{+2N}+  \log {\dot Y_{1,1}}{Y_{1,1}}{\dot Y_{1,-1}}{Y_{1,-1}}+ 2 \log \left(1+Y_{0} \right)\,,\\
[\log{Y_0}]_{-2N}= & \left[ \log\frac{\dot Y_{1,\pm1}}{Y_{1,\pm1}}\right]_{- 2 N}
- \log {\dot Y_{1,1}}{Y_{1,1}}{\dot Y_{1,-1}}{Y_{1,-1}}- 2 \log \left(1+Y_{0} \right)\\
&+\left[ \log  \left(1+Y_{1,1} \right)\left(1+Y_{1,-1} \right) \left( 1+\frac{1}{\dot Y_{1,1}} \right)\left( 1+\frac{1}{\dot Y_{1,-1}} \right) \right]_{-2N}\,.
\eal

\paragraph{Massive Y functions.}
For the massive Y functions it is convenient to introduce the auxiliary quantities $\Delta_1$ and $\bar\Delta_1$, corresponding to the discontinuities of $Y_{1,0}$ and $\dot Y_{1,0}$ across the first long branch cut:
\beq
\label{def:Delta_1}
\Delta_1(u)\equiv [\log Y^+_{1,0}]_0,\quad \bar\Delta_1(u)\equiv[\log \dot Y^+_{1,0}]_0\,.
\eeq
In \cite{Cavaglia:2026scj}, a very simple expression was found for these quantities. Setting the twist parameter to zero and using the T-hook notation defined at the beginning of this section, we obtain
\bal
\label{eq:Delta1_local}
\Delta_1=&- \log Y_0  + \log  (1+Y_0) + \log \left(1+  Y_{1,1} \right) \left(1+  Y_{1,-1} \right) -c \,,\\
\bar{\Delta}_1=&- \log Y_0  + \log (1+Y_0) +  \log  \prod_{\a=1,2}\left(1+\dot{Y}_{1,1} \right) \left(1+\dot{Y}_{1,-1} \right)+c\,,
\eal
where $c$ is a constant (independent of $u$) taking the form:
\beq
\label{def:constant_c_final}
c=-\frac{1}{8}\left(\log{\frac{\dot Y_{1,1}}{Y_{1,1}}}( 2)+\log{\frac{\dot Y_{1,-1}}{Y_{1,-1}}}( 2)+\log{\frac{\dot Y_{1,1}}{Y_{1,1}}}(-2)+\log{\frac{\dot Y_{1,-1}}{Y_{1,-1}}}(-2)\right)\,.
\eeq

In the following derivation, it will also be useful to have at hand some simple consequences of these relations. From \eqref{def:Delta_1}, we immediately get the following identities for the discontinuities of $\Delta_1$ and $\bar{\Delta}_1$ at higher cuts with $N\in \mathbb{N}^+$:
\bal
\label{eq:D1_2N_disc_mag_sh_nonLR}
[\Delta_1]_{+ 2N} =& \left[ \log Y_{1,\pm1}\dot Y_{1,\pm1} \right]_{+ 2 N}-  \log Y_{1,1}\dot Y_{1,1}Y_{1,-1}\dot Y_{1,-1}- 2 \log \left(1+Y_{0} \right)\\
&+ \left[ \log  (1+Y_0) \right]_{+ 2N} + \left[ \log  \left(1+  Y_{1,1} \right)\left(1+  Y_{1,-1} \right) \right]_{+ 2N} \,,\\
[\Delta_1]_{-2N} =&  \left[ \log\frac{\dot Y_{1,\pm1}}{Y_{1,\pm1}}\right]_{- 2 N}
+ \log {\dot Y_{1,1}}{Y_{1,1}}{\dot Y_{1,-1}}{Y_{1,-1}}+ 2 \log \left(1+Y_{0} \right)\\
&-\left[ \log   \left( 1+\frac{1}{\dot Y_{1,1}} \right)\left( 1+\frac{1}{\dot Y_{1,-1}} \right) \right]_{-2N}+ \left[ \log  (1+Y_0) \right]_{- 2N} \,.
\eal
\bal
\label{eq:D1bar_2N_disc_mag_sh_nonLR}
[\bar\Delta_1]_{+ 2N} =&  \left[ \log Y_{1,\pm1}\dot Y_{1,\pm1} \right]_{+ 2 N}-  \log Y_{1,1}\dot Y_{1,1}Y_{1,-1}\dot Y_{1,-1}- 2 \log \left(1+Y_{0} \right)\\
&+ \left[ \log  (1+Y_0) \right]_{+ 2N} + \left[ \log  \left(1+  \dot Y_{1,1} \right)\left(1+  \dot Y_{1,-1} \right) \right]_{+ 2N} \,,\\
[\bar\Delta_1]_{-2N} =&  \left[ \log\frac{\dot Y_{1,\pm1}}{Y_{1,\pm1}}\right]_{- 2 N}
+ \log {\dot Y_{1,1}}{Y_{1,1}}{\dot Y_{1,-1}}{Y_{1,-1}}+ 2 \log \left(1+Y_{0} \right)\\
&-\left[ \log  \left(1+Y_{1,1} \right)\left(1+Y_{1,-1} \right)  \right]_{-2N}+[\log \dot Y_{1,1} \dot Y_{1,-1}]_{-2N}+ \left[ \log  (1+Y_0) \right]_{- 2N}  \,.
\eal
The following combinations of discontinuities on the real axis will be particularly useful in the subsequent analysis:
\bal
\label{eq:disc_massiveY_realaxis}
\{\Delta_1+\bar{\Delta}_1\}_0=&-2[\log{\dot Y_{1,1} \dot Y_{1,-1}}]_0-2[\log (1+Y_0)]_0\\
&+[\log{(1+Y_{1,1})(1+Y_{1,-1})(1+\dot Y_{1,1})(1+\dot Y_{1,-1})}]_0\,,\\
[\Delta_1-\bar{\Delta}_1]_0=&\left[\log \frac{(1+Y_{1,1}) (1+Y_{1,-1})}{(1+\dot{Y}_{1,1}) (1+\dot{Y}_{1,-1}) }\right]_0\,,\\
\{\Delta_1-\bar{\Delta}_1\}_0=&\left\{\log\frac{1+Y_{1,1}}{1+\dot Y_{1,1}}\frac{1+Y_{1,-1}}{1+\dot Y_{1,-1}}\right\}_0-4c\, ,\\
[\Delta_1+\bar{\Delta}_1]_0=&2[\log(1+Y_0^{-1})]_0+\log[(1+Y_{1,1})(1+Y_{1,-1})(1+\dot Y_{1,1})(1+\dot Y_{1,-1})]_0\,.
\eal
These identities follow directly from \eqref{eq:Delta1_local}. In deriving the first relation, we also made use of the explicit expression \eqref{Y0symmdisc} for the symmetrised discontinuity of $Y_0$.

Finally, we record the discontinuities of the massive Y functions across the cuts at $\Im(u)=-1/h$, which are the analogues of \eqref{eq:Delta1_local}:
\bal
\label{eq:disc_lnY1_m1_Tsection}
\left[ \log  Y_{1,0} \right]_{- 1}&=& \log Y_0   - \log  (1+Y_0) -  \log  \left(1+{Y_{1,1}} \right)^\gamma \left(1+{Y_{1,-1}} \right)^\gamma+ c \,,\\
[ \log  \dot Y_{1,0} ]_{- 1}&=& \log Y_0   - \log  (1+Y_0) -  \log  (1+{\dot Y_{1,1}} )^\gamma(1+{\dot Y_{1,-1}} )^\gamma -c \,.
\eal

\subsection{The T-system}
\label{sec:T-system}
\begin{figure}[htbp]
    \centering

    \begin{subfigure}{0.49\linewidth}
        \centering
        \includegraphics[width=\linewidth]{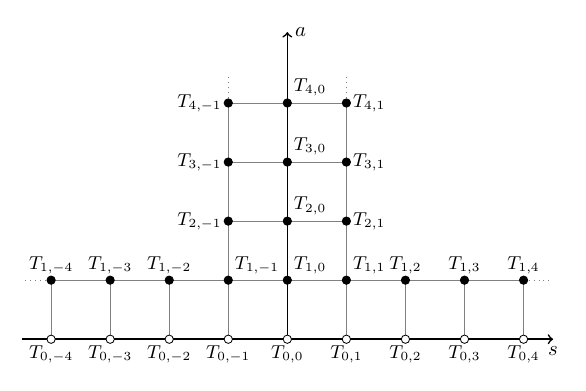}
    \end{subfigure}
    \hfill
    \begin{subfigure}{0.49\linewidth}
        \centering
        \includegraphics[width=\linewidth]{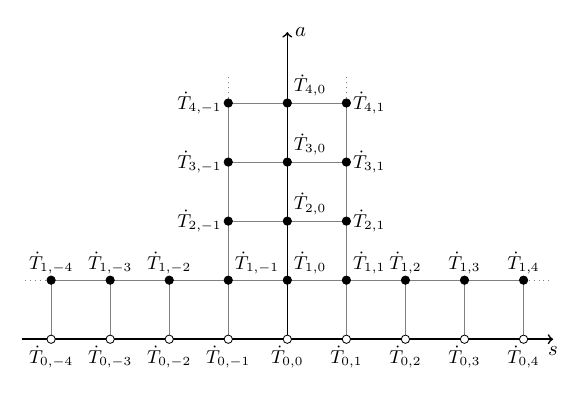}
    \end{subfigure}

    \caption{The two $\text{PSU}(1,1|2)$ T-hooks with the T-functions drawn on top of them.}
    \label{fig:Thooks_withT}
\end{figure}
Having recalled the Y-system and the associated discontinuity relations,
we now turn to the first step in the derivation of the Quantum Spectral
Curve, namely the introduction of the T-system associated with two
$\mathrm{PSU}(1,1|2)$ T-hooks. Each of these structures is closely
related to the integrable structure underlying the Hubbard-model
Quantum Spectral Curve~\cite{Cavaglia:2015nta,Ekhammar:2021pys}. We
denote by $T_{a,s}$ and $\dot T_{a,s}$ the T-functions of the undotted
and dotted T-hooks, respectively. The non-vanishing nodes of the undotted T-hook are
\begin{eqnarray}
&&T_{0,s}\;, \;\;\;\; -\infty\leq s\leq +\infty,\\
&&T_{1,s}\;, \;\;\;\; -\infty\leq s\leq +\infty,\\
&& T_{a,s}\;, \;\;\; -1 \leq s \leq +1 , \;\; a\geq 2 .
\end{eqnarray}
The dotted T-hook is defined analogously. The two T-hooks are shown in figure~\ref{fig:Thooks_withT}. On each of them, the T-functions satisfy the Hirota bilinear equations
\begin{align}
\label{def:Hirotaeq}
T^+_{a,s} T^-_{a,s} &= T_{a,s+1} T_{a,s-1} +T_{a+1,s} T_{a-1,s}\, ,\\
\label{def:Hirotaeq2}
\dot{T}^+_{a,s} \dot{T}^-_{a,s} &= \dot{T}_{a,s+1} \dot{T}_{a,s-1} +\dot{T}_{a+1,s}\dot{ T}_{a-1,s}\, .
\end{align}
These equations are supplemented by the boundary conditions
\beqa
&T_{a,s}=0\; \text{for } a<0 \text{ and for all } s,\quad \quad T_{a,s}=0\;\text{for } |s|>1 \text{ and }a>1\,,\\
&\dot T_{a,s}=0\; \text{for } a<0 \text{ and for all } s,\quad \quad \dot T_{a,s}=0\;\text{for } |s|>1 \text{ and }a>1\,.
\eeqa
The Y functions introduced in the previous section are recovered from the T-functions through
\begin{eqnarray}
\label{def:YsasTs}
Y_{a,s} = \frac{T_{a, s+1} T_{a,s-1}}{T_{a+1, s} T_{a-1,s}} ,\quad \dot Y_{a,s} = \frac{\dot T_{a, s+1} \dot T_{a,s-1}}{\dot T_{a+1, s} \dot T_{a-1,s}}\,.   
\end{eqnarray}
This parametrisation is well defined for all nodes with four nearest neighbours in the T-hook, so that no singular denominators arise. Moreover, the Hirota equations~\eqref{def:Hirotaeq}, \eqref{def:Hirotaeq2} imply that the functions defined in \eqref{def:YsasTs} automatically satisfy the Y-system equations~\eqref{def:Ysystem}, \eqref{def:Ysystem2}.

A key feature of the parametrisation \eqref{def:YsasTs} is that it does not uniquely determine the T-functions. Instead, they are defined only up to the gauge transformation

\beq
 \label{def:gauge_transf_T}
T_{a,s}\rightarrow g_1^{[a+s]}g_2^{[a-s]}g_3^{[-a+s]}g_4^{[-a-s]}T_{a,s}\,,
 \eeq
where $g_i(u)$ are arbitrary functions. These gauge transformations leave the Hirota equations \eqref{def:Hirotaeq} invariant and, furthermore, it is easy to see from \eqref{def:YsasTs} that Y functions are also invariant under  them. In the following sections, we will exploit this freedom to construct gauges with analyticity properties adapted to the different steps of the derivation of the Quantum Spectral Curve.

\begin{framed}
\textbf{Convention.}
Unless stated otherwise, we take all shifts in the T-system on the mirror Riemann section. Equivalently, all branch cuts are regarded as long, extending through infinity, so that for $\Re(u)\in(-2,2)$ no long branch cut is crossed when performing a shift.
\end{framed}

\section{Discontinuities of T-functions}
\label{sec:Tgauges}

In this section, we will use the discontinuities of the Y functions listed in section \ref{sec:Y-system} to derive analytical properties of the T-functions. This will lead to the existence of two distinguished gauge choices, denoted respectively by
$\bT$ and $\wT$ (and their dotted version),  where the analytic properties manifest very simply. This will lay the foundation for the derivation of the QSC in the next section.

\paragraph{The LR symmetric sector.}In this paper, we will assume an additional symmetry in each T-hook, which we can call Left-Right (LR) symmetry in analogy with the AdS$_5$ terminology. This means that at the level of T-functions, we are only considering states for which
\beq 
\label{def:LRsymmetry}
T_{a,s}=T_{a,-s}\quad  \text{ and } \quad \dot T_{a,s}=\dot T_{a,-s}\,,\eeq
valid for any gauge choice. We leave the analysis of non-LR symmetric states for future work.

In contrast, we do not assume the invariance of the solution under the swap of the two PSU(1,1|2) T-systems, which corresponds to the \emph{chiral symmetry} introduced in \cite{Cavaglia:2026scj}: our derivation covers states that are not invariant under this transformation. 
\paragraph{Vertical-friendly gauges. } 
We start our analysis by assuming that we are in a gauge where T-functions, denoted by $\mathcal{T}_{a,s}$ and $\dot{\mathcal{T}}_{a,s}$, have analyticity strips \emph{as wide as possible} in the vertical part of the T-hooks.
The maximal
 width of these analyticity strips is dictated by the analyticity strips of the Y functions.
 It is a feature of T-hooks, first observed in \cite{Gromov:2011cx}, that we cannot have maximal analyticity strips for all T-functions at the same time. For this reason, here we impose these maximal analyticity strips only along the vertical band, demanding
\begin{eqnarray}\label{eq:vertstrip}
\mathcal{T}_{a \geq 0, 0} \in \mathcal{A}_{a+1} , \;\;\; \mathcal{T}_{a \geq 1, \pm 1} \in \mathcal{A}_{a},
\end{eqnarray}
where $\mathcal{A}_k$ is  the set of functions analytic in the strip $| \text{Im}(u) | < k/h$. We believe that, given the analyticity strips of Y functions, it would be possible to give a constructive proof for the existence of a family of gauges satisfying (\ref{eq:vertstrip}). However, here we simply take this as given: indeed this is a very natural assumption, suggested by the analogy with the AdS$_5$ case~\cite{Gromov:2011cx}.

Starting from any gauge $\mathcal{T}$ satisfying these conditions, in the next section we will build the two special gauges $\bT$, $\wT$, whose properties are determined by exploiting the discontinuities of the Y functions written in section \ref{sec:Y-system}. These gauges will enjoy similar analytical properties to those found in the T-systems of AdS$_5$ \cite{Gromov:2011cx} and AdS$_4$ \cite{Cavaglia:2013lgg}.

\subsection{The `black' gauge $\bT$ and its properties}
\label{sec:black_gauge_LR_sym}

Starting from any gauge $\mathcal{T}$ characterised by having the analyticity strips in~\eqref{eq:vertstrip}, we can take the discontinuities of the auxiliary Y functions on the cuts at $\pm 2N$ \eqref{eq:disc1} and \eqref{eq:disc2} and translate them into T-functions using \eqref{def:YsasTs}. 

Let us perform this calculation for equation \eqref{eq:disc1} explicitly for the cuts at $-2N$.
After using the map \eqref{def:YsasTs} and Hirota equations \eqref{def:Hirotaeq}, the discontinuity relation becomes
\beq
\label{eq:disc_Yaux_as_T1}
\left[ \log \frac{\dot{\mT}_{1,2} \dot{\mT}_{1,0} \mT_{1,2} \mT_{1,0} }{\dot{\mT}_{0,1} \dot{\mT}_{2,1} \mT_{0,1} \mT_{2,1}} \right]_{-2N} = \sum^N_{n=1} \left[ \log \frac{\dot{\mT}^+_{n,0} \dot{\mT}^-_{n,0} \mT^+_{n,0} \mT^-_{n,0} }{\dot{\mT}_{n+1,0} \dot{\mT}_{n-1,0} \mT_{n+1,0} \mT_{n-1,0}} \right]_{-2N+n}\,.
\eeq
Writing the sum explicitly, it is easy to see that
most of the terms telescope, and we end up with
\bal
\text{RHS} &= \left[ \log \frac{ \dot{\mT}_{1,0} \mT_{1,0} \dot{\mT}^{N+1}_{N,0} \mT^{N+1}_{N,0} }{ \dot{\mT}^+_{0,0} \mT^+_{0,0} \dot{\mT}^{N}_{N+1,0} \mT^{N}_{N+1,0}} \right]_{-2N} \,.
\eal
Because of the analyticity strips of the T-functions in this gauge, the functions $\mT_{N,0}^{N+1}$ do not have a branch cut at $\Im(u)=-2N$ in either wing, and thus the expression above is equal to
\bal
\text{RHS} &= \left[ \log \frac{ \dot{\mT}_{1,0} \mT_{1,0}  }{ \dot{\mT}^+_{0,0} \mT^+_{0,0} } \right]_{-2N} \,.
\eal
Plugging this formula into the RHS of~\eqref{eq:disc_Yaux_as_T1}, we obtain
\bal
\label{eq:disc_Yaux_as_T2}
\left[ \log \frac{\dot{\mT}_{1,2}  \mT_{1,2} \dot{\mT}^+_{0,0} \mT^+_{0,0} }{\dot{\mT}_{0,1} \dot{\mT}_{2,1} \mT_{0,1} \mT_{2,1}} \right]_{-2N} = 0 \,. 
\eal
An analogous calculation can be repeated for all the discontinuities in \eqref{eq:disc1} and \eqref{eq:disc2}, giving the following four sequences of identities:
\bal
\label{eq:Tdisc_left_from_Yaux_disc}
&\left[  \log\frac{\mT_{1,2} \, \mT_{0,0}^{\mp}  }{ \mT_{2,1} \, \mT_{0,1}   } + \log\frac{\dot{\mT}_{1,2} \, \dot{\mT}_{0,0}^{\mp}  }{ \dot{\mT}_{2,1} \, \dot{\mT}_{0,1}   } \right]_{\pm 2 N} = 0 ,\;N = 1,2,3,4 ,\dots \\
&\left[  \log\dfrac{\mT_{1,2} \, \mT_{0,0}^{\mp}  }{ \mT_{2,1} \, \mT_{0,1}   } -\log\dfrac{\dot{\mT}_{1,2} \, \dot{\mT}_{0,0}^{\mp}  }{ \dot{\mT}_{2,1} \, \dot{\mT}_{0,1}   } \right]_{\pm 2 N} = 0 ,\;N = 1,2,3,4 ,\dots .
\eal
These conditions imply that we can define the following four functions (two for each T-hook):
\beqa
\label{def:smallbpm}
&&\mathbf{b}_{\uparrow}(u) \equiv \frac{\mT_{1,2}(u) \, \mT_{0,0}^{+}(u)  }{ \mT_{2,1}(u) \, \mT_{0,1}(u)   } , \qquad \mathbf{b}_{\downarrow}(u) \equiv \frac{\mT_{1,2}(u) \, \mT_{0,0}^{-}(u)  }{ \mT_{2,1}(u) \, \mT_{0,1}(u)   }~,
\\
\label{def:smalldbpm}
&&\dot{\mathbf{b}}_{\uparrow}(u) \equiv \frac{\dot{\mT}_{1,2}(u) \, \dot{\mT}_{0,0}^{+}(u)  }{ \dot{\mT}_{2,1}(u) \, \dot{\mT}_{0,1}(u)   } \,, \qquad
\dot{\mathbf{b}}_{\downarrow}(u) \equiv \frac{\dot{\mT}_{1,2}(u) \, \dot{\mT}_{0,0}^{-}(u)  }{ \dot{\mT}_{2,1}(u) \, \dot{\mT}_{0,1}(u)   } \,,
\eeqa
where $\mathbf{b}_{\uparrow}$ and $\dot{\mathbf{b}}_{\uparrow}$ are analytic in the lower half-plane (LHPA), and $\mathbf{b}_{\downarrow}$ and $\dot{\mathbf{b}}_{\downarrow}$ are analytic in the upper half-plane (UHPA).

\paragraph{The change of gauge to $\bT$.}
{Using the analyticity properties of $\mathbf{b}_{\uparrow },\,\mathbf{b}_{\downarrow}$ and $\mathbf{\dot{b}}_{\uparrow},\,\mathbf{\dot{b}}_{\downarrow}$, it is possible to show that there is a special gauge where the functions in~\eqref{def:smallbpm} and~\eqref{def:smalldbpm} are equal to $1$, while the analytic properties of all the T-functions are preserved. Let us show how to obtain this gauge transformation:}
we use the functions \eqref{def:smallbpm} to define two ``change-of-gauge'' functions $F_{\downarrow}, \,\dot{F}_{\downarrow}$ through the Upper-Half Plane Analytic (UHPA) solution of the finite difference equations
\beq
\label{def:Fdown_eq}
\left(F_{\downarrow}^+/F_{\downarrow}^-\right)^{2} \equiv \mathbf{b}_{\downarrow}(u) ,\;\;\;\; \left(\dot{F}^+_{\downarrow}/\dot{F}_{\downarrow}^-\right)^{2} \equiv \dot{\mathbf{b}}_{\downarrow}(u) .
\eeq
We also define $F_{\uparrow}, \,\dot{F}_{\uparrow}$ through the Lower-Half Plane Analytic (LHPA) solution of another difference equation:
\beq
\label{def:Fup_eq}
\left(F_{\uparrow}^+/F_{\uparrow}^-\right)^{2} \equiv \mathbf{b}_{\uparrow}^{-1}(u) ,\;\;\;\; \left(\dot{F}_{\uparrow}^+/\dot{F}_{\uparrow}^-\right)^{2} \equiv \dot{\mathbf{b}}_{\uparrow}^{-1}(u) ,
\eeq
{where we demand that the solutions have non-exponential asymptotics at infinity, and are analytic for $\Im(u) > -1/h$ and $\Im(u) < 1/h$ respectively. This fixes them up to an unimportant multiplicative constant.}

Then the gauge $\bT$ is defined via the functions $F$ as follows:
\beqa
\label{def:bT_gauge}
&&\bT_{a,s} \equiv F_{\downarrow}^{[a+s]} F_{\downarrow}^{[a-s]} {F}_{\uparrow}^{[-a+s]}  {F}_{\uparrow}^{[-a-s]} \mathcal{T}_{a,s}\,,\\&& \dot \bT_{a,s} \equiv \dot{F}_{\downarrow}^{[a+s]} \dot{F}_{\downarrow}^{[a-s]} {\dot{F}}_{\uparrow}^{[-a+s]}  {\dot{F}}_{\uparrow}^{[-a-s]} \dot{\mathcal{T}}_{a,s}\, .
\eeqa
This transformation has the same form as \eqref{def:gauge_transf_T}, and thus by definition preserves the Hirota equations \eqref{def:Hirotaeq}.

The T-functions in the $\bT$-gauge have the same analyticity strips as the T-functions in any $\mT$-gauge, defined in \eqref{eq:vertstrip}. In fact, since $F_\downarrow$ is analytic above $-i/h$ and $F_\uparrow$ is analytic below $i/h$, the branch cuts of these functions do not change the analyticity strips of the T-functions \eqref{eq:vertstrip} due to shifts found in \eqref{def:bT_gauge}. Thus we have that:
\begin{eqnarray}\label{eq:vertstrip_balckT}
\bT_{a \geq 0, 0} \in \mathcal{A}_{a+1} , \;\;\; \bT_{a \geq 1, \pm 1} \in \mathcal{A}_{a}.
\end{eqnarray}
\paragraph{Group theoretical properties.}
The properties  making the  gauge $\bT$ 
 special can be deduced from the following identities
\beqa
\label{eq:black_gauge_identities}
\frac{\bT_{1,2}(u) \, \bT_{0,0}^{-}(u)  }{ \bT_{2,1}(u) \, \bT_{0,1}(u)   } &=& \frac{\dot{\bT}_{1,2}(u) \, \dot{\bT}_{0,0}^{-}(u)  }{ \dot{\bT}_{2,1}(u) \, \dot{\bT}_{0,1}(u)   } = 1 ,\\
\label{eq:black_gauge_identities2}\frac{\bT_{1,2}(u) \, \bT_{0,0}^{+}(u)  }{ \bT_{2,1}(u) \, \bT_{0,1}(u)   } &=& \frac{\dot{\bT}_{1,2}(u) \, \dot{\bT}_{0,0}^{+}(u)  }{ \dot{\bT}_{2,1}(u) \, \dot{\bT}_{0,1}(u)   } = 1 ,
\eeqa
which can be verified immediately by changing to the previous gauge and using the difference equations defining $F_{\downarrow}$, $\dot{F}_{\downarrow}$, $F_{\uparrow}$, $\dot{F}_{\uparrow}$ in~\eqref{def:Fdown_eq} and~\eqref{def:Fup_eq}.

Now, taking the ratio of the LHS of the two equations in  \eqref{eq:black_gauge_identities} and \eqref{eq:black_gauge_identities2}, we find
\beq
\bT_{0,0}^+=\bT_{0,0}^-\, ,
\eeq
while from the ratio of the second terms in the same equations we have the same identity on the dotted wing. Thus, we have found that in the $\bT$ gauge there are two special quantities, $\bT_{0,0}$ and $\dot\bT_{0,0}$, that are periodic on the sheet with mirror cuts, and we call them $\mu$ and $\dot\mu$: \footnote{To avoid confusion, we stress that $\mu$ is not the twist parameter appearing in the TBA equations defined in~\cite{Frolov:2021bwp}, which we set to zero. From now on, the label $\mu$ always refers to the mirror periodic function defined by~\eqref{eq:mu_first_def}.}
\beq
\label{eq:mu_first_def}
\mu \equiv \mathbf{T}_{0,0}^+ \,,\;\;\;\;\dot\mu \equiv \dot{\mathbf{T}}_{0,0}^+\,.
\eeq
These functions, which are mirror periodic, will become part of the so-called $\bP \bm{\mu}$-system formulation of the QSC.

From this fact and the Hirota equations \eqref{def:Hirotaeq}, we can derive the remaining properties of the $\bT$ gauge. First, we consider the Hirota equations for $(a,s)=(0,0)$ in each T-hook, which thanks to LR symmetry \eqref{def:LRsymmetry} are:
\beq
\bT^+_{0,0} \bT^-_{0,0} = \bT_{0,1}^2 ,\quad \dot\bT^+_{0,0} \dot\bT^-_{0,0} = \dot\bT_{0,1}^2 .
\eeq
Using the mirror periodicity property for $\bT_{0,0}^+\equiv \mu$ and for $\dot\bT_{0,0}^+\equiv \dot\mu$, we get that:\footnote{The plus sign in these identities can always be fixed since, even if we had $\bT_{0,1} = -\mu$, we could use the residual gauge freedom $\bT_{a,s}\rightarrow (-1)^s \bT_{a,s}$ to change the sign. We choose the plus sign, which is consistent with the character limit of the T-functions, i.e. the fact that at large $u$ the T-functions must become characters of \psu. 
}
\beq \bT_{0,1}=\mu\,,\quad \dot\bT_{0,1}=\dot\mu\,.\eeq
Furthermore, we can plug the identities we have just derived in \eqref{eq:black_gauge_identities}, obtaining
\beq
\label{eq:group_theoretical}
\bT_{1,2}  =  \bT_{2,1} \,,\quad \dot\bT_{1,2}  =  \dot\bT_{2,1} \,.
\eeq
Finally, from the mirror periodicity of the $\mu$ functions and \eqref{eq:group_theoretical}, we can use the Hirota equations to derive the following identities, 
which in analogy with \cite{Gromov:2011cx} we call  \emph{group-theoretical properties}:
\beqa
\label{eq:group_theoretical2}
\bT_{1,n\geq 2} &=  \bT_{n\geq 2,1}  ,\;\;\;\;&\dot\bT_{1,n \geq 2} =  \dot\bT_{n \geq 2,1}  , \\
\label{eq:group_theoretical2bis}
 \bT_{0, n}   &= \bT_{0,0}^{[+n]},\;\;\;\; &\dot\bT_{0, n}   = \dot\bT_{0,0}^{[+n]}.
\eeqa
\paragraph{The discontinuity of $\mu$.}
 We now want to compute the discontinuity of $\mu$ and $\dot{\mu}$ on the real axis;  this is the only independent discontinuity for these functions, since the mirror periodicity implies that $[\log \mu]_{\pm 2N}=[\log \mu]_0$. In the $\bT$ gauge, using the parametrisation \eqref{def:YsasTs} and the group-theoretical properties \eqref{eq:group_theoretical2}, we have that
\beq
\label{def:Y11prod_ratio}
Y_{1,1} \dot{Y}_{1,1} = \frac{ \bT_{1,0}\dot{\bT}_{1,0}}{
\bT_{0,1} \dot{\bT}_{0,1}}= \frac{ \bT_{1,0}\dot{\bT}_{1,0}}{
\mu \dot\mu}\,,\qquad \frac{Y_{1,1}}{\dot Y_{1,1}}=\frac{\bT_{1,0}\,\dot \bT_{0,1}}{\dot\bT_{1,0}\, \bT_{0,1}}=\frac{\bT_{1,0}}{\dot\bT_{1,0}}\frac{\dot \mu}{\mu}.
\eeq
Since the quantities $\bT_{1,0}$, $\dot \bT_{1,0}$ and $Y_{1,1}/\dot{Y}_{1,1}$ do not have cuts on the real axis, by taking the discontinuity on the real axis of \eqref{def:Y11prod_ratio} we get:
\beq
\label{eq:disc_mu_andY1}
\left[\log{Y_{1,1} \dot{Y}_{1,1} }\right]_0 = -\left[\log \mu \dot{\mu}  \right]_0 \,,\qquad [\log \mu]_0=[\log \dot \mu]_0.
\eeq
The second equation implies that the ratio of $\mu$ and $\dot\mu$  does not have branch points in the whole complex plane. This ratio is a periodic function, with power-like asymptotics at infinity, \footnote{All the Y functions in the TBA have this type of asymptotic behaviour at infinity. We can then construct T-functions with the same property. } which could in principle have zeros and poles; however, we will make the minimal assumption that no such zeros or poles are present, so that this function must in fact be a constant. Furthermore, we are always free to rescale the T-functions in each wing by an arbitrary constant. Using this freedom, we can impose that:
\beq 
\label{eq:equal_mu}
\mu=  \dot \mu\,.
\eeq
The first equation in \eqref{eq:disc_mu_andY1} then becomes
\beq
\label{eq:disc_Yaux_realaxis}
[\log Y_{1,1}]_0=[\log \dot Y_{1,1}]_0=-[\log \mu]_0\,. \eeq 
The last identity we can obtain, which will play a fundamental role in the next section, comes from using \eqref{eq:disc_mu_andY1} in the second equation in~\eqref{eq:Yaux_product}:
\beq\label{eq:magic}
2 \log(1 + Y_0) + 2 \log( Y_{1,1} \dot{Y}_{1,1} ) = -2 [\log \mu ]_0 .
\eeq

\subsection{The `white' gauge $\wT$ and its properties}
\label{sec:white_T_gauge}
The gauge $\wT$ is obtained via a gauge transformation applied to the $\bT$ gauge, defined as follows:
\beq
\mathbb{T}_{a,s} = \mathbf{T}_{a,s} (\mu^{[a+s-1]})^{a-1},\;\;\;\;\dot{\mathbb{T}}_{a,s} = \dot{\mathbf{T}}_{a,s} (\mu^{[a+s-1]})^{a-1}.
\eeq
This is a gauge transformation since it leaves Y functions invariant, thanks to the mirror periodicity of the $\mu$ functions under a shift of $i/h$; furthermore, it can be easily proven that the T-system (i.e. the Hirota equations \eqref{def:Hirotaeq}) is invariant under it.

From its definition, we see immediately that in the $\wT$-gauge the T-functions on the bottom row of the T-hooks are trivial: 
\beq\label{eq:cornerT}
\wT_{0,s} = 1,\qquad \dot\wT_{0,s} = 1.
\eeq
Moreover, given the group-theoretical properties \eqref{eq:group_theoretical2}-\eqref{eq:group_theoretical2bis}, in this gauge we have that: 
$$\mathbb{T}_{1,n} =\mathbf{T}_{1,n}  = \mathbf{T}_{n,1}\,,\;\;\;\dot{\mathbb{T}}_{1,n} =\dot{\mathbf{T}}_{1,n}  = \dot{\mathbf{T}}_{n,1}\,.$$
A consequence of this is that in this new gauge the $\wT_{1,s}$ functions, moving along the horizontal arms of the T-hook, possess wide analyticity strips: 
\beq \wT_{1,  s\geq 1} \in \mathcal{A}_{s}\,,\;\;\;\;\dot{\wT}_{1,  s\geq 1} \in \mathcal{A}_{s}\,.
\eeq
In contrast, the $\wT_{a\geq 2,1}$ functions in the vertical part of the T-hook do not have wide analyticity strips anymore, since by definition we have that 
\beq
\mathbb{T}_{a,1}=\bT_{a,1}(\mu^{[a]})^{a-1}\,,\,\,\dot\wT_{a,1}=\dot\bT_{a,1}(\mu^{[a]})^{a-1}\,,\eeq
and the factors $\mu$ on the RHS create an infinite ladder of branch cuts at $
\Im(u)=\pm 2 \frac{N}{h},\;N=0,1,2\dots$.

\paragraph{The magic sheet.}The most important property of the $\wT$ gauge for our construction is that the T-functions in the horizontal bands of the two T-hooks $\wT_{1,s\geq1},\,\dot\wT_{1,s\geq1}$ possess a Riemann sheet where they only have four branch points, which we connect using short branch cuts. This Riemann sheet corresponds to the physical kinematics and was dubbed the ``magic sheet'' in the AdS$_5$ literature \cite{Gromov:2011cx, Gromov:2013pga}.

In particular, we can define the T-functions in the magic sheet $\hat{\mathbb{T}}_{1,s}$ (for $s=1$ and $s=2$) as seen in section \ref{sec:notations}, i.e. they are defined to agree with the original function around the real axis, and outside this strip they are defined via analytic continuation while keeping the branch cuts at $\Im(u) = \pm  s/h$ short. Explicitly, this gives
\begin{equation}
\label{eq:def_T1s_seq12}
\hat{\mathbb{T}}_{1,s\geq 1 }(u) \equiv
\begin{cases}
\mathbb{T}_{1,s} \,\qquad\qquad\qquad\;\;\,\; \hspace{7mm} \text{if} \ -s/h<\Im (u) < + s/h \,,\\
 \wT^{\gamma_+}_{1,s} \equiv ((\wT^{[+s]})^{\gamma} )^{[-s]}\, \hspace{7mm} \text{if} \ \Im (u) >s/h , \\
  \wT^{\gamma_-}_{1,s} \equiv ((\wT^{[-s]})^{\bar{\gamma}} )^{[+s]}\, \hspace{7mm} \text{if} \ \Im (u) <-s/h \,. 
\end{cases}
\end{equation} 
{A crucial property is that, in this Riemann section, these functions do not have other branch cuts beyond the ones closest to the real axis, i.e. the functions $\hat\wT_{1,s}, \,\hat{\dot \wT}_{1,s}$  have only two short branch cuts, at $(-2 \pm \frac{i s}{h},+2\pm\frac{i s}{h} )$. 
We prove these properties in appendix \ref{app:magic_sheet_LRsymm}, where we show explicitly that this holds for $\hat{\wT}_{1,1}$ and $\hat{\wT}_{1,2}$. Importantly, this encodes the information contained in the discontinuity relations \eqref{eq:D1_2N_disc_mag_sh_nonLR}.

{
Once $\hat{\mathbb{T}}_{1,1}$ and $\hat{\mathbb{T}}_{1,2}$ are given, we can define $\hat{\wT}_{1,s\geq 3}$ recursively via the Hirota equations, which, thanks to the wide analyticity strips around the real axis, also hold in this section:
\begin{equation}
\label{eq:Hirota_for_hat_T}
\hat{\mathbb{T}}_{1,s}= \frac{\hat{\mathbb{T}}^+_{1,s-1} \hat{\mathbb{T}}^-_{1,s-1}}{\hat{\mathbb{T}}_{1,s-2}} \,, \qquad s=3, \, 4, \, \dots \, .
\end{equation}
From this expression, we see that all these functions have only two short branch cuts.

\subsection{Connecting the two T-hooks }
So far, the properties of the gauges $\bT$ and $\wT$ (including the important magic sheet property of the latter) have utilised the discontinuities of the auxiliary Y functions \eqref{eq:Yaux_product}, \eqref{eq:disc1}, \eqref{eq:disc2} and those of the quantities $\Delta_1$ and $\bar{\Delta}_1$ (related to the massive Y functions) on the cuts lower and upper halves of the complex plane~\eqref{eq:D1_2N_disc_mag_sh_nonLR}, \eqref{eq:D1bar_2N_disc_mag_sh_nonLR}.
However, we still have further discontinuities to impose, namely those of massless Y functions, and those of $\Delta_1$ and $\bar{\Delta}_1$ on the real axis. In this section, we show that the latter discontinuities, found in equations \eqref{eq:disc_massiveY_realaxis}, impose non-trivial relations among the T-functions in the two T-hooks $\hat\wT_{1,s}$ and $\hat{\dot\wT}_{1,s}$.

In order to see this, we need to rephrase \eqref{eq:disc_massiveY_realaxis} in terms of T-functions. 
We start from the first equation, and after continuing it along $\bar\gamma$, we obtain, after using \eqref{eq:discYpmnew1} and some rearranging:
\beqa
\label{eq:deltaplusdeltabar_real}
(\{\Delta_1+\bar{\Delta}_1\}_0)^{\bar\gamma}=&&+4\log{\frac{\dot Y_{1,1}}{\dot Y_{1,1}^{\bar\gamma}}}+\log{\frac{\dot Y_{1,1}^{\bar\gamma}}{\dot Y_{1,1}^\gamma}}+\log{\frac{ Y_{1,1}^{\bar\gamma}}{ Y_{1,1}^\gamma}}
\\
&&+2\log{(1+Y_{1,1}^{\bar\gamma})(1+\dot Y^{\bar\gamma}_{1,1})}-2\log{(1+ Y_{1,1})(1+\dot Y_{1,1})}\,.
\eeqa
First, we want to write the RHS of \eqref{eq:deltaplusdeltabar_real} in terms of T-functions; thanks to the identity~\eqref{eq:disc_Yaux_realaxis}, and using the fact that $\bT_{2,1}=\bT_{1,2}$ does not have a cut on the real axis, we get
\beq
\label{eq:delta1adisc_gammabar2}
(\{\Delta_1+\bar{\Delta}_1\}_0)^{\bar\gamma}=2\log\mu^{\gamma}-2\log{\mu^{\bar\gamma}}+2\left(\left[\log{{\bT_{1,1}^+\bT_{1,1}^-}}\right]_0\right)^{\bar\gamma}+2\left(\left[\log{{\dot\bT_{1,1}^+\dot\bT_{1,1}^-}}\right]_0\right)^{\bar\gamma}\,.
\eeq
Then, we express the LHS of \eqref{eq:deltaplusdeltabar_real} in terms of T-functions, obtaining
\beq
(\{\Delta_1+\bar{\Delta}_1\}_0)^{\bar\gamma}\equiv\log{(Y_{1,0}^+)^{\bar\gamma}}-\log{(Y_{1,0}^+)^{\gamma}}+\log{(\dot Y_{1,0}^+)^{\bar\gamma}}-\log{(\dot Y_{1,0}^+)^{\gamma}},
\eeq
which can be rewritten, plugging in $Y_{1,0} = \bT_{1,1}^2/(\mu^+ \bT_{2,0})$, as
\beq\label{eq:delta1adisc_gammabar3}
2 \log{(\bT_{1,1}^+\dot \bT_{1,1}^+)^{\bar\gamma}}-2\log{(\bT_{1,1}^+\dot \bT_{1,1}^+)^{\gamma}}+2\log \mu^{\gamma}-2\log{\mu^{\bar\gamma}}\,,
\eeq
with the quantities $(\bT_{2,0}^+)^{\gamma}$ and $(\bT_{2,0}^+)^{\bar{\gamma}}$ cancelling out due to the analyticity strips of this function. 
Thus, equating \eqref{eq:delta1adisc_gammabar2} with \eqref{eq:delta1adisc_gammabar3} we find:
\beq
\label{eq:massless_constraint}
2([\log \bT_{1,1}^-]_0)^{\bar\gamma}-2[\log \bT_{1,1}^+]_0+2([\log \dot\bT_{1,1}^-]_0)^{\bar\gamma}-2[\log \dot\bT_{1,1}^+]_0=0\,.
\eeq
For the second equation of \eqref{eq:disc_massiveY_realaxis}, we can express its RHS directly in terms of T-functions, getting
\beq
[\Delta_1-\bar{\Delta}_1]_0=2\left[\log{\frac{1+Y_{1,1}}{1+\dot{Y}_{1,1}}}\right]_0=2[\log{\bT_{1,1}^+\bT_{1,1}^-}]_0-2[\log{\dot\bT_{1,1}^+\dot\bT_{1,1}^-}]_0\,.
\eeq
On the other hand, the  LHS in terms of T-functions is
\beqa
[\Delta_1-\bar{\Delta}_1]_0&=&\log{Y_{1,0}^+}-2\log{(Y_{1,0}^+)^{\gamma}}+\log{(Y_{1,0}^+)^{\gamma\gamma}}-\log{\dot Y_{1,0}^+}+2\log{(\dot Y_{1,0}^+)^{\gamma}}-\log{(\dot Y_{1,0}^+)^{\gamma\gamma}}\nonumber\\
&=& 2 \log{\bT_{1,1}^+\over \dot \bT_{1,1}^+}-4\log{\left(\bT_{1,1}^+\over \dot \bT_{1,1}^+\right)^{\gamma}}+2\log{\left(\bT_{1,1}^+\over \dot \bT_{1,1}^+\right)^{\gamma\gamma}}\,.
\eeqa
Equating the two definitions, we find
\beq
\label{eq:massless_disc2}
2\left[\log \frac{\bT_{1,1}^-}{\dot \bT_{1,1}^-}\right]_0=-2\left(\left[\log \frac{\bT_{1,1}^+}{\dot \bT_{1,1}^+}\right]_0\right)^\gamma\,.
\eeq
Finally, to combine the two relations we have just found, we take $\bar{\gamma}$ of \eqref{eq:massless_disc2}  and respectively sum and subtract it from \eqref{eq:massless_constraint}, obtaining the following constraints:
\beq
\label{eq:relating_discontinuities_T}
[\log\bT_{1,1}^+]_0=\left([\log\dot{\bT}_{1,1}^-]_0\right)^{\bar\gamma}\,,\quad [\log\dot\bT_{1,1}^+]_0=\left([{\log\bT}_{1,1}^-]_0\right)^{\bar\gamma}\,.
\eeq

These relations will link the two T-hooks, allowing us to parametrise the T-functions of both using a common set of Q-functions.
\paragraph{Discontinuity containing the constant $c$.}
We now wish to impose the third equation in~\eqref{eq:disc_massiveY_realaxis}. This yields:\footnote{Here and in the equation below, shifts are in the mirror section.}
\beqa
\label{eq:constant_C_inT}
&&2\log(\bT_{1,1}^{+})^{\bar\gamma}(\dot\bT_{1,1}^{+})^{\gamma}-2\log(\dot\bT_{1,1}^{+})^{\bar\gamma}(\bT_{1,1}^{+})^{\gamma}\\&&=2\left\{\log{\frac{\bT_{1,1}^+\bT_{1,1}^-}{\bT_{1,2}}}\right\}^{\bar\gamma}_0-2\left\{\log{\frac{\dot\bT_{1,1}^+\dot\bT_{1,1}^-}{\dot\bT_{1,2}}}\right\}^{\bar\gamma}_0-4c\,,
\eeqa
or equivalently
\beq\label{eq:constant_C_inT2}
2 \log\frac{\dot\bT_{1,1}^{+} \dot\bT_{1,1}^{-} (\dot\bT_{1,1}^{+})^{\gamma} (\dot\bT_{1,1}^{-})^{\bar{\gamma}}}{\bT_{1,1}^{+} \bT_{1,1}^{-} (\bT_{1,1}^{+})^{\gamma} (\bT_{1,1}^{-})^{\bar{\gamma}}}+4\log\frac{\bT_{1,2}}{\dot\bT_{1,2}} =-4c,
\eeq
where the constant $c$, defined in \eqref{def:constant_c_final}, can be rewritten in terms of T-functions as
\beq\label{eq:cconstT}
c=\frac{1}{4}\left( \left. \log \frac{\bT_{1,0}}{\dot\bT_{1,0}}\right|_{u=2}+\left. \log \frac{\bT_{1,0}}{\dot\bT_{1,0}}  \right|_{u=-2}\right)\,.
\eeq
We will see that these relations encode a further constraint for the QSC, valid at the branch points on the real axis $u=\pm2$.

\subsection{The role of massless Y functions}
In this section we impose the remaining discontinuities, i.e. those of the massless Y function $Y_0$ \eqref{Y0symmdisc} and \eqref{eq:lnY0_disc_p2N}, on our T-system. We will prove some nice properties of the $\mu$ function, and also find some relations that will be extremely important in our derivation of the Quantum Spectral Curve.
\paragraph{Massless exact Bethe roots. } First of all, using the relations \eqref{eq:discYpmnew1}, we can express $Y_0$ in terms of the auxiliary Y functions evaluated on two different sheets as follows:
\beq
\label{eq:Y0_as_Ts}
1+Y_0 = \frac{1}{Y_{1,1}\dot Y^{\gamma}_{1,1}}=\frac{1}{Y^{\gamma}_{1,1}\dot Y_{1,1}}\,.
\eeq
Plugging the parametrisation for Y functions \eqref{def:YsasTs} in the $\bT$ gauge, these two expressions give the equivalent result:
\beq
\label{eq:Y0_as_Ts2}
1+Y_0 = \frac{\mu \mu^{\gamma}}{\mathbf{T}_{1,0}\dot{\mathbf{ T}}_{1,0}}\,.
\eeq
From the point of view of excited-state TBA, exact massless Bethe roots $\{u_j\}$ would be identified with solutions to the `exact' Bethe equations (see \cite{Brollo:2023pkl,Brollo:2023rgp,Brollo:2025rgp}, where these equations were studied in the tensionless limit)
\beq\label{eq:stringexcited}
Y_0^{\bar{\gamma}}(u_j) = -1 ,
\eeq
where the position of the root should satisfy $u_j \in (-2, 2)$, and the continuation along the path $\bar{\gamma}$ is needed to move to the string kinematical region.

The equation above gives $1 + Y_0^{\bar{\gamma}} = \mu^{\bar{\gamma}} \mu /(\mathbf{T}_{1,0}\dot{\mathbf{ T}}_{1,0})$, therefore implying that such exact massless Bethe roots correspond to zeros of $\mu$ or $\mu^{\bar\gamma}$. 
 When the massless roots are in the real range $u_j\in (-2, 2)$ (which we expect at least for the simplest solutions), they thus correspond to zeros of the $\mu$ function on the two sides of a short cut, as found in \cite{Ekhammar:2024kzp} in the ABA limit. 

\paragraph{Massless modes and non-quadratic cuts. }
We can also use \eqref{eq:Y0_as_Ts2} 
to deduce a relation between the presence of massless modes in the system, i.e. a non-vanishing $Y_0$, and the non-quadratic nature of branch points. In fact, since $\bT_{1,0}$ and $\dot{\bT}_{1,0}$ do not have a cut on the real axis, by studying the monodromy of $1+Y_0$ on the real branch points using \eqref{eq:Y0_as_Ts2}, we find:
\beq
\label{eq:Y0_mu_nonquadratic}
[\log (1+Y_0)]_0=\log \mu-\log\mu^{\gamma\gamma}\,.
\eeq
This equation implies that $\mu$ possesses non-quadratic branch cuts as a direct consequence of the presence of massless Y-functions. 

Let us now move on to translate the discontinuities \eqref{Y0symmdisc}-\eqref{eq:lnY0_disc_p2N} in terms of T-functions. 

\paragraph{Discontinuities of $Y_0$ on the branch cuts in the UHP.}
Taking the discontinuity at $\Im(u)=2N/h$ of \eqref{eq:Y0_as_Ts2}, we get:
\beq
[\log{Y_0}]_{+2N}=[\log(\mu \mu^{\gamma}-\bT_{1,0}\dot{\bT}_{1,0}) ]_{+2N}
-[\log{\bT_{1,0}\dot{\bT}_{1,0}]_{+2N}}\,.
\eeq
We can compare this with the same quantity obtained from the TBA \eqref{eq:lnY0_disc_p2N}, which, using  relation \eqref{eq:magic} can be written as:
\bal
[\log{Y_0}]_{+2N}
=-[\log{Y_{1,1}\dot{Y}_{1,1}}]_{+2N}-2[\log \mu]_{0}=-[\log \mathbf{T}_{1,0}\dot{\mathbf{T}}_{1,0}]_{+2N}
\,.
\eal
Equaling the two expressions for $[\log{Y_0}]_{+2N}$ we find that:
\beq
\label{eq:masslessdisc_UHP}
[\log{(\mu \mu^{\gamma}-\bT_{1,0}\dot{\bT}_{1,0}})]_{+2N}=0 , \;\; N\in \mathbb{N}^+,
\eeq
i.e. the quantity $\log (\mu \mu^{\gamma}-\bT_{1,0}\dot{\bT}_{1,0})$ has no branch cuts in the UHP. 
\paragraph{Discontinuities of $Y_0$ on the branch cuts in the LHP.}The same calculation can be performed in the LHP to obtain a similar result.

The relevant TBA expression that we need to convert in terms of T-functions is the second line in \eqref{eq:lnY0_disc_p2N}. Using  relation \eqref{eq:magic} this is:
\bal
[\log{Y_0}]_{-2N}
=-[\log \mathbf{T}_{1,0}\dot{\mathbf{T}}_{1,0}]_{-2N}+2\left[\log\frac{\bT_{1,1}^+\bT_{1,1}^-}{\bT_{2,1}}\frac{\dot\bT_{1,1}^+\dot\bT_{1,1}^-}{\dot\bT_{2,1}}\right]_{-2N}
\,.
\eal
Some of the terms on the RHS of the equation above are zero; in fact, we can repeatedly use Hirota's equations to get that:
\bal
\frac{\mathbf{T}^+_{1,1}}{\mathbf{T}_{2,1}} = \frac{\mathbf{T}^{[+2]}_{2,1}}{\mathbf{T}^+_{3,1}} = \dots = \frac{\mathbf{T}^{[+N+1]}_{N+1, 1}}{\mathbf{T}^{[N]}_{N+2, 1}} \,,
\eal
which from the properties of the $\bT$ gauge has no cuts at $\Im(u)=-2N/h$. The same is of course true in the dotted wing, leaving us with:
\bal
[\log{Y_0}]_{-2N}
=-[\log \mathbf{T}_{1,0}\dot{\mathbf{T}}_{1,0}]_{-2N}+2[\log \bT_{1,1}^- \dot \bT_{1,1}^-]_{-2N}
\,.
\eal
Equalling this with the discontinuity at $\Im(u)=-2N/h$ of \eqref{eq:Y0_as_Ts2}, we get:
\beq
\label{eq:masslessdisc_LHP}
[\log{(\mu \mu^{\gamma}-\bT_{1,0}\dot{\bT}_{1,0}})]_{-2N}=2[\log \bT_{1,1}^- \dot \bT_{1,1}^-]_{-2N}
\eeq
i.e. the quantity $\log (\mu \mu^{\gamma}-\bT_{1,0}\dot{\bT}_{1,0})$ has the same discontinuities along the branch cuts in the LHP as $(\log \bT_{1,1}^- \dot \bT_{1,1}^-)^2$.

\paragraph{Discontinuities of $Y_0$ on the real axis.}
We can take the symmetrised discontinuity of \eqref{eq:Y0_as_Ts2} to get:
\beq
\{\log{Y_0}\}_{0}=\{ \log (\mu \mu^{\gamma}-\bT_{1,0}\dot{\bT}_{1,0} )\}_{0}
-2\log{\bT_{1,0}\dot{\bT}_{1,0}}\,.
\eeq
We want to compare this expression with \eqref{Y0symmdisc}. In terms of T-functions, it becomes, after a series of simplifications:
\bal
&\{\log Y_0 \}_0 =-2\log{\bT_{1,0}\dot{\bT}_{1,0}}+2\log{\left(\frac{\bT_{1,1}^+\bT_{1,1}^-}{\bT_{1,2}}\right)^\gamma}+2\log{\left(\frac{\dot\bT_{1,1}^+\dot\bT_{1,1}^-}{\dot\bT_{1,2}}\right)^\gamma}\,.
\eal
Equating the two RHS, we get:
\beq
\label{eq:massless_disc_realaxis}
\{\log (\mu \mu^{\gamma}-\bT_{1,0}\dot{\bT}_{1,0}) \}_{0}=2\log{\left(\frac{\bT_{1,1}^+\bT_{1,1}^-}{\bT_{1,2}}\right)^\gamma}+2\log{\left(\frac{\dot\bT_{1,1}^+\dot\bT_{1,1}^-}{\dot\bT_{1,2}}\right)^\gamma}\,.
\eeq
These massless discontinuities will be crucial to close the \adsst\,$\bP\bm{\mu}$-system.

\section{From the T-system to the $\bP$\texorpdfstring{$\bm{\mu}$}{m}-system}
\label{sec:Pmusystem}
In this section, we rephrase the T-functions in terms of simpler building blocks - the Q-functions - which have simple analytical properties and are the main objects appearing in the QSC. 

Our starting point is the so-called Wronskian solution to the T-system  \cite{Tsuboi:2011iz}, which lets us rephrase any T-function in the T-hook in terms of Q-functions of a Q-system (see Appendix \ref{app:Qsysgen} for a brief review of Q-systems and some of their properties). By imposing the known analytical properties of the T-functions in the `magic sheet' of the last section, we will find the analytical properties of these Q-functions. These functions will be linked together with some jump matrices (of which one component, i.e. the mirror-periodic function $\mu$, already appeared) into a self-contained set of functional equations, the   $\bP\bm{\mu}$-system. 
\paragraph{Parametrising the horizontal part of the T-hooks.}

{
As discussed at the end of section~\ref{sec:white_T_gauge} and proved in appendix~\ref{app:magic_sheet_LRsymm}, each of the function $\hat{\mathbb{T}}_{1,s\geq 1 }$ has two short cuts at $(-2+\frac{i}{h}s, +2+\frac{i}{h}s)$ and $(-2-\frac{i}{h}s, +2-\frac{i}{h}s)$, with $s=1, 2, \dots$.
Because of this we can parameterise $\hat{\wT}_{1,1}$ and $\hat{\wT}_{1,2}$ in terms of four more elementary functions, $\bP_1$, $\dot{\bP}_1$, $\bR_1$ and $\dot\bR_1$, all having a single short cut $(-2,+2)$, as follows

\bal
\hat{\wT}_{1,1} = \bP_1^{[+1]} \dot\bP_1^{[-1]} \,, \qquad \hat{\wT}_{1,2} = \bR_1^{[+2]} \dot\bR_1^{[-2]} \,,
\eal
In general we could expect that $(\bR_1, \dot\bR_1) \ne (\bP_1, \dot\bP_1)$. However, it can be shown that these functions are actually connected. Indeed, from the Hirota equation we know that
\bal
\hat{\wT}_{1,3} = \frac{\hat{\wT}^+_{1,2}\hat{\wT}^-_{1,2}}{\hat{\wT}_{1,1}} = \frac{\bR^{[+3]}_1 \dot\bR^{[-1]}_1 \bR^{[+1]}_1 \dot\bR^{[-3]}_1}{\bP^{[+1]}_1 \dot\bP^{[-1]}_1} \,.
\eal
The requirement that $\hat\wT_{1,3}$ has no cuts at $\Im(u)=i/h$ implies that:
\begin{equation}
\hat{\wT}^{[+], \gamma}_{1,3} = \hat{\wT}^{[+]}_{1,3} \implies \left(\frac{\dot\bR_1}{\dot\bP_1}\right)^\gamma = \frac{\dot\bR_1}{\dot\bP_1} \,.
\end{equation}
Similarly, the requirement that is thas no cuts at $\Im(u)=-i/h$ implies:
\begin{equation}
\hat{\wT}^{[-], \gamma}_{1,3} = \hat{\wT}^{[-]}_{1,3} \implies \left(\frac{\bR_1}{\bP_1}\right)^\gamma = \frac{\bR_1}{\bP_1} \,.
\end{equation}
From these two conditions, we see that it must hold
\begin{equation}
\bR_1 = \bP_1 \,, \qquad \dot\bR_1 = \dot\bP_1 .
\end{equation}
 The ratio of these functions then has no cuts. With a possible redefinition of $\mathbf{P}_1$, $\dot{\mathbf{P}}_1$ which does not change the $\mathbb{T}$ functions, it is  always possible to remove this ratio and choose
\beq
\label{def:wronskian_solution_12}
\hat{\wT}_{1,2}= \bP_1^{[+2]}\dot{\bP}_1^{[-2]} \,.
\eeq
By iterating the Hirota equation, we then reach the universal parameterisation
\beq
\hat{\wT}_{1,s\geq 1}=\bP_1^{[+s]}\dot{\bP}_1^{[-s]} \,.
\eeq
 This parameterisation on the horizontal parts of the T-hooks in terms of products of shifted Q-functions is also what we expect by the Wronskian solution to the \psu\, T-system of \cite{Tsuboi:2011iz}. Repeating the same argument on the dotted wing, which we parametrise via an independent set of functions $\{\mathcal{P}_1,\,\dot{\mathcal{P}}_1\}$, we end up with the following expressions:
\beq
\label{def:wronskian_solution_LR}
\hat{\wT}_{1,s\geq 1}=\bP_1^{[+s]}\dot{\bP}_1^{[-s]},\quad \hat{\dot{\wT}}_{1,s\geq 1}=\dot{\mathcal{P}}_1^{[+s]}{\mathcal{P}}_1^{[-s]}\,.
\eeq
Like $\bP_1$ and $\dot{\bP}_1$, $\mathcal{P}_1$ and $\dot{\mathcal{P}}_1$ have a single short cut on the real line. We show the cut of these functions in figure~\ref{fig:Pfunc-main}.
}
When we cross this short cut either via the path $\gamma$ or $\bar\gamma$ (see figure \ref{fig:analytic_cont_short}), we end in new Riemann sheets characterised by an infinite series of branch points at $\Im(u)=2i N/h, \,N\in \mathbb{Z}$, induced from the cut structure of the T-functions outside the magic sheet. 
Our conventional choice will be to define the hidden sheets of $\bP$ functions with `short' cuts, as can be seen in figure \ref{fig:Pfunc-continued}. 
 
 Finally, from the properties of Y functions solving the TBA equations, we will also assume that $\bP,\,\dot\bP,\,\mathcal{P},\,\dot{\mathcal{P}}$ have power-like asymptotics at large $u$, and, based on the other examples of QSC, we extend this assumption to the hidden sheets. We will shortly identify these functions with some of the Q-functions appearing in the QSC. 
\begin{figure}[ht]
    \centering

    \begin{subfigure}{\textwidth}
        \centering
        \includegraphics[
            page=1,
            width=0.58\textwidth
        ]{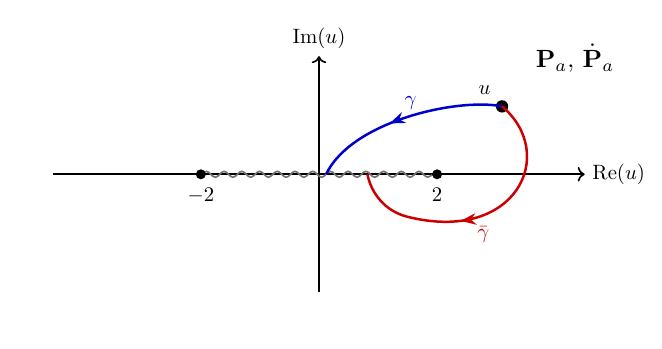}
        \caption{The main Riemann sheet of $\bP$ functions,
                 with a single short branch cut.}
        \label{fig:Pfunc-main}
    \end{subfigure}

    \par\medskip

    \begin{subfigure}{\textwidth}
        \centering
        \includegraphics[
            page=2,
            width=0.47\textwidth
        ]{img/Pfunc_analytic_cont.pdf}
        \hfill
        \includegraphics[
            page=3,
            width=0.47\textwidth
        ]{img/Pfunc_analytic_cont.pdf}
        \caption{The Riemann sheets obtained after analytic
                 continuation along the paths $\gamma$ (in blue)
                 and $\bar\gamma$ (in red). Since the branch cuts of the $\bP$ functions are not quadratic, these are two different sheets, both containing an infinite ladder of short branch cuts equally spaced on the imaginary axis at distance $2 i n/h$ ($n \in \mathbb{Z}$) from the real axis.}
        \label{fig:Pfunc-continued}
    \end{subfigure}
    \caption{Analytic properties of $\bP$ functions.}
    \label{fig:Pfunc}
\end{figure}

In the next section, we will impose on the parametrisation \eqref{def:wronskian_solution_LR} the discontinuities and the analytical properties of the $\hat{\wT}$ functions found in the last section (see eq.~\eqref{eq:relating_discontinuities_T}).

\subsection{Connecting the Q-functions in the two wings}
Our parametrisation \eqref{def:wronskian_solution_LR} contains, at this stage, four independent functions. Using the discontinuity condition \eqref{eq:relating_discontinuities_T} for T-functions, we will now relate them and end up with only \emph{two} independent functions, namely $\bP_1$ and $\dot\bP_1$.

The first equation in \eqref{eq:relating_discontinuities_T}, when translated in terms of the Riemann section with short cuts $\hat\wT$, reads
\beq
(\log\hat{\wT}^{+}_{1,1} )^{\bar\gamma} - \log\hat{\wT}^{+}_{1,1} = (\log\hat{\dot\wT}^{-}_{1,1})^{\bar\gamma} - \log\hat{\dot\wT}^{-}_{1,1} ,
\eeq
and plugging the parametrisation \eqref{def:wronskian_solution_LR} in it, we find
\beq
\log\dot{\bP}_1^{\bar\gamma} - \log\dot{\bP}_1 =\log\dot{\mathcal{P}}_1^{\bar\gamma} -\log\dot{\mathcal{P}}_1 ,
\eeq
where we used the fact that on this sheet $\dot\bP_1$ and $\dot{\mathcal{P}}_1$ do not have branch cuts away from the real axis. 
{This relation tells us that the functions $\log\dot{\bP}_1$ and $\log{\dot{\mathcal{P}}}_1$ have the same discontinuity across the single short cut on their main Riemann sheet.
 Thus, the ratio $\dot{\bP}_1/\dot{\mathcal{P}}_1$ is a function with no cuts and power-like asymptotics, which could be at most meromorphic. We will assume the simplest possibility, i.e. that this function is constant. 
A completely symmetric relation from the second identity in \eqref{eq:relating_discontinuities_T} allows us to relate $\bP_1$ and $\mathcal{P}_1$: again, their ratio has no cuts and we make the simplifying assumption that it is a constant in $u$.

 Thus, we have found that  the T-functions of both wings can be parametrised in terms of \emph{only two} independent Q-functions, $\bP_1$ and $\dot\bP_1$, and in principle two constants, which however will appear as a product, so we will replace them by a single constant $\AA$: 
\beq
\label{def:wronskian_solution_LR_final}
\hat{\wT}_{1,s\geq 1}=\bP_1^{[+s]}\dot{\bP}_1^{[-s]},\quad \hat{\dot{\wT}}_{1,s\geq 1}=\AA\dot{\bP}_1^{[+s]}{\bP}_1^{[-s]}\,.
\eeq
{ One can rescale the dotted T system by $1/\AA$ and remove such a constant, but this changes the condition~\eqref{eq:equal_mu} and such a constant appears back in the ratio $\dot{\mu}/\mu$. Because of this, we do not have any more gauge freedom to rescale away $\AA$, and we will keep this constant in the remaining part of the paper.}
{We can now determine $\bT_{1,0}$ and $\dot\bT_{1,0}$ in terms of $\bP$ and $\mu$ functions using the Hirota equation for $(a,s)=(1,1)$, which in the undotted wing is:
\beq
\label{eq:HirotaT10}
\wT_{1,1}^+\wT_{1,1}^-=\wT_{1,0}\wT_{1,2}+\wT_{0,1}\wT_{2,1}=\wT_{1,2}(\bT_{1,0}+\mu)\,,
\eeq
where in the second equality we have used the group-theoretical properties \eqref{eq:group_theoretical}, which in the $\wT$ gauge imply $\wT_{2,1}= \mu\wT_{1,2}$.  Plugging into this equation the parametrisation \eqref{def:wronskian_solution_LR_final}, 
 and being careful with the shifts (which in the Hirota equation are meant to be performed on the mirror section, with $|\Re(u)<2|$), we can extract:
\beq\label{eq:T10eqs} \bT_{1,0}=\dot\bP_1^{\bar\gamma}\bP_1-\mu\,,
 \;\;\;\dot\bT_{1,0} = \AA\bP_1^{\bar\gamma}\dot\bP_1-\mu\,. 
\eeq
 More precisely, we start with $u$ just above the segment $(-2,+2)$, where $\wT_{1,1}=\hat{\wT}_{1,1}$. Then parameterise the function $\hat{\wT}_{1,1}$ as in~\eqref{def:wronskian_solution_LR_final} and derive the expression for $\bT_{1,0}$. This expression is then valid for $u$ just above the real line, with $|\Re(u)|<2$. From this point, we can analytically continue~\eqref{eq:T10eqs} to obtain an expression for $\bT_{1,0}$ on the full complex plane. The same argument applies on the dotted wing.

\paragraph{A   condition on behaviour at the branch points. } 
With the new parametrization we have introduced, we find that the discontinuity equation \eqref{eq:constant_C_inT2} simplifies dramatically. In fact,  plugging  \eqref{def:wronskian_solution_LR_final} into this relation, we find a drastic simplification on the LHS, and we obtain the simple constraint 
\begin{equation}
\label{eq:constraint_c}
\AA^4 = e^{-4c} ,
\end{equation}%\eqref{eq:constant_C_inT2} 
namely, the constant $c$ appearing in the TBA discontinuity equations is related to $\AA$. 

On the other hand, we also have a definition of the constant $c$ in terms of T-functions, given in \eqref{eq:cconstT}. This condition now takes the form:
\begin{equation}\label{eq:regularityzeta}
\AA^{4} = \left. \left(\frac{\mu-\AA \dot\bP_1\bP_1^{\bar\gamma}}{\mu-\bP_1\dot\bP_1^{\bar\gamma}}\right)\right|_{u=+2}\times \left.\left( \frac{\mu-\AA\dot\bP_1\bP_1^{\bar\gamma}}{\mu-\bP_1\dot\bP_1^{\bar\gamma}}\right)\right|_{u = -2}.
\end{equation}
For $\AA =1$,  this condition is simply satisfied if we ask that the $\bP$ functions are regular at the branch points $\pm 2$: i.e., they assume a finite limit for $u\rightarrow \pm 2$ along any direction, implying
\beq \label{def:regularity_Q}\bP^{\bar\gamma}_1(\pm2)=\bP_1(\pm 2), \quad \dot\bP^{\bar\gamma}_1(\pm2)=\dot\bP_1(\pm 2).\eeq 
The \emph{regularity property} \eqref{def:regularity_Q} is assumed in all known QSC's. In particular, it was argued in \cite{Cavaglia:2022xld} that it should constitute a necessary quantisation condition in the AdS$_3$ case. The same work also found numerically that this condition is associated with a specific structure of the so-called ``gluing matrix'' (see Appendix \ref{app:Qsysgen2} for its definition). It is noteworthy that, in our treatment, this property seems to  emerge naturally from the discontinuity relations of the TBA, at least for a value of the constant $\AA$. 

For $\AA \neq 1$, the condition (\ref{eq:regularityzeta}) takes a more complicated form and is not automatically satisfied when we assume \eqref{def:regularity_Q}. In section \ref{sec:matchQSC0}, we will rephrase this condition in terms of variables appearing in the QSC and further comment on it.

\paragraph{Massless Y function in terms of $\bP$ and $\mu$ functions.} To close this subsection, let us provide an expression for the Y-function $Y_0$ in terms of the new emerging quantities, which will eventually be building blocks of the QSC.  Starting from \eqref{eq:Y0_as_Ts2}, and using the expression for $\bT_{1,0}$ and $\dot\bT_{1,0}$  given in \eqref{eq:T10eqs}, we find:
\beq
\label{eq:Y0asPm}
1+Y_0=\frac{\mu \mu^\gamma}{(\dot \bP_1^{\bar\gamma}\bP_1-\mu)( \AA\bP_1^{\bar\gamma}\dot\bP_1-\mu)}\,.
\eeq

\subsection{Starting to uncover the $\bP$\texorpdfstring{$\bm{\mu}$}{m}-system}
Having established the correct parametrisation of T-functions in terms of Q-functions, we can now derive the $\bP\bm{\mu}$-system, equivalent to the full Quantum Spectral Curve, from various analytical properties of the T-functions.

\paragraph{Parametrising the T-hooks with $\bP$ and $\mu$. }
To derive the $\bP\bm{\mu}$-system, we first need to parametrise all T-functions on the two T-hooks in terms of three quantities, which are $\bP_1,\,\dot\bP_1$ and $\mu$ (and their analytic continuation on other sheets). This can be done via the Hirota equations \eqref{def:Hirotaeq} and our parametrisation for some of the T-functions \eqref{def:wronskian_solution_LR_final}, \eqref{eq:cornerT}, as well as the relation $\bT_{2,1} = \bT_{1,2}$, which translates into $\wT_{2,1} = \mu \wT_{1,2}$. From this seed, via Hirota, we can compute all remaining T-functions. {In the previous section, we have already done so for $\bT_{1,0}$ and $\dot\bT_{1,0}$, obtaining the expressions \eqref{eq:T10eqs}.}

The same process can be repeated to fix any $\bT_{a,0}$ in terms of $\bP$ and $\mu$ functions, and can be iterated quite efficiently via a modified version of the Mathematica code found in Appendix A of \cite{Gromov:2014caa}.
In the next sections, we will impose that the expressions for $\bT_{a,0}$ we have found possess the wide analyticity bands in \eqref{eq:vertstrip}. This gives nontrivial constraints on the $\bP$ and $\mu$ functions.

\paragraph{Equations for $\mu$ from analyticity of $\bT_{1,0}$.  }
By assumption~\eqref{eq:vertstrip}, $\bT_{1,0}$ and $\dot\bT_{1,0}$ do not have branch cuts on the real axis. However,
from the RHS of the expressions \eqref{eq:T10eqs}, the absence of such a cut is not at all evident: in fact, we have a combination of different functions which all do possess branch points on the real axis! Enforcing \eqref{eq:vertstrip} on the RHS therefore yields nontrivial relations between the $\bP$ and the $\mu$ function, which turn out to be exactly two of the equations of the $\bP\bm{\mu}$-system.

In fact, imposing that $\bT_{1,0}=(\bT_{1,0})^\gamma$ and $\dot\bT_{1,0}=(\dot\bT_{1,0})^\gamma$, we find the following equations constraining the monodromy of $\mu$ functions around their branch points on the first sheet:
\beq
\label{eq:gluing_mu}
\mu^\gamma-\mu=\bP_1^\gamma \dot{\bP}_1-\bP_1 \dot{\bP}^{\bar\gamma}_1=\AA(\dot\bP_1^\gamma {\bP}_1-\dot\bP_1 {\bP}^{\bar\gamma}_1)\,.
\eeq

\paragraph{Imposing the analyticity strips of $\bT_{a,0}$. }
 Next, we want to impose the appropriate analyticity strip for $\bT_{a\geq 2,0}$. 
 We start with the following expressions:\footnote{Here, the $\bP$ functions should be evaluated on the Riemann section with short cuts, and the $\mu$ should be understood on the Riemann section with long cuts.}
\begin{small}
\beqa
\mathbf{T}_{2,0}&=&\frac{1}{\mu^+}\left[\bP_1^+ (\dot{\bP}_1^{\bar\gamma})^+(\bP_1^{\gamma})^-\dot{\bP}^-_1-\mu^+(\bP_1^+(\dot{\bP}_1^{\bar\gamma} )^++ (\bP_1^{\gamma})^-\dot{\bP}_1^-)+(\mu^+)^2-(\bP_1^+\dot{\bP}^-_1)^2\right]\,,\\
\dot{\mathbf{T}}_{2,0}&=&\frac{1}{\mu^+}\left[\AA^2\dot\bP_1^+ ({\bP}_1^{\bar\gamma} )^+(\dot\bP_1^\gamma)^-{\bP}^-_1-\mu^+\AA(\dot\bP_1^+({\bP}_1^{\bar\gamma} )^{+} + (\dot\bP_1^\gamma)^{-}{\bP}_1^-)+(\mu^+)^2-\AA^2(\dot\bP_1^+{\bP}^-_1)^2\right]\,,
\eeqa
\end{small}
which arise from application of the Hirota equation (where the shifts are taken in the mirror section) at the node $(1,0)$.  
Because of the structure of the shifts, the first potential branch points appear at $\Im(u)=\pm 1/h$. We should impose trivial monodromy at these branch points, since the analyticity strip of $\bT_{2,0}$ and $\dot\bT_{2,0}$ should have width $3/h$. 

After some simplifications, imposing that there is no cut at $\Im(u)=1/h$ results in the following conditions:
\beqa
\label{eq:no_cuts_T20}
\mu \,(\dot{\bP}_1^{\bar\gamma,\,[+2]})^{\gamma}&=&{\mu^{\gamma}}(\dot{\bP}_1^{\bar\gamma})^{[+2]}+\bP_1^{[+2]}\,\frac{{\mu^{\gamma}}(\dot{\bP}_1^{\bar\gamma})^2-\mu(\dot{\bP}_1)^2}{\mu-\bP_1\dot{\bP}_1^{\bar\gamma}}\,,\\
\label{eq:no_cuts_T20bis}
\mu \,({\bP}_1^{\bar\gamma,\,[+2]})^{\gamma}&=&{\mu^\gamma}({\bP}_1^{\bar\gamma})^{[+2]}+\AA\dot\bP_1^{[+2]}\,\frac{{\mu^\gamma}({\bP}_1^{\bar\gamma})^2-\mu({\bP}_1)^2}{\mu-\AA\dot\bP_1{\bP}_1^{\bar\gamma}}\,.
\eeqa
In these expressions, $(\dot{\bP}_1^{\bar\gamma,\,[+2]})^{\gamma}$ means that we first go through the short cut on the real axis from below, entering a sheet where the Q-functions have, potentially, an infinite ladder of further branch points. Then we shift up by $2i/h$ and go around the branch point located there following the path $\gamma$.

It is possible to repeat this process for any $\bT_{a,0}$; while the resulting expressions are quite bulky, imposing the absence of cuts of $\bT_{a,0}^{[+(a-1)]}$ and $\dot\bT_{a,0}^{[+(a-1)]}$ results in remarkably simple conditions, \footnote{Imposing that there are no branch cuts below these in the UHP gives no new constraints. We verified these equations explicitly for a few values of $a$, but we believe they should hold for any $a\geq1$.} which are a generalisation of \eqref{eq:no_cuts_T20} and~\eqref{eq:no_cuts_T20bis}:
\beqa
\label{eq:no_cuts_P}
\mu \,(\dot{\bP}_1^{\bar\gamma,\,[+2N]})^{\gamma}&=&{\mu^{\gamma}}(\dot{\bP}_1^{\bar\gamma})^{[+2N]}+\bP_1^{[+2N]}\,\frac{{\mu^{\gamma}}(\dot{\bP}_1^{\bar\gamma})^2-\mu(\dot{\bP}_1)^2}{\mu-\bP_1\dot{\bP}_1^{\bar\gamma}}\,,\\
\label{eq:no_cuts_P2}
\mu \,({\bP}_1^{\bar\gamma,\,[+2N]})^{\gamma}&=&{\mu^\gamma}({\bP}_1^{\bar\gamma})^{[+2N]}+\AA\dot\bP_1^{[+2N]}\,\frac{{\mu^\gamma}({\bP}_1^{\bar\gamma})^2-\mu({\bP}_1)^2}{\mu-\AA\dot\bP_1{\bP}_1^{\bar\gamma}}\,.
\eeqa
\paragraph{Enter new $\bP$ functions. }
Equations \eqref{eq:no_cuts_P} and \eqref{eq:no_cuts_P2} can be rephrased in a much nicer way. In fact, we first define two new Q-functions as follows:
\beq
\label{def:P3}
\bP_{2}\equiv  \frac{1}{\mu} \dot{\bP}_1^{\bar\gamma}+\frac{\lambda}{\mu} \bP_1\,,\quad \dot\bP_{2}\equiv  \AA\left(\frac{1}{\mu} {\bP}_1^{\bar\gamma}+\frac{\dot \lambda}{\mu} \dot{\bP}_1\right)\,,
\eeq
where $\lambda$ and $\dot\lambda$ are two new mirror periodic functions, i.e. functions with a sheet having long branch cuts where they are periodic with period $i$, i.e. $\lambda^{++}=\lambda$ and $
\dot\lambda^{++}=\dot\lambda$. {In a moment, we will specify these functions conveniently by giving the values of their discontinuities.  The point is that we can choose them in such a way that the new Q functions $\bP_2$ and $\dot{\bP}_2$ have a single cut on the physical Riemann sheet. }
In fact, let us consider the condition that the new Q-functions \eqref{def:P3} should not have branch cuts in the UHP. Given their definitions, this would mean that the following relations are satisfied:
\beqa
0&=&{\mu^{\gamma}}(\dot{\bP}_1^{\bar\gamma})^{[+2N]}-\mu{(\dot{\bP}_1^{\bar\gamma,[+2N]})^{\gamma}}{}+\left({{\mu^\gamma} \lambda
}{} -{\mu{\lambda^\gamma}
}{}\right)\bP_1^{[+2N]},\\
0&=&{\mu^{\gamma}}({\bP}_1^{\bar\gamma})^{[+2N]}-\mu{({\bP}_1^{\bar\gamma,[+2N]})^{\gamma}}{}+\left({{\mu^{\gamma}} \dot\lambda
}{} -{\mu{\dot\lambda^\gamma}
}{}\right)\dot\bP_1^{[+2N]}\,.
\eeqa
These equations have a very similar form to \eqref{eq:no_cuts_P} and \eqref{eq:no_cuts_P2}; in fact, they are the same equations, provided that the new mirror periodic functions have the following analytic continuations across their cuts: 
\beq\label{eq:partiallambdagluing}
\frac{ \lambda
}{\mu} -\frac{{\lambda^{\gamma}}
}{ \mu^{\gamma}}=\frac{\frac{(\dot{\bP}_1^{\bar\gamma})^2}{\mu}-\frac{(\dot{\bP}_1)^2}{\mu^{\gamma}}}{\mu-\bP_1\dot{\bP}_1^{\bar\gamma}}\,, \qquad \frac{ \dot\lambda
}{\mu} -\frac{{\dot\lambda^\gamma}
}{ \mu^\gamma}=\AA\frac{\frac{({\bP}_1^{\bar\gamma})^2}{\mu}-\frac{({\bP}_1)^2}{\mu^\gamma}}{\mu-\AA\dot\bP_1{\bP}_1^{\bar\gamma}}\,.
\eeq
{Therefore, with this choice we find that two new single-cut $\bP$ functions have emerged naturally, encoding the analyticity strips of the $\bT$ functions. }

\paragraph{New $\bP\bm{\mu}$ relations. }
Now, we can massage these relations to get closer to the $\bP\bm{\mu}$-system form. Taking the first (second)  of the relations \eqref{eq:partiallambdagluing}, and plugging the expression for $\mu^\gamma$ coming from the first (second) of the  equations \eqref{eq:gluing_mu}, we obtain respectively:
\beqa
{\lambda^{\gamma}}&=&\lambda+(\bP_1^\gamma \dot\bP_1-\bP_1\dot{\bP}_1^{\bar\gamma})\frac{\lambda}{\mu}+\frac{\dot{\bP}_1^2-(\dot{\bP}_1^{\bar\gamma})^2}{\mu}+\frac{\dot{\bP}_1\dot{\bP}_1^{\bar\gamma}(\bP_1 \dot{\bP}_1-\bP_1^{\gamma}\dot{\bP}_1^{\bar\gamma})}{\mu(\mu-\bP_1\dot{\bP}_1^{\bar\gamma})}\,,\\
{\dot\lambda^{\gamma}}&=&\dot\lambda+\AA(\dot{\bP}_1^\gamma \bP_1-\dot{\bP}_1{\bP}_1^{\bar\gamma})\frac{\dot{\lambda}}{\mu}+\AA \frac{{\bP}_1^2-({\bP}_1^{\bar\gamma})^2}{\mu}+\AA^2\frac{{\bP}_1{\bP}_1^{\bar\gamma}(\dot{\bP}_1 {\bP}_1-\dot{\bP}_1^{\gamma}{\bP}_1^{\bar\gamma})}{\mu(\mu-\AA\dot{\bP}_1{\bP}_1^{\bar\gamma})}\,.
\eeqa
Finally, using the definitions  \eqref{def:P3} and their analytic continuation along $\gamma$, it is possible to rephrase them in a much simpler form:
\beqa
\label{eq:gluing_lambda}
\lambda^\gamma-\lambda&=&\bP_2^{\gamma}\dot{\bP}_1 -\bP_2\dot{\bP}_1^{\bar\gamma} \,,\\
\label{eq:gluing_lambdabis}
\dot{\lambda}^\gamma-\dot{\lambda}&=&\dot{\bP}_2^{\gamma}{\bP}_1 - \dot{\bP}_2{\bP}_1^{\bar\gamma}\,.
\eeqa
Given the form of these relations, we will identify the mirror periodic functions $\lambda$ and $\dot\lambda$ as two more elements of the matrix $\mu_{a}^{\;\dot b}$ introduced in the $\bP\bm{\mu}$-system of \cite{Cavaglia:2021eqr, Ekhammar:2021pys}.

Furthermore, the definitions \eqref{def:P3} can be interpreted as two further relations of the $\bP\bm{\mu}$ system:
\beq
\label{eq:gluing_P1}
\dot{\bP}_1^{\bar\gamma}=-\lambda \bP_1+\mu \bP_2\,,\quad {\bP}_1^{\bar\gamma}=-\dot\lambda \dot\bP_1+\frac{\mu}{ \AA} \dot\bP_2\,.
\eeq
To recap the results of this section, we have found that the condition that $\bT_{a,0}$ and $\dot\bT_{a,0}$ have wide analyticity strips is satisfied if the two new Q-functions defined in \eqref{def:P3} have no cuts in the UHP and if the functions $\lambda$ and $\dot\lambda$ used in their definition satisfy \eqref{eq:gluing_lambda}.

\paragraph{What about the lower half-plane?}
{
We can also ask what conditions we get when imposing that the $\bT_{a,0}$ functions do not have branch cuts in the LHP. We find similar expressions as those in \eqref{eq:no_cuts_P}, which are given by
 imposing that $\bT_{a,0}^{[-a+1]}$ has no cuts. This implies the following constraints for the $\bP$ functions:
\beqa
\label{eq:equation1_LHP}
\mu(\bP_1^{\gamma,[-2N]})^{\bar\gamma}&=&\mu^{\bar\gamma} \bP_1^{\gamma,[-2N]}+\dot\bP_1^{[-2N]}\frac{\mu^{\bar\gamma} \bP_1^2-\mu(\bP_1^{\bar\gamma})^2}{\mu-\bP_1\dot\bP_1^{\bar\gamma}}\,,\\
\mu(\dot\bP_1^{\gamma,[-2N]})^{\bar\gamma}&=&\mu^{\bar\gamma} \dot\bP_1^{\gamma,[-2N]}+\AA\bP_1^{[-2N]}\frac{\mu^{\bar\gamma} \dot\bP_1^2-\mu(\dot\bP_1^{\bar\gamma})^2}{\mu-\AA\dot\bP_1\bP_1^{\bar\gamma}}\,.
\eeqa
We have checked for a few values of $a$ that these conditions, together with \eqref{eq:no_cuts_P}, are enough for the $\bT_{a,0}$ functions to have the required wide analyticity strips. 

Furthermore, we will see that they also imply that $\bP_2$ and $\dot{\bP}_2$ have a single short cut; namely, they are also free of cuts in the lower half-plane. We postpone this proof to appendix \ref{app:LHP}, since it will be convenient first to deduce some other useful identities. 
}

\subsection{Closing the $\bP$\texorpdfstring{$\bm{\mu}$}{m}-system thanks to  massless Y functions}
So far, we have identified a total of four Q-functions $\bP_1,\,\bP_2, \,\dot\bP_1,\,\dot\bP_2$ and three mirror-periodic functions $\mu,\,\lambda,\,\dot\lambda$, which serve to describe their monodromies around branch points.  We also know a few relations connecting these different functions on different sheets, in particular \eqref{eq:gluing_P1}, the \emph{two} relations describing the discontinuity of $\mu$  \eqref{eq:gluing_mu}, and a similar equation each for $\lambda$ and $\dot\lambda$, i.e.  \eqref{eq:gluing_lambda}, \eqref{eq:gluing_lambdabis}. However, the aforementioned equations involve $\bP_2^\gamma$ and $\dot\bP_2^{\bar\gamma}$, and to close the system we still need some independent expressions for these quantities. In this section we will see that the discontinuities of the massless Y-functions \eqref{eq:masslessdisc_UHP} and \eqref{eq:massless_disc_realaxis}, rephrased in terms of Q-functions, provide exactly the ingredients that are needed to close the emerging $\bP\bm{\mu}$-system.

\paragraph{A key quantity... }
The main object appearing in the discontinuities of $Y_0$ \eqref{eq:masslessdisc_UHP} and \eqref{eq:massless_disc_realaxis} is the combination $\mu \mu^{\gamma}-\bT_{1,0}\dot\bT_{1,0}$. In fact, we have seen that the discontinuities of the massless Y function try to tell us something about the cuts of this quantity in the complex plane. Before decoding what this something actually is, we notice that this quantity
  can be expressed in two equivalent ways using \eqref{eq:T10eqs} and \eqref{eq:gluing_mu}. We get:
\beq
\label{eq:massless_gluing_LHS}
\mu \mu^{\gamma}-\bT_{1,0}\dot\bT_{1,0}=\mu (\AA\bP_1 \dot{\bP}_1^\gamma+\bP_1 \dot{\bP}_1^{\bar\gamma})-\AA \bP_1 \dot{\bP}_1\bP_1^{\bar\gamma} \dot{\bP}_1^{\bar\gamma}=\mu (\dot\bP_1 {\bP}_1^\gamma+\AA\dot\bP_1 {\bP}_1^{\bar\gamma})-\AA\bP_1 \dot{\bP}_1\bP_1^{\bar\gamma} \dot{\bP}_1^{\bar\gamma}\,,
\eeq
{where the two expressions above are equal due to the second equality in \eqref{eq:gluing_mu}. }

\paragraph{...  and a key identity. }We will now prove that the key quantity \eqref{eq:massless_gluing_LHS} is actually a very simple object. The equations \eqref{eq:masslessdisc_UHP} tell us that  $\mu \mu^{\gamma}-\bT_{1,0}\dot\bT_{1,0}$  has no cuts in the upper half plane. 
Then, the same will be true for the following quantity
\beq
\label{def:mathcalF}
\mathcal{F} \equiv \frac{\mu \mu^{\gamma}-\bT_{1,0}\dot\bT_{1,0} }{\AA (\bP_1 \dot{\bP}_1 )^2 } \,,
\eeq
since the $\bP$ and $\dot{\bP}$ functions have a single cut (and $\sigma$, we recall, is a constant).  
 What can we say about the cut of $\mathcal{F}$ on the real axis and its behaviour in the lower half-plane?

To inspect the cut on the real axis, we consider the massless discontinuity \eqref{eq:massless_disc_realaxis} together with the last of the massive discontinuities in equation \eqref{eq:disc_massiveY_realaxis}}. Translating them in terms of T-functions first (using \eqref{def:YsasTs} and \eqref{eq:Y0_as_Ts2}) and then in terms of Q-functions,
we find that they become, respectively
\beqa
\label{eq:calF_cond1}
\{\log(\mu \mu^{\gamma}-\bT_{1,0}\dot\bT_{1,0})\}_0&=&2\log{\AA\bP_1\bP_1^\gamma \dot{\bP}_1  \dot{\bP}_1^\gamma}
,\\
 \big[\log(\mu\mu^{\gamma}-\bT_{1,0}\dot\bT_{1,0})\big]_0&=&2\log{\dfrac{\bP_1\dot{\bP}_1}{ \bP_1^\gamma \dot{\bP}_1^\gamma}}
,
\eeqa
which can be rewritten as the very simple statements for $\mathcal{F}$:
\beq
\label{eq:Fcutsreal}
\mathcal{F}\mathcal{F}^\gamma=1\,,\qquad \frac{\mathcal{F}}{\mathcal{F}^\gamma}=1  \,.
\eeq
These conditions can only be solved simultaneously by
\beq
\mathcal{F}(u)\equiv \,\ss, \qquad \ss = \pm 1 \,.
\eeq
Finally, we have identities \eqref{eq:masslessdisc_LHP}, describing the cuts of the quantity  $\log(\mu \mu^{\gamma}-\bT_{1,0}\dot\bT_{1,0})$ in the LHP. 
Plugging the $\bP$ function expression for the last term of \eqref{eq:masslessdisc_LHP}, these relations become\footnote{Here, the appearance of $\bP^{\gamma}$ and $\dot\bP^{\gamma}$ is due to the fact that these functions have their primary definition with short cuts, while here we are working on the long-cuts section.}
\beq
\label{eq:masslessdisc_LHP2}
[\log{(\mu \mu^{\gamma}-\bT_{1,0}\dot{\bT}_{1,0}})]_{-2N}=2[\log (\AA\bP_1^{\gamma} \dot\bP_1^\gamma)]_{-2N},
\eeq
which tells us precisely that $\mathcal{F}$ also has no cuts when continued (using the long-cuts section) into the lower half-plane! Moreover, the same is true also if we use a section with a short cut on the real axis, thanks to the property \eqref{eq:Fcutsreal}.

All of this leads to the following key identity:
\beq
\label{eq:massless_pmu_start}
\mu\mu^\gamma-\bT_{1,0}\dot{\bT}_{1,0}=\ss\, \AA (\bP_1\dot{\bP}_1)^2,\qquad \ss=\pm1\,.
\eeq
It is quite remarkable that the simple equation \eqref{eq:massless_pmu_start} encapsulates all the massless discontinuities of the TBA. 
This equation is the cornerstone which we will use to derive all the remaining equations of the $\bP\bm{\mu}$-system.

\paragraph{Simplicity assumption on the parameters $\AA$, $\ss$. } In closing the derivation, we find several indications that the parameters $\ss$ and $\AA$ should be fixed to very simple values, i.e. $\ss=\AA = \pm 1$. While we did not find a completely rigorous proof, let us present some of this evidence and motivate this assumption, which we will take from now on.

We will start by computing $\bP$ and $\dot\bP$ on the next Riemann sheet, following the analytic continuation $\gamma$. 
To do this, we replace   $\mu^\gamma$ with its two possible expressions \eqref{eq:massless_gluing_LHS} in the LHS of \eqref{eq:massless_pmu_start}. In the two resulting expressions, we can furthermore use \eqref{eq:gluing_P1} to eliminate the quantities $\bP_1^{\bar\gamma}$ and $\dot\bP_1^{\bar\gamma}$. In this way, we obtain two equations that depend solely on $\bP_1^\gamma$, $\dot\bP_1^\gamma$ and other quantities evaluated on the first sheet. Solving them to determine $\bP_1^\gamma$, $\dot\bP_1^\gamma$ in terms of quantities on the first sheet, we get
\beqa
\label{eq:gluing_P1gamma}
\bP_1^\gamma&=& \AA\dot\lambda\dot\bP_1-\mu\dot\bP_2-\bP_1(\bP_1(\lambda\dot\bP_2-\kappa \dot\bP_1)+\bP_2(\AA\dot\lambda\dot\bP_1-\mu\dot\bP_2))\,,\\
\label{eq:gluing_P1gammabis}
\dot\bP_1^\gamma&=&\AA^{-1}\left(\lambda\bP_1-\mu\bP_2-\dot\bP_1(\dot\bP_1(\AA\dot\lambda\bP_2-\kappa \bP_1)+\dot\bP_2(\lambda\bP_1-\mu\bP_2))\right) \,,
\eeqa
where we have defined another mirror periodic function $\kappa$ as:
\beq
\label{def:kappa}
\kappa(u)\equiv \AA \frac{ \ss+\lambda(u) \dot\lambda(u)}{\mu(u)}\,.
\eeq
As a cross-check, we can verify that the two alternative equations for $\mu - \mu^{\gamma}$ in \eqref{eq:gluing_mu} are identical when expressed in terms of quantities on the first sheet via \eqref{eq:gluing_P1} and \eqref{eq:gluing_P1gamma}.

From \eqref{eq:gluing_P1gamma}, \eqref{eq:gluing_P1gammabis}, we can obtain  similar expressions for $\bP^{\gamma}_2$ and $\dot\bP^{\gamma}_2$. In fact, by continuing along $\gamma$ the RHS of the two expressions \eqref{def:P3} and using \eqref{eq:gluing_P1gamma}-\eqref{eq:gluing_P1gammabis} along with \eqref{eq:gluing_mu},  \eqref{eq:gluing_lambda}-\eqref{eq:gluing_lambdabis} to convert the resulting expression to first-sheet quantities, we obtain:
\beqa
\label{eq:gluing_P2gamma}
\bP_2^\gamma&=& \kappa\dot\bP_1-\lambda\dot\bP_2-\bP_2(\bP_2(\AA\dot\lambda\dot\bP_1-\mu \dot\bP_2)+\bP_1(\lambda\dot\bP_2-\kappa\dot\bP_1))+R_1\,,\\
\label{eq:gluing_P2gammabis}
\dot\bP_2^\gamma&=&\AA^{-1}\left(\kappa\bP_1-\AA\dot\lambda\bP_2-\dot\bP_2(\dot\bP_2(\lambda\bP_1-\mu \bP_2)+\dot\bP_1(\AA\dot\lambda\bP_2-\kappa\bP_1))\right)+R_2 \,.
\eeqa
{In these expressions, $R_1$ and $R_2$ are rational expressions in the $\bP$ and $\mu$ functions:
\beq
R_1=-(1-\AA\ss)\frac{\dot \bP_1}{\bT_{1,0} },\quad R_2=- (\AA-\ss)\frac{\bP_1}{\dot\bT_{1,0}}\,.
\eeq
We notice that in general these expressions could introduce poles on a second Riemann sheet of $\bP_2$ and $\dot \bP_2$, corresponding to zeros of the T functions in their denominators. We expect that these functions will play the role of Q-functions in the QSC, and that they should not have such poles, on any Riemann sheet, similar to the other known QSC's for holographic theories. 

From now on, we will \emph{assume}  that these poles are cancelled in the simplest possible way, namely by setting 
\begin{equation}\label{eq:constantassumption}
\ss=\AA, \;\;\; \AA = \pm 1,
\end{equation}
which leads to $R_1 = R_2 = 0$. 

\paragraph{More $\bP\bm{\mu}$ equations. }

Let us consider the $\kappa$ function defined in \eqref{def:kappa}, which thanks to the above choice simplifies to
\beq
\label{def:kappa1}
\kappa\equiv \frac{1+\AA \lambda \dot\lambda}{\mu}\,.
\eeq
This function satisfies mirror periodicity (which is evident from its definition) and two equations describing its discontinuities, analogous to those in \eqref{eq:gluing_mu}. To deduce them, we compute from the definition:
\beq
\kappa^\gamma-\kappa\equiv \frac{1+\AA\lambda^\gamma \dot\lambda^\gamma}{\mu^\gamma}-\frac{1+\AA \lambda \dot\lambda}{\mu}\,,
\eeq
and rewrite the RHS. Using~\eqref{eq:gluing_lambda}, \eqref{eq:gluing_lambdabis} to substitute $\lambda^\gamma$ and $\dot\lambda^\gamma$, we find the following two equivalent expressions by using either the first or the second of \eqref{eq:gluing_mu} to substitute $\mu^\gamma$:
\beq
\label{eq:gluing_kappa_ini}
\kappa^\gamma-\kappa=\bP_2^\gamma\dot\bP_2 -\bP_2  (\AA\dot\lambda \bP_2-\kappa \bP_1)=\AA\dot\bP_2^\gamma \bP_2- \dot\bP_2 (\lambda \dot\bP_2-\kappa \dot\bP_1)\,,
\eeq
where $\bP_2^\gamma,\dot\bP_2^\gamma$ are given respectively by \eqref{eq:gluing_P2gamma} and \eqref{eq:gluing_P2gammabis}.

Now, let us prove that the quantities appearing in round brackets in the equations above are nothing but $\bP_2^{\bar\gamma}$ and $\dot\bP_2^{\bar\gamma}$ respectively. In fact, if we analytically continue them along $\gamma$, we find respectively:
\beqa
\AA \dot\lambda^\gamma \bP^\gamma_2-\kappa^\gamma \bP_1^\gamma\,,\qquad 
\lambda^\gamma \dot\bP^\gamma_2-\kappa^\gamma \dot\bP_1^\gamma\,.
\eeqa
If, in the first and second equations, respectively, we use the second and first forms in \eqref{eq:gluing_kappa_ini} to substitute $\kappa^\gamma$, and then replace $\bP_2^{\gamma}$ and $\dot\bP_2^{\gamma}$ using \eqref{eq:gluing_P2gamma},\eqref{eq:gluing_P2gammabis} (where $R_1=R_2=0)$. After some dramatic simplifications, we are left with
\beqa
\AA \dot\lambda^\gamma \bP^\gamma_2-\kappa^\gamma \bP_1^\gamma=\dot\bP_2 ,\\
\lambda^\gamma \dot\bP^\gamma_2-\kappa^\gamma \dot\bP_1^\gamma=\AA \bP_2 .
\eeqa
Thus, we found that
\beq\label{eq:gluing_P2gammabar}
\dot\bP_2^{\bar\gamma}=\AA \dot\lambda \bP_2-\kappa \bP_1\,\;\;\;
\bP_2^{\bar\gamma} =\AA \left(\lambda \dot\bP_2-\kappa \dot\bP_1\right)\,\,.
\eeq
The expressions above also allow us to rewrite \eqref{eq:gluing_kappa_ini} in their final forms:
\beqa
\label{eq:gluing_kappa}
\kappa^\gamma-\kappa=\bP_2^\gamma \dot\bP_2-\bP_2 \dot\bP_2^{\bar\gamma} =\AA \left( \dot\bP_2^\gamma \bP_2- \dot\bP_2 \bP_2^{\bar\gamma} \right)\,.
\eeqa

\paragraph{A further constraint.}
Reshuffling some of these equations, a constraint emerges as a consistency condition. One way to obtain it is to start from equations~\eqref{eq:gluing_P1gamma}, \eqref{eq:gluing_P1gammabis}, recognising in their RHS the combinations appearing in \eqref{eq:gluing_P2gammabar} and  \eqref{eq:gluing_P1}.  Then one gets the expressions:
\beqa\label{eq:Pmualternative5}
\frac{\bP_1^\gamma}{ \bP_1} +\frac{\AA\bP_1^{\bar\gamma}}{\bP_1}&=&\AA(\bP_2 \bP_1^{\bar\gamma}-\bP_1 \bP_2^{\bar\gamma})\,,\\
\frac{\AA \dot\bP_1^\gamma}{\dot\bP_1} +\frac{\dot\bP_1^{\bar\gamma}}{\dot\bP_1}&=&\dot\bP_2 \dot\bP_1^{\bar\gamma}-\dot\bP_1 \dot \bP_2^{\bar\gamma}\,.
\eeqa
Due to {the second equality in \eqref{eq:gluing_mu}}, the LHS of these expressions must be equal, and so it follows that the following constraint must be true:
\beq
\AA(\bP_2 \bP_1^{\bar\gamma}-\bP_1 \bP_2^{\bar\gamma})=\dot\bP_2 \dot\bP_1^{\bar\gamma}-\dot\bP_1 \dot \bP_2^{\bar\gamma}\,.
\eeq

\paragraph{Alternative forms of the equations for $\lambda,\dot\lambda$.} 
Thanks to \eqref{eq:gluing_P2gammabar}, and the already known \eqref{eq:gluing_P1}, \eqref{eq:gluing_P1gamma}-\eqref{eq:gluing_P1gammabis} and \eqref{eq:gluing_P2gamma}-\eqref{eq:gluing_P2gammabis}, we can now also prove, by direct substitution, that the following alternative forms of \eqref{eq:gluing_lambda} are also true:
\beqa
\label{eq:gluing_lambda_final}
\lambda^\gamma-\lambda&=&\AA(\dot\bP_1^\gamma \bP_2-\dot\bP_1\bP_2^{\bar\gamma}) \,,\\\
\dot{\lambda}^\gamma-\dot{\lambda}&=&
{\AA}{(\bP_1^\gamma \dot\bP_2-\bP_1\dot\bP_2^{\bar\gamma})}\,.
\eeqa
To recap, we found two alternative equations for the discontinuities of each of the four mirror-periodic functions $\mu,\lambda,\dot\lambda,\kappa$. 
Together with the monodromy equations for each of the four Q-functions $\bP_1,\bP_2,\dot\bP_1,\dot\bP_2$, this constitutes the $\bP\bm{\mu}$-system of the \adsst\, Quantum Spectral Curve, which we now rewrite in a more covariant form.

\subsection{Putting together the $\bP{\mu}$ system}
\label{sec:matchQSC0}

\paragraph{Summary of the monodromy equations and constraints. }
To recap the results of this section, 
we have parametrised the T-system and its analytic properties in terms of 4 Q-functions with a single cut on the real axis, denoted as $\bP_1,\,\bP_2$ and $\dot\bP_1,\,\dot\bP_2$, plus 4 mirror periodic functions denoted as $\mu,\,\lambda,\,\dot\lambda,\,\kappa$, which are not all independent since $\kappa= \dfrac{1+\AA\lambda\dot\lambda}{\mu}$. The constant $\AA$ appearing in the equations below is assumed to be $\pm 1$, which is the only choice simplifying the construction and guaranteeing absence of poles on all sheets for the $\bP$ functions. 

For the analytic continuation of the single-cut Q-functions, we found
\beqa
\label{eq:gluing_finalP}
\dot{\bP}_1^{\bar\gamma}=\mu \bP_2-\lambda \bP_1\,,&\quad& {\bP}_1^{\bar\gamma}=
{\AA}{\mu} \dot\bP_2-\dot\lambda \dot\bP_1\,,\\
\dot\bP_2^{\bar\gamma}= \AA\dot\lambda \bP_2-\kappa \bP_1\,,&\quad& \bP_2^{\bar\gamma}= 
{\AA}\left(\lambda \dot\bP_2-\kappa \dot\bP_1\right)\,,\label{eq:gluing_finalPbis}
\eeqa
while for the mirror periodic functions we have found
\beqa
\label{eq:gluing_finalmu}
\mu^\gamma-\mu&=&\bP_1^\gamma \dot{\bP}_1-\bP_1 \dot{\bP}^{\bar\gamma}_1=\AA(\dot\bP_1^\gamma {\bP}_1-\dot\bP_1 {\bP}^{\bar\gamma}_1)\,,\\
\lambda^\gamma-\lambda&=&\bP_2^{\gamma}\dot{\bP}_1 -\bP_2\dot{\bP}_1^{\bar\gamma}=\AA(\dot\bP_1^\gamma \bP_2-\dot\bP_1\bP_2^{\bar\gamma}) \,,\\\
\dot{\lambda}^\gamma-\dot{\lambda}&=&\dot{\bP}_2^{\gamma}{\bP}_1 - \dot{\bP}_2{\bP}_1^{\bar\gamma}=
{\AA}
{(\bP_1^\gamma \dot\bP_2-\bP_1\dot\bP_2^{\bar\gamma})}\,,\\
\kappa^\gamma-\kappa&=&\bP_2^\gamma \dot\bP_2-\bP_2 \dot\bP_2^{\bar\gamma} =\AA(\dot\bP_2^\gamma \bP_2- \dot\bP_2 \bP_2^{\bar\gamma})\,.\label{eq:gluing_finalmulast}
\eeqa
We also obtained analytic continuation along $\gamma$ for the $\bP$ functions \eqref{eq:gluing_P1gammabis}, \eqref{eq:gluing_P2gammabis}, which can be further simplified using \eqref{eq:gluing_finalP} and \eqref{eq:gluing_finalPbis}; even  though the relevant expressions can be obtained from the set of equations above, for the benefit of the reader we write them down explicitly:
\beqa\label{eq:Pgammma1}
\bP_1^\gamma&=&\AA( -\bP_1^{\bar\gamma}-\bP_1(\bP_1\bP_2^{\bar\gamma}-\bP_2\bP_1^{\bar\gamma}))\,,\\
\dot\bP_1^\gamma&=&
{\AA}(-\dot\bP_1^{\bar\gamma}-\dot\bP_1(\dot\bP_1\dot\bP_2^{\bar\gamma}-\dot\bP_2\dot\bP_1^{\bar\gamma}) )\,,\\
\bP_2^\gamma&=& \AA(-\bP_2^{\bar\gamma}-\bP_2(\bP_1\bP_2^{\bar\gamma}-\bP_2\bP_1^{\bar\gamma}))\,,\\
\dot\bP_2^\gamma&=&
{\AA}(-\dot\bP_2^{\bar\gamma}-\dot\bP_2(\dot\bP_1\dot\bP_2^{\bar\gamma}-\dot\bP_2\dot\bP_1^{\bar\gamma})) \,.\label{eq:Pgammma4}
\eeqa
Finally, we have obtained the following constraints:
\beq
\label{eq:constraint_final}
\AA(\bP_2 \bP_1^{\bar\gamma}-\bP_1 \bP_2^{\bar\gamma})=\dot\bP_2 \dot\bP_1^{\bar\gamma}-\dot\bP_1 \dot \bP_2^{\bar\gamma}\,,
\eeq
as well as a condition concerning behaviour at the branch points: 
\begin{equation}\label{eq:regularityzeta2}
1 = \left. \left(\frac{\mu-\AA \dot\bP_1\bP_1^{\bar\gamma}}{\mu-\bP_1\dot\bP_1^{\bar\gamma}}\right)\right|_{u=+2}\times \left.\left( \frac{\mu-\AA\dot\bP_1\bP_1^{\bar\gamma}}{\mu-\bP_1\dot\bP_1^{\bar\gamma}}\right)\right|_{u = -2}.
\end{equation} 
\paragraph{Matrix notation. }
We can put our equations in a form matching the one of the $\bP\bm{\mu}$-system of the Quantum Spectral Curve for \adsst\, in the LR symmetric sector, formulated in \cite{Cavaglia:2021eqr, Ekhammar:2021pys}. That construction featured  4  Q-functions with a single short cut, which we will identify with our $\bP_1,\,\bP_2,\, \dot\bP_1,\,\dot\bP_2$, and two independent $2 \times 2$ $\bm{\mu}$ matrices, which, following the notation of \cite{Cavaglia:2021eqr},  we denote as follows
\beqa
\label{def:gluing_QSC}
\bm{\mu}=\begin{pmatrix}\mu_{1}^{\;\dot 1} & \mu_{1}^{\;\dot 2}\\ \mu_{2}^{\;\dot 1} & \mu_{2}^{\;\dot 2}\end{pmatrix}\,, \qquad \dot{\bm{\mu}}=\begin{pmatrix}\mu_{\dot1}^{\; 1} & \mu_{\dot1}^{\; 2}\\ \mu_{\dot 2}^{\; 1} & \mu_{\dot 2}^{\; 2}\end{pmatrix}\,.
\eeqa
Now, we make the following identifications:
\beq
{\AA}\mu_1^{\;\dot 1}=-\mu_{\dot 2}^{\;2}=\dot\lambda\,,\quad \mu_1^{\;\dot 2}=\AA\mu_{\dot 1}^{\;2}=-\mu\,,\quad \mu_2^{\;\dot 1}=\AA\mu_{\dot 2}^{\;1}=\kappa\,,\quad \mu_2^{\;\dot 2}=-\AA\mu_{\dot 1}^{\;1}=-\lambda\, ,
\eeq
which gives the following form for these matrices
\beqa
\label{def:mu_matrix_final}
\bm{\mu}=\begin{pmatrix}\AA\dot\lambda & -\mu\\ \kappa & -\lambda\end{pmatrix}\,, \qquad \dot{\bm{\mu}}=\begin{pmatrix}
{\AA}{\lambda} & -
{\AA}{\mu}\\ 
{\AA}{\kappa} & -\dot\lambda\end{pmatrix}\,.
\eeqa
With our definition of $\kappa$, it is obvious that both matrices have unit determinant:
\beq\label{eq:unitdetmu}
\det\bm{\mu}=\det\dot{\bm{\mu}}=1.
\eeq
Moreover, the monodromy equations \eqref{eq:gluing_finalP}-\eqref{eq:gluing_finalPbis} can be rewritten compactly as
\beq
\label{def:QSCgluing_P}
(\dot\bP_{\dot a})^{\bar \gamma}=\bP_b\,\mu^{b}_{\;\dot a}\,,\qquad (\bP_{ a})^{\bar \gamma}=\dot\bP_{\dot b}\,\mu^{\dot b}_{\; a}\, ,
\eeq
where all indices run from $1$ to $2$, repeated indices are summed over, and the quantities $\mu^{b}_{\;\dot a}$ and $\mu^{\dot b}_{\; a}$ are, by definition, the inverse transpose matrices, i.e. 
\beq
\mu_{a}^{\;\dot b} \mu^b_{\; \dot b} = \delta_a^b,\;\;\;\mu_{\dot a}^{\; b} \mu^{\dot b}_{\;  b} = \delta_{\dot a}^{\dot b}\,.
\eeq
Since we are dealing with $2 \times 2$ matrices with unit determinant, these are related to the mirror periodic functions \eqref{def:gluing_QSC} by raising indices with the Levi-Civita tensor \footnote{Our convention is $\epsilon_{12} = 1$ and $\epsilon^{12} = -1$. } 
\beq\label{eq:muepsilon}
\mu^b_{\;\dot a}=\eps^{b c}\mu_c^{\;\dot d}\,\eps_{\dot d\dot a}\,,\qquad \mu^{\dot b}_{\; a}=\eps^{\dot b \dot c}\mu_{\dot c}^{\;d}\,\eps_{ da}
\,.
\eeq
Therefore, the identities \eqref{def:QSCgluing_P} above can also be rewritten as
\beq
\label{def:QSCgluing_P2}
(\dot\bP_{\dot a})^{\bar \gamma}=(\bm{\mu}^{-1} )_{\dot a}^{\; b} \,\bP_b\,,\qquad (\bP_{ a})^{\bar \gamma}=(\dot{\bm{\mu}}^{-1} )_{ a}^{\; \dot b} \,\dot\bP_{\dot b}\,.
\eeq
Furthermore, the discontinuities of the mirror-periodic functions in \eqref{eq:gluing_finalmu}-\eqref{eq:gluing_finalmulast} can be written as
\beq
\label{def:QSCgluing_mu1}
(\mu_{a}^{\;\dot b})^{\gamma}-\mu_{a}^{\;\dot b}=\bP_a(\dot\bP^{\dot b})^{\bar\gamma}-(\bP_a)^{\gamma}\dot\bP^{\dot b}\,,\;\;\;
(\dot\mu_{\dot a}^{\; b})^{\gamma}-\dot\mu_{\dot a}^{\; b}=\dot\bP_a(\bP^{\dot b})^{\bar\gamma}-(\dot\bP_a)^{\gamma}\bP^{\dot b}\,,
\eeq
where we have defined the $\bP$ functions with raised indices using the $\epsilon$ tensor: 
\beq\label{eq:LRrestriction}
\bP^{ a}\equiv \eps^{ a  b}\bP_{ b},\;\;\;\dot\bP^{\dot a}\equiv \eps^{\dot a \dot b}\bP_{\dot b}.
\eeq
Notice that the appearance of the same mirror-periodic functions inside the matrices $\bm{\mu}$ and $\dot{\bm{\mu}}$ is the reason why we find two independent forms for each of the four discontinuities \eqref{eq:gluing_finalmu}-\eqref{eq:gluing_finalmulast}.

Now, the equations \eqref{def:QSCgluing_P2}, \eqref{def:QSCgluing_mu1}
with the constraint \eqref{eq:unitdetmu} perfectly match the core identities of the $\bP\bm{\mu}$-system of \cite{Cavaglia:2021eqr, Ekhammar:2021pys}. Further, the identification of $\bP$ functions with upper and lower indices in  \eqref{eq:LRrestriction} marks the projection to the LR-symmetric sector, which we should have expected since we started from the same symmetry assumption in the T-system.

However,  from the form \eqref{def:mu_matrix_final} we also see a new, nontrivial constraint embedded in our construction:
\beq
\label{def:muandmudot}
\bm{\mu} = -\AA\dot{\bm{\mu}}^{-1}\,,
\eeq
i.e. $
\mu_a^{\; \dot b} = -\AA\epsilon_{ab} \mu_{\dot c}^{\; b} \epsilon^{\dot{c} \dot{b}}$. This relation was not noted before in the proposal for the QSC of \cite{Cavaglia:2021eqr, Ekhammar:2021pys}. 
 It implies that, at least for the LR-symmetric states considered here, there is only one independent $\bm{\mu}$ matrix. This has interesting consequences, which we discuss in the Conclusions. 

We can also see that the relations \eqref{eq:Pgammma1}-\eqref{eq:Pgammma4} can be rewritten, in the new notation, as 
\beq\label{eq:Pgammalast}
\bP_a^{\gamma} = (\delta_a^b + \bP_a \bP^b ) \, \mu_b^{\; \dot b} \, \dot\bP_{\dot b} ,\;\;\;\; \dot\bP_{\dot a}^{\gamma} = (\delta_{\dot a}^{\dot b} + \dot\bP_{\dot a} \dot\bP^{\dot b} ) \, \mu_{\dot b}^{\;  b} \, \bP_{ b} ,
\eeq
which also match perfectly the results of \cite{Cavaglia:2021eqr, Ekhammar:2021pys}. In fact, \eqref{eq:Pgammalast} can be derived from \eqref{def:QSCgluing_P}, \eqref{def:QSCgluing_mu1}, with simple manipulations which can be found in \cite{Cavaglia:2021eqr}.

{The equations above can also be rewritten in a simpler form
\beq 
\bP_a^\gamma=(\mu_R)_a^{\;\dot b}\dot\bP_{\dot b}\,,\quad \dot\bP_{\dot a}^\gamma=(\dot\mu_R)_{\dot a}^{\; b}\bP_{ b}\,,
\eeq by defining two matrices which are free of cuts on the real axis  \cite{Cavaglia:2021eqr, Ekhammar:2021pys}:
\beq
\label{def:muR_matrices}
(\mu_R)_a^{\;\dot b}\equiv(\delta_a^b + \bP_a \bP^b ) \, \mu_b^{\; \dot b}
\,,\qquad (\dot \mu_R)_{\dot a}^{\;b}\equiv (\delta_{\dot a}^{\dot b} + \dot\bP_{\dot a} \dot\bP^{\dot b} ) \, \mu_{\dot b}^{\;  b} \,.\eeq
Notice that two of their matrix elements are exactly equal to two T-functions, as can be readily verified from the definition \eqref{def:muR_matrices} and \eqref{eq:T10eqs}, namely:
\beq
{(\mu_R)}_1^{\,\dot 2}=\bT_{1,0}\,,\qquad {(\dot \mu_R)}_{\dot 1}^{\; 2}=\dot\bT_{1,0}\,.
\eeq
It would be interesting to check whether the remaining matrix elements of \eqref{def:muR_matrices} also admit an interpretation in terms of T-functions.

We also notice that the constraint on the behaviour at the branch points \eqref{eq:regularityzeta} can be reformulated cleanly in terms of these matrix elements as follows:
\begin{equation}\label{eq:bpconstraint2}
\frac{(\mu_R)_1^{\;\dot 2}}{(\mu_R)_{\dot 1}^{\; 2}}\bigg|_{u=+2}= \frac{(\mu_R)_{\dot 1}^{\; 2}}{(\mu_R)_1^{\;\dot 2}}\bigg|_{u=-2}.
    \end{equation}    
This constraint poses a non-trivial condition on QSC solutions that should hold for any state. Finally, we have also found the constraint \eqref{eq:constraint_c} for the constant $c$ appearing in the TBA equations, which relates it to $\AA$ and under our simplifying assumptions becomes simply:
\beq
e^{-4c}=1\,.
\eeq

\paragraph{A reformulation of the algebraic constraint. }
Finally, we note that the constraint $\bP^a \bP_a^{\bar\gamma}=\dot\bP^a \dot\bP_a^{\bar\gamma}$ found in \eqref{eq:constraint_final}, {which can also be written in a more elementary way as
\beq\label{eq:constraintsimpler}
\bP_a^{\bar\gamma} \epsilon^{ab} \bP_b =\AA \dot\bP_{\dot a}^{\bar\gamma} \epsilon^{\dot{a} \dot{b} } \dot\bP_{\dot b},
\eeq
}
is also an automatic consequence of \eqref{def:muandmudot} combined with the other $\bP\bm{\mu}$ equations. In fact,
\beq
\bP^a \bP_a^{\bar\gamma}=\bP_b \dot\bP_{\dot b} \epsilon^{\dot{b} \dot{a} } \mu_{\dot a}^{\; b}=\dot \bP^{\dot a} \bP_{ b} \epsilon_{\dot{a}\dot{c}} \mu_a^{\; \dot{c}} \epsilon^{ab}=\AA\dot \bP^{\dot a} \dot\bP_{\dot a}^{\bar\gamma}\,,
\eeq
where in the first and last equalities we have used \eqref{def:QSCgluing_P} and \eqref{eq:muepsilon}, while the second equality in addition uses the LR projection and the constraint \eqref{def:muandmudot}.

 Note that \eqref{eq:constraintsimpler} is similar, but not the same, as the equation 
\beq\label{eq:constraintsimpler2}
\bP_a^{\bar\gamma} \epsilon^{ab} \bP_b =\dot\bP_{\dot a}\epsilon^{\dot{a} \dot{b} } \dot\bP_{\dot b}^{\gamma},
\eeq
which is a constraint following from the rest of the $\bP\bm{\mu}$-system, already found in \cite{Cavaglia:2021eqr}. \footnote{In that paper, the constraint is written in terms of $\bQ$ functions. However, by generalising the arguments there, we verified that it also holds for the $\bP$ functions. It holds in any sector and can be proved independently of \eqref{def:muandmudot}. 
}
In fact, thanks to the condition \eqref{def:muandmudot}, \emph{they should both hold at the same time!} Since the RHS of \eqref{eq:constraintsimpler2} is the analytic continuation through the cut of the RHS of \eqref{eq:constraintsimpler}, this implies that the quantity defined by the LHS of \eqref{eq:constraintsimpler2} can have different properties depending on the value of $\AA$:
\begin{enumerate}
    \item 
For $\AA=1$, it must be a function where the cut on the real axis completely cancels, and the constraint can be reformulated as follows:
\beq\label{eq:constraintpowerful}
\bP_a^{\bar\gamma} \epsilon^{ab} \bP_b \;\;\;\; \text{     
          has no cut on the real axis.}
\eeq
\item For $\AA=-1$, the constraint can be reformulated as follows:
\beq\label{eq:constraintpowerful2}
\bP_a^{\bar\gamma} \epsilon^{ab} \bP_b \;\;\;\; \text{     
          changes sign across the cut on the real axis.}
\eeq 
\end{enumerate}

\subsection{From the $\bP {\mu}$-system to the full QSC}
In this paper, we have recovered the $\bP\bm{\mu}$-system formulation of the QSC, which already contains all the information. The way to go from this formulation to the full structure of the analytic Q-system presented in \cite{Cavaglia:2021eqr,Ekhammar:2021pys} is well understood. We briefly sketch how this works. The main steps are essentially equivalent to reading section 3.3 of \cite{Cavaglia:2021eqr} in reverse. 

First, one can embed the single-cut $\bP$ functions into two copies of the PSU(1,1|2) Q-system, which is summarised in appendix \ref{app:Qsysgen}. The relevant identifications are $\bP_a \equiv Q_{a|\emptyset}$, $\dot{\bP}_{\dot a} \equiv \dot{Q}_{\dot a | \emptyset}$, where we are keeping the same convention of marking with dots Q-functions in one of the PSU(1,1|2) wings. 

Assuming $Q_{\emptyset|\emptyset} = Q_{12|12} = 1$ in each wing, we can then use the functional relation of the Q-system to define all remaining Q-functions in terms of the $\bP$ functions implicitly.\footnote{In some cases, this involves solving finite difference equations in the complex rapidity domain. }
Then we find that some of the Q functions constructed this way also have notable properties. In particular, the functions $\bQ_k \equiv Q_{\emptyset|k}$ and their dotted counterparts can be defined as functions with a Riemann sheet and a single \emph{long} branch cut. The proof of this statement follows from the $\bP \bm{\mu}$-system in a standard way, and we summarise it in appendix \ref{app:Qsysgen2}. 

 The conjecture for the QSC in \cite{Cavaglia:2021eqr,Ekhammar:2021pys} was strongly based on working out the compatibility between the fact that the $\bP$ functions admit a Riemann section with a single short cut, while the $\bQ$ functions admit a sheet with a single long cut. This led to some \emph{gluing equations}, which contain constant \emph{gluing matrices} serving to connect the $\bQ$ functions in the two wings on two sides of the cut on the real axis. There, the $\bP \bm{\mu}$-system was found as a consequence. 
 However, one can take the opposite route and recover the gluing equations from the $\bP\bm{\mu}$-system, as we review explicitly in appendix \ref{app:Qsysgen2}.

\section{Conclusions}
\label{sec:conclusions}
In this paper, we have completed the derivation, started in our previous  work  \cite{Cavaglia:2026scj}, of the \adsst\, Quantum Spectral for the LR-symmetric sector, starting from the ground state  Thermodynamic Bethe Ansatz~\cite{Frolov:2021bwp}. 

 We used the same logic followed previously in the case of the $\mathcal{N}$=4 SYM and ABJM theory spectral problems: first the TBA was reformulated as a set of universal functional equations, and then the latter were simplified in terms of a finite number of fundamental functions, obtaining the $\bP\bm{\mu}$-system, equivalent to the QSC. 
 Our derivation assumes a discrete symmetry of the T-functions, corresponding to the so-called Left-Right symmetric sector (which includes all states for which QSC solutions have been studied so far). At the same time, our argument is fully general with respect to the other $\mathbb{Z}_2$ symmetry of the system, i.e. the one reflecting the chiral symmetry of the dual 2D CFT. 
 
 Our results agree with the QSC equations proposed in \cite{Cavaglia:2021eqr,Ekhammar:2021pys},  with the addition of previously unknown constraints, on which we comment below. 
One of these constraints depends on the discrete value of the constant $\AA = \pm 1$ and therefore distinguishes two ``sectors''. 

 Along the way, we have built the T-system for this model, and found a novel parametrisation of its T-functions in terms of the Q-functions appearing in the Quantum Spectral Curve. In particular, our results show how to build a solution of excited-state TBA equations out of any numerical solution of the QSC, such as the ones found in \cite{Cavaglia:2022xld, Ekhammar:2026ykk}. This makes the link between the two formulations completely transparent. This double description could help pinpoint interesting families of states.

One of the extra constraints we found is the algebraic condition on the $\bm{\mu}$ and $\dot{\bm{\mu}}$ matrices, \eqref{def:muandmudot}, which had not been written before in \cite{Cavaglia:2022xld, Ekhammar:2026ykk}.  
We expect this constraint to hold at least for LR-symmetric states. It would be interesting to see if it could be realised in more generality, but this should be checked carefully, since our derivation started from the ground-state TBA equations and therefore in principle only covers states connected to the ground state by analytic continuation. 
Preliminary comparison with the numerical solutions of the QSC\footnote{
We thank Nikolay Gromov and Charles Thull for communications.} suggests that this constraint might be enforced dynamically on the solutions of the QSC, when an important assumption is made: the off-diagonality of the so-called gluing matrices 
(see appendix \ref{app:Qsysgen2} for a review of their definition).
Until now, no simple argument for this off-diagonal nature of the gluing matrix had been found, but there were several indications that this assumption is needed, and it was used to solve the QSC in \cite{Cavaglia:2022xld, Ekhammar:2026ykk}.  
 The constraint \eqref{def:muandmudot} works as the missing piece of the puzzle, since it can be verified\footnote{We thank Simon Ekhammar for sharing this result~\cite{Simon_note}.} that this extra assumption allows to prove that the gluing matrices should be off-diagonal (as well as establishing spin quantisation).
 The other constraint concerns the behaviour of the solutions of the QSC at the branch points, i.e. \eqref{eq:bpconstraint2}. It would be interesting to verify if this too is enforced on the solutions of the QSC once a specific form of the gluing matrix is assumed. 

Our derivation also makes the role of massless modes explicit in the Quantum Spectral Curve. 
On one hand, the presence of the massless Y functions and the associated discontinuities were crucial to close the derivation. 
 On the other hand, we have shown that massless mode excitations correspond to zeros of the $\mu$ function above and below its cut on the real axis, which confirms the picture emerging in the ABA limit \cite{Ekhammar:2024kzp}. We also provided a recipe to compute the massless Y function using QSC quantities in \eqref{eq:Y0asPm}, and showed that the non-quadratic nature of the cuts of the QSC is deeply linked to the presence of a nontrivial massless Y function, cf. \eqref{eq:Y0_mu_nonquadratic}. This resonates with the intuition of \cite{Cavaglia:2022xld, Ekhammar:2026ykk} that the complicated nature of the cuts in this system is related to the massless modes.

\paragraph{Future directions.} 

\begin{itemize}

\item \textit{Mixed-flux generalisation.} Currently, a QSC formulation for strings on
  \adsst\, with mixed RR and NSNS
  fluxes is not available. It would be natural to extend
  the analysis of this paper and of \cite{Cavaglia:2026scj} to this setting, building on the mirror TBA of~\cite{Frolov:2025tda}. This would yield QSC equations valid at any point of the moduli space, and it would let us draw a comparison with the perturbative results of the dual CFT~\cite{Gaberdiel:2015uca,Hampton:2018ygz,Guo:2019ady,Guo:2020gxm,Lima:2020boh,Lima:2020kek,Lima:2020urq,Benjamin:2021zkn,Gaberdiel:2024nge,Fabri:2025rok}, which have been obtained for the case of one unit of NSNS flux and perturbatively small string tension.

\item \textit{Massless states in the QSC.} 
  We have shown how to identify real massless excitations in the physical channel from the QSC perspective.   It would be interesting to construct solutions with these massless excitations and study the behaviour of massless states at intermediate tension --- something so far only done at small tension using the mirror TBA~\cite{Brollo:2023pkl,Brollo:2023rgp,Brollo:2025rgp}. This might help identify the dual CFT of pure-RR \adsst\ strings.

\item \textit{The }AdS$_3\times$S$^3\times$S$^3\times$S$^1$\textit{ model.} {For this different AdS$_3$ model a worldsheet S-matrix has been proposed in \cite{Babichenko:2009dk, OhlssonSax:2011ms,Borsato:2012ss, Borsato:2015mma, Frolov:2025syj}. A Quantum Spectral Curve has been conjectured at the $\alpha=1/2$ point in \cite{Cavaglia:2025icd, Chernikov:2025jko}, and in principle this could be generalised to the full $D(2,1;\alpha)$ case as sketched in \cite{Cavaglia:2025icd}. It would be interesting to see if these QSC proposals can be consolidated by linking them to the TBA approach as done in this paper.} 
\end{itemize}
\section*{Acknowledgements}

We thank S.~Ekhammar, S.~Frolov, N.~Gromov, B.~Stefa\'nski Jr, C.~Thull and A. Torrielli for many helpful discussions. We also thank S.~Ekhammar for sharing his unpublished calculation \cite{Simon_note} with us, and are particularly grateful to N. Gromov and C. Thull for inspecting the constraint \eqref{def:muandmudot} against numerical QSC data. This analysis sparked us to consider the existence of two  sectors related to  $\ss=\AA = \pm 1 $.
NP, DP and AS thank the organisers and participants of the workshop ``Higher-dimensional Integrability and Holography'', held in Buggerru, Italy, and of the ``Workshop on Higher-d Integrability'' held in Favignana, Italy, for stimulating discussions that contributed to this work.
AC, NP and RT participate in the project HORIZON-MSCA-2023-SE-01-101182937-HeI.
AS gratefully acknowledges support from the CARIPARO Foundation under grant No.~68079. This work was also partially supported by the INFN initiatives GAST, SFT, and STEFI.
DP acknowledges support from the Deutsche Forschungsgemeinschaft (DFG, German Research Foundation) through SFB 1624 (Project No.~506632645). NP acknowledges support from the Italian Ministry of University and Research (MUR)
under the FIS grant BootBeyond (CUP: D53C24005470001), and from the PRIN project “2022ABPBEY” CUP-
E53D23002220006.

\newpage
\appendix
\section{Proof of the magic sheet properties of $\wT_{1,s}$}
\label{app:magic_sheet_LRsymm}
In this appendix, we prove the magic sheet properties of the T-functions \eqref{eq:def_T1s_seq12} in the LR symmetric sector.

We will need to prove this property separately for $\hat{\wT}_{1,1}$, $\hat{\wT}_{1,2}$ and $\hat{\wT}_{1,s\geq 3}$ and $\hat{\dot\wT}_{1,1}$, $\hat{\dot\wT}_{1,2}$ and $\hat{\dot\wT}_{1,s\geq 3}$. We will analyse the undotted wing explicitly only, since for the dotted wing most of the proofs are identical.
\subsection*{Magic sheet property for $\hat{\wT}_{1,1}$}
We start by proving the magic sheet property for $\hat{\wT}_{1,1}$, which will be done by utilising some of the discontinuity relations written in section \ref{sec:Y-system}.
\paragraph{Short cuts at $\Im(u)=\pm i/h$.}
From the analytic properties of the $\bT$ gauge \eqref{eq:vertstrip}, $\wT_{1,1}=\bT_{1,1}$ does not possess any cut in the region $-1/h<\Im(u)<1/h$, and thus by the definition \eqref{eq:def_T1s_seq12} the same is true for $\hat{\wT}_{1,1}$. 

At $\Im(u)=\pm 1/h$, for $|u|>2$ we have
\begin{equation}
\hat{\mathbb{T}}_{1,1}(u+\frac{i}{h}+i 0)=\mathbb{T}_{1,1}^\gamma(u+\frac{i}{h}+i 0)=\mathbb{T}_{1,1}(u+\frac{i}{h}-i 0)=\hat{\mathbb{T}}_{1,1}(u+\frac{i}{h}-i 0).
\end{equation}
This means that $\hat{\wT}_{1,1}$ has no `long' cuts at $(-\infty\pm i/h,-2\pm i/h)\cup (2\pm i/h,+\infty\pm i/h) $.

For $|u|<2$ we have instead
\begin{equation}
\hat{\mathbb{T}}_{1,1}(u+\frac{i}{h}+i 0)=\mathbb{T}_{1,1}^\gamma(u+\frac{i}{h}+i 0)=\mathbb{T}_{1,1}^\gamma(u+\frac{i}{h}-i 0) \ne \hat{\mathbb{T}}_{1,1}(u+\frac{i}{h}-i 0) \,.
\end{equation}
Therefore, the interval $(-2+ \frac{i}{h}, +2+ \frac{i}{h})$ is a `short' cut of $\hat{\mathbb{T}}_{1,1}(u)$.
\paragraph{Absence of further cuts in the UHP.}
We now wish to prove that $\hat\wT_{1,1}$ does not possess further cuts above $\Im(u)=1/h$. Given the definition \eqref{eq:def_T1s_seq12}, this is equivalent to proving that:
\beq
[\log (\wT_{1,1}^+)^\gamma]_{+2N}=[\log(\bT_{1,1}^+)^\gamma]_{+2N}=0\,.
\eeq
This can be proven with the help of the discontinuity relations for the cuts in the UHP for the function $\Delta_1$ in \eqref{eq:D1_2N_disc_mag_sh_nonLR}. Let us recast them as follows by using the identity \eqref{eq:magic}:
\beq
[ \Delta_1 - \log\left(Y_{1,1} \dot{Y}_{1,1}\right) ]_{2 N} = 2 [\log \mu ]_0 +\left[ \log(1 + Y_0 ) \right]_{+2N}  + 2\left[ \log(1 + Y_{1,1}) \right]_{+2N}  .
\eeq
 Writing $Y_{1,1}$ on the RHS in terms of T-functions and using (\ref{eq:magic}) again, we get:
\beqa
\left[ \Delta_1 + [\log \mu]_0  \right]_{+2N} = + 2[\log \mu ]_0 +\left[  2 \log\frac{\mathbf{T}_{1,1}^{+} \mathbf{T}_{1,1}^{-}}{\mathbf{T}_{0,1} \mathbf{T}_{2,1} } \right]_{+2N} .
\eeqa
Moreover,  using the fact that $\mathbf{T}_{0,1}\equiv\mu$ is mirror periodic and therefore $[\log \mu]_{+2N}=[\log \mu]_0$, we can further simplify to:
\beqa
\label{eq:delta1_UHPcuts1}
\left[ \Delta_1 + [\log \mu]_0  \right]_{+2N} = +\left[  2 \log\frac{\mathbf{T}_{1,1}^{+} \mathbf{T}_{1,1}^{-}}{\mathbf{T}_{2,1} } \right]_{+2N} .
\eeqa
Now we use the definition of $\Delta_1$ \eqref{def:Delta_1} and take its discontinuity along the cut at $\Im(u)=+2N$, getting that
\beq
\label{eq:delta1_UHPcuts2}
\left[ \Delta_1 + [\log \mu]_0  \right]_{+2N} = \left[ 2 \log\mathbf{T}_{1,1}^+ - 2\log(\mathbf{T}_{1,1}^+)^\gamma \right]_{+2N}.
\eeq
Equating the RHS of \eqref{eq:delta1_UHPcuts1} and \eqref{eq:delta1_UHPcuts2}, we find:
\beq
\left[ 2\log(\mathbf{T}_{1,1}^+)^\gamma \right]_{+2N} = \left[ -2\log\frac{\mathbf{T}_{1,1}^-}{\mathbf{T}_{2,1}}  \right]_{+2N}.
\eeq
We now prove that the RHS of this expression is zero; to see this, we use the Hirota equations \eqref{def:Hirotaeq} repeatedly along a vertical edge of the T-hook, i.e. $\bT_{a,1}^+\bT_{a,1}^-=\bT_{a+1,1}\bT_{a-1,1}$ for $a\geq 2$. There is a telescoping effect, and we can write
\beq
\label{eq:telescoping_for_T11}
\frac{\mathbf{T}_{1,1}^-}{\mathbf{T}_{2,1}}  = \frac{\mathbf{T}_{2,1}^{--}}{\mathbf{T}_{3,1}^{-}} = \frac{\mathbf{T}_{3,1}^{[-3]}}{\mathbf{T}_{4,1}^{[-2]}} =\dots =  \frac{\mathbf{T}_{N+1,1}^{[-N-1]}}{\mathbf{T}_{N+2,1}^{[-N ]}},
\eeq
and the combination on the RHS does not have a cut at $\Im(u) = 2 N/h$, since we are in the $\bT$ gauge and so $
\bT_{a,1}$ are analytic for $|\Im(u)|<a/h$.

Therefore, we proved that:
\beq
\label{eq:T11_nocuts_UHP}
\left[ 2\log(\mathbf{T}_{1,1}^+)^\gamma\right]_{+2N} = 0 ,
\eeq
and thus $\hat{\wT}_{1,1}$ has no cuts above $\Im(u)=1/h$.

\paragraph{Absence of further cuts in the LHP.}
Finally, we wish to prove that $\hat\wT_{1,1}$ does not possess further cuts below $\Im(u)=-1/h$. Given the definition \eqref{eq:def_T1s_seq12}, this is equivalent to proving that:
\beq
[\log(\wT_{1,1}^-)^{\bar\gamma}]_{-2N}=[\log(\bT_{1,1}^-)^{\bar\gamma}]_{-2N}=0\,.
\eeq
In order to do so, we need to take the first discontinuity in \eqref{eq:disc_lnY1_m1_Tsection}, continue it below the long cut along $\bar\gamma$, and compute the discontinuities of the resulting object at $\Im(u)=-2N/h$. 
First, we find:
\beq
([\log Y_{1,0}]_{-1})^{\bar\gamma}=-\log Y_0-\frac{1}{2}\log Y_{1,1}\dot Y_{1,1} Y_{1,1}^{\bar\gamma}\dot Y_{1,1}^{\bar\gamma}+2\log \dot Y_{1,1}^{\bar\gamma}+2\log\left(1+\frac{1}{\dot Y_{1,1}}\right)+c\,,
\eeq
where we have used \eqref{Y0symmdisc} and \eqref{eq:discYpmnew1} to express $Y_0^{\bar\gamma}$ in terms of $Y_0$ and other quantities.

Then, we take the discontinuity of the quantity above at $-2iN/h$, 
using \eqref{eq:lnY0_disc_p2N}, \eqref{eq:discYpmnew1} and \eqref{eq:magic}, and find after many simplifications
\bal
\label{eq:Y1_disc_atm1_partial}
\left[(\left[ \log  Y_{1,0} \right]_{- 1})^{\bar\gamma}\right]_{-2N}&= 
\frac{3}{2} \left[ \log \frac{\dot{Y}_{1,1}^{\bar\gamma}}{\dot{Y}_{1,1} } \right]_{-2N} +\frac{1}{2} \left[ \log \frac{Y_{1,1}}{Y_{1,1}^{\bar\gamma} } \right]_{-2N} \\
&  +  \log  \frac{Y_{1,1}\dot{Y}_{1,1}}{Y_{1,1}^\gamma\dot{Y}_{1,1}^\gamma}  - 2 \left[ \log  \left(1+ Y_{1,1} \right) \right]_{-2N} \,.
\eal
Then using again \eqref{eq:discYpmnew1} we can convert \eqref{eq:magic} into the following identities:
\bal
[\log \mu]_0=-\log \frac{Y_{1,1}}{Y_{1,1}^{\bar\gamma}}=-\log \frac{\dot{Y}_{1,1}}{\dot{Y}_{1,1}^{\bar\gamma}}\,.
\eal
Plugging these into \eqref{eq:Y1_disc_atm1_partial} we obtain:
\bal
\label{eq:lnY1_m1_m2N_Ysyst}
\left[(\left[ \log  Y_{1,0} \right]_{- 1})^{\bar\gamma}\right]_{-2N}&= - [ [\log \mu]_0^{\bar\gamma} ]_{-2N}   -  2 [\log \mu]_0   - 2 \left[ \log  \left(1+ Y_{1,1} \right) \right]_{-2N} \,.
\eal
Let us compute the last quantity on the RHS of the expression above. By using \eqref{def:YsasTs} in the $\bT$ gauge we find
\bal
&2 \left[\log  \left(1+Y_{1,1} \right) \right]_{-2N} = 2 \left[ \log \frac{ \mathbf{T}^-_{1,1}}{ \mu} \right]_{-2N}+2\left[ \log \frac{\mathbf{T}^+_{1,1} }{\mathbf{T}_{2,1}}  \right]_{-2N}
\eal
Using the Hirota equations \eqref{def:Hirotaeq} repeatedly on the second term, we obtain
\bal
\frac{\mathbf{T}^+_{11}}{\mathbf{T}_{21}} = \frac{\mathbf{T}^{[+2]}_{21}}{\mathbf{T}^+_{31}} = \dots = \frac{\mathbf{T}^{[+N+1]}_{N+1, 1}}{\mathbf{T}^{[N]}_{N+2, 1}} \,,
\eal
which from the properties of the $\bT$ gauge has no cuts at $\Im(u)=-2N/h$. Therefore this piece vanishes and~\eqref{eq:lnY1_m1_m2N_Ysyst} is then equal to
\bal
\label{eq:Y1_disc_atm1_partial_2}
\left[(\left[ \log  Y_{1,0} \right]_{- 1})^{\bar\gamma}\right]_{-2N}&= - [ [\log \mu]_0^{\bar\gamma} ]_{-2N}      - 2 \left[ \log  \bT_{1,1}^-  \right]_{-2N} \,.
\eal
We want to compare the RHS of this expression with  $\left[(\left[ \log  Y_1 \right]_{- 1})^{\bar\gamma}\right]_{-2N}$ expressed in terms of T-functions. In particular, we have that (recall that $\bT^{-}_{2,0}$ does not have a cut on the real axis):
\bal
& \left[ \log  Y_{1,0} \right]_{- 1}^{\bar\gamma}\equiv \log  (Y^-_{1,0})^{\bar\gamma}- \log  Y^-_{1,0}=2\log {(\bT_{1,1}^-)^{\bar\gamma}}-2\log {\bT_{1,1}^-}-[\log \mu]_0^{\bar\gamma}\,.
\eal
Therefore, taking the discontinuity at $-2N$ we find:
\bal
& [\left[ \log  Y_{1,0} \right]_{- 1}^{\bar\gamma}]_{-2N}=2[\log {(\bT_{1,1}^-)^{\bar\gamma}}]_{-2N}-2[\log {\bT_{1,1}^-}]_{-2N}-\left[[\log \mu]_0^{\bar\gamma}\right]_{-2N}\,.
\eal
Finally, we equate the RHS of the equation above with that of \eqref{eq:Y1_disc_atm1_partial_2}, finding that:
\beq
\label{eq:T11_nocuts_LHP}
2[\log {(\bT_{1,1}^-)^{\bar\gamma}}]_{-2N}=0\,.
\eeq
As argued at the beginning of this section, this condition means that $\hat{\wT}_{1,1}$ has no cuts below $\Im(u)=-1/h$.
\paragraph{Conclusion.} In conclusion, we proved that $\hat{\wT}_{1,1}$ defined in \eqref{eq:def_T1s_seq12} possesses only two short cuts at $\Im(u)=\pm1/h$, and no further cuts in the complex plane.
\subsection*{Magic sheet property for $\hat{\wT}_{1,2}$}
We now prove the magic sheet property for $\hat{\wT}_{1,2}$. This corresponds to checking that $(\wT_{1,2}^{++})^\gamma$ and $(\wT_{1,2}^{--})^{\bar\gamma}$ have no cuts in the UHP and LHP, respectively.

We start by noting that $\mathbb{T}_{1,3}$ has no cuts at $\Im(u)=\pm 1/h$, since $\mathbb{T}_{1,3}=\mathbf{T}_{1,3} = \mathbf{T}_{3,1}$ and therefore it has the analyticity strips in \eqref{eq:vertstrip}.
 Then, for the UHP cuts, we have that
 \beq
 1 = \frac{(\mathbb{T}_{1,3}^+)^{\gamma}}{\mathbb{T}_{1,3}^+} .
 \eeq
Using the Hirota equations \eqref{def:Hirotaeq} $\mathbb{T}_{1,3} = \mathbb{T}_{1,2}^+ \mathbb{T}_{1,2}^-/\mathbb{T}_{1,1}$ we find that
\beq
1 = \frac{(\mathbb{T}_{1,2}^{++})^{\gamma}}{ \mathbb{T}_{1,2}^{++}} \frac{\mathbb{T}_{1,1}^{+}}{( \mathbb{T}_{1,1}^{+} )^{\gamma}} \rightarrow (\mathbb{T}_{1,2}^{++})^{\gamma} = \underbrace{ (\mathbb{T}_{1,1}^{+})^{\gamma}}_{\text{safe}} \, \frac{\mathbb{T}_{1,2}^{++}}{\mathbb{T}_{1,1}^+},
\eeq
where we see that the ``safe'' part has no cuts in the upper half-plane, because this is exactly \eqref{eq:T11_nocuts_UHP}.
Remarkably,  the remaining ratio of two T-functions also has no cuts in the upper half-plane. In fact, we can repeatedly use the Hirota equations on it to get
$$
\frac{\mathbb{T}_{1,2}^{++}}{\mathbb{T}_{1,1}^+} = \frac{\mathbb{T}_{1,3}^+}{\mathbb{T}_{1,2}} = \frac{\mathbb{T}_{1,4} }{\mathbb{T}_{1,3}^-} = \dots \frac{\mathbb{T}_{1,n}^{[4-n]}}{\mathbb{T}_{1,n-1}^{[3-n]}}.
$$
The last expression manifestly has no cut at distance $2 n  - 6$ from the real axis, if $n\geq 3$ (recall that $\bT_{s,1}=\bT_{1,s}=\wT_{1,s}$ and the analyticity properties \eqref{eq:vertstrip}). Since we can use this rewriting for any positive $N$, we conclude that  $
(\mathbb{T}_{1,2}^{++})^{\gamma} 
$
 has no cuts in the upper half-plane. The proof regarding $(\mathbb{T}_{1,2}^{--})^{\bar{\gamma}}$ in the lower half plane is almost identical, except that it uses the property of $\wT_{1,1}$ given in \eqref{eq:T11_nocuts_LHP}.

 \subsection*{Magic sheet property for $\hat{\wT}_{1,s \ge 3}$}

Finally, for $\hat{\wT}_{1,s\geq 3}$, we need to start from the property that $\wT_{1,s+1}$ has no cuts at $\Im(u)=\pm (s-2)/h$, i.e. $(\wT_{1,s}^{[+s-2]})^\gamma=\wT_{1,s}^{[+s-2]}$ and $(\wT_{1,s}^{[-s+2]})^{\bar\gamma}=\wT_{1,s}^{[-s+2]}$. 
Then, repeating the same steps that we did above, we can easily prove that 
\beq [\log (\wT^{[+s]}_{1,s})^{\gamma}]_{+2N}=0,\qquad [\log (\wT^{[-s]}_{1,s})^{\bar\gamma}]_{-2N}=0.\eeq

\subsection*{Magic sheet property for $\hat{\dot{\wT}}_{1,s}$}

In this section, we prove the magic-gauge properties in the dotted wing. Although chiral symmetry (introduced in \cite{Cavaglia:2026scj}) suggests that the properties established in one T-hook should also hold in the other, for clarity we present the corresponding arguments explicitly rather than relying on this symmetry alone.

Starting from $\hat{\dot \wT}_{1,1}$, we need to prove the properties:
\beq
[(\hat{\dot \wT}^+_{1,1})^\gamma]_{+2N}=0,\quad [(\hat{\dot \wT}_{1,1}^-)^{\bar\gamma}]_{-2N}=0\,.
\eeq
For the first property, we will need the first discontinuity equation in \eqref{eq:D1bar_2N_disc_mag_sh_nonLR}, which translates to
\beq
[ \bar{\Delta}_1 - \log\left(Y_{1,1} \dot{Y}_{1,1}\right) ]_{2 N} = 2 [\log \mu ]_0 +\left[ \log(1 + Y_0 ) \right]_{+2N}  + 2\left[ \log(1 + \dot Y_{1,1}) \right]_{+2N}  .
\eeq
This is identical to the one in the other wing except for the last term, where we find $\dot Y_{1,1}$ instead of $Y_{1,1}$. Therefore, it is immediate to recast it in the following form:
\beqa
\label{eq:Deltabarabove2}
\left[ \bar{\Delta}_1 + [\log \mu]_0  \right]_{+2N} = +\left[  2 \log\frac{\dot{\mathbf{ T}}_{1,1}^{+} \dot{\mathbf{ T}}_{1,1}^{-}}{\dot{\mathbf{ T}}_{2,1} } \right]_{+2N} .
\eeqa
From the definition of $\bar{\Delta}_1$ it follows that:
\beq\label{eq:Deltabarabove}
\bar{\Delta}_1 = \left[ 2\log{ \dot{\mathbf{T}}_{1,1}^+ }- \log{ \mu} \right]_0 = 2 \log \dot{\mathbf{T}}^{+}_{1,1} - 2 \log (\dot{\mathbf{T}}_{1,1}^+)^\gamma-\left[\log\mu \right]_0\,.
\eeq
Equating the RHS of \eqref{eq:Deltabarabove} and \eqref{eq:Deltabarabove2}, and using the T-system like in equation \eqref{eq:telescoping_for_T11}, it is immediate to prove that:
\beq
\left[ 2\log(\dot{\mathbf{T}}_{1,1}^+)^\gamma \right]_{+2N} = 0 .
\eeq
For the cuts in the LHP, we need to start from \eqref{eq:D1bar_2N_disc_mag_sh_nonLR}. Showing that $\left[ 2\log(\dot{\mathbf{T}}_{1,1}^-)^{\bar\gamma} \right]_{-2N} = 0$ from here follows the same steps as in the undotted wing and is left to the reader.

Regarding $\hat{\dot \wT}_{1,s\geq 2}$, the proofs for the magic sheet properties are the same as those for the undotted wing and therefore we will not repeat them here.

\section{Properties of $\bP$ functions in the lower half plane }\label{app:LHP}
In section \ref{sec:Pmusystem}, we have constructed two new functions, named $\bP_2$ and $\dot\bP_2$, and showed that the absence of branch cuts for these functions in the UHP corresponds to the absence of branch cuts for the T-functions $\bT_{a,0}$ at $\Im(u)=(a-1)/h$.

In this appendix, we will show that the fact that the absence of cuts in the LHP of the same Q-functions is equivalent to \eqref{eq:equation1_LHP}: overall, this means that the wide analyticity strips of the $\bT_{a,0}$ functions \eqref{eq:vertstrip_balckT} is guaranteed if $\bP_2$ and $\dot\bP_2$ have a single cut on the real axis.

To do so, we will start from the quantities $\bP_{2}^{\bar\gamma}$ and $\dot\bP_{2}^{\bar\gamma}$, evaluated with a slightly positive imaginary part, which are defined in terms of quantities in the first Riemann sheet via equations \eqref{eq:gluing_P2gammabar}. For the scope of this section, however, we want to express them in a slightly different form:
\beq
\label{eq:P2gammabar_alt}
\dot\bP_2^{\bar\gamma}=-\frac{1}{\mu}\bP_1 +\AA\frac{\dot\lambda}{\mu}\dot\bP_1^{\bar\gamma}\,,\qquad \bP_2^{\bar\gamma}=-\AA\frac{1}{\mu}\dot\bP_1 +\AA\frac{\lambda}{\mu}\bP_1^{\bar\gamma}\,.
\eeq
The equivalence of this expression for $\bP_{2}^{\bar\gamma}$ and $\dot\bP_{2}^{\bar\gamma}$ with their definition \eqref{eq:gluing_P2gammabar} can be verified immediately using \eqref{def:kappa} and \eqref{eq:gluing_P1}.

Now we take \eqref{eq:P2gammabar_alt} and drag it to the LHP by a small shift in the mirror kinematics: the $\bP$ functions will be continued along $\gamma$, while the mirror periodic functions will be left invariant. What we find is therefore a definition of $\bP_2$ and $\dot\bP_2$ valid with a small imaginary \emph{negative} shift, i.e.:
\beq
\dot\bP_2|_{\Im(u)<0}=-\frac{1}{\mu}\bP_1^\gamma +\AA\frac{\dot\lambda}{\mu}\dot\bP_1\,,\qquad \bP_2|_{\Im(u)<0}=-\AA\frac{1}{\mu}\dot\bP^\gamma_1 +\AA\frac{\lambda}{\mu}\bP_1\,.
\eeq
We can now impose that these quantities have no branch cuts in the LHP. This implies the following equation:
\beqa
\label{eq:equation2_LHP}
&&0=\mu^{\bar\gamma} \bP_1^{\gamma,[-2N]}-\mu(\bP_1^{\gamma,[-2N]})^{\bar\gamma}+\AA(-\mu^{\bar\gamma} \dot\lambda+\mu \dot\lambda^{\bar\gamma})\dot\bP_1^{[-2N]}\,, \\&&0=\mu^{\bar\gamma} \dot\bP_1^{\gamma,[-2N]}-\mu(\dot\bP_1^{\gamma,[-2N]})^{\bar\gamma}+(-\mu^{\bar\gamma} \lambda+\mu \lambda^{\bar\gamma})\bP_1^{[-2N]}\,.
\eeqa
Note that these have a very similar structure as \eqref{eq:equation1_LHP}, i.e. the requirement that $\bT_{a,0}^{[-a+1]}$ have no cuts. The two sets of equations are actually equivalent provided that the functions $\lambda$ and $\dot\lambda$ satisfy:
\beqa
\label{eq:lambdaLHP}
\frac{\dot\lambda^{\bar\gamma}}{\mu^{\bar\gamma}}-\frac{\dot\lambda}{\mu}&=&\AA\frac{1}{\mu-\bP_1\dot{\bP}_1^{\bar\gamma}}\left(\frac{\bP_1^2}{\mu}-\frac{(\bP_1^{\bar\gamma})^2}{\mu^{\bar\gamma}}\right)\,,\\\frac{\lambda^{\bar\gamma}}{\mu^{\bar\gamma}}-\frac{\lambda}{\mu}&=&\AA\frac{1}{\mu-\AA\dot\bP_1{\bP}_1^{\bar\gamma}}\left(\frac{\dot\bP_1^2}{\mu}-\frac{(\dot\bP_1^{\bar\gamma})^2}{\mu^{\bar\gamma}}\right)\,.
\eeqa
Substituting $\mu^{\bar\gamma}$ via \eqref{eq:gluing_mu}, we may rewrite these equations in the simpler forms:
\beqa\label{eq:discxi}
\dot\lambda^{\bar\gamma}-\dot\lambda&=&\AA(\bP_1^ {\bar\gamma}\dot\bP_{2}^{\bar\gamma\bar\gamma} - \bP_1 \dot\bP_{2}^{\bar\gamma})\,,\\
\lambda^{\bar\gamma}-\lambda&=&\AA(\dot\bP_1^ {\bar\gamma}\bP_{2}^{\bar\gamma\bar\gamma} - \dot\bP_1 \bP_{2}^{\bar\gamma})\,,
\eeqa
which after being continued along $\gamma$ are exactly the same equations we found in \eqref{eq:gluing_lambda_final}: equations \eqref{eq:lambdaLHP} are already satisfied without the need to impose any further constraint.

Therefore, we conclude that if the functions $\bP_2$ and $\dot\bP_2$, defined via \eqref{def:P3}, can be continued to a Riemann sheet with a single short cut on the real axis, and the functions $\lambda$ and $\dot\lambda$ satisfy equations \eqref{eq:gluing_lambda} and \eqref{eq:gluing_lambda_final}, the $\bT_{a,0}$ functions have the required analyticity strips defined in \eqref{eq:vertstrip_balckT}.

\section{Quick review of Q-systems}\label{app:Qsysgen}
In this appendix, we write down some general properties about Q-systems and the Q-functions contained in them.

\subsection*{The Q-system.} Given a Lie superalgebra $\mathfrak{g}$, a $\mathfrak{g}$ Q-system is a set of Q-functions, depending on the spectral parameter $u$, satisfying a set of functional equations known as QQ-relations. In general, a $\mathfrak{g}$ Q-system can describe any integrable model with Yangian symmetry $Y(\mathfrak{g})$, with the simplest example being a supersymmetric spin chain. To describe a specific model, one needs to impose suitable analytical conditions on the Q-functions.

In this paper, we consider two copies of a \psu\, Q-system. Each copy contains a total $2^{4}=16$ Q-functions, which we denote by:
\begin{small}
\beq
Q_{\emptyset|\emptyset}, \;\;Q_{a|\emptyset}\equiv \bP_a, \;\; Q_{\emptyset|i}\equiv \mathbf{Q}_i,\;\;Q_{a|i},\;\;Q_{ab|\emptyset},\;\; Q_{\emptyset|ij},\;\;Q_{a|ij}\equiv \epsilon_{ab}\bP^b, \;\;Q_{ab|i}\equiv \epsilon_{ij}\mathbf{Q}^j,\;\;Q_{ab|ij}  .
\eeq
\end{small}
The indices before and after the separator are called bosonic and fermionic,
respectively. Each group of indices is totally antisymmetric, and every
index takes the values $1,2$. We use
$\epsilon_{12}=-\epsilon^{12}=1$. The QQ-relations are:
\beqa
\label{def:QQrels}
Q_{\emptyset|I}Q_{ab|I}&=&Q_{a|I}^+Q_{b|I}^--Q_{a|I}^-Q_{b|I}^+\,,\\
Q_{A|\emptyset}Q_{A|ij}&=&Q_{A|i}^+Q_{A|j}^--Q_{A|i}^-Q_{A|j}^+\,,\\
Q_{Aa|i}Q_{A|Ii}&=&Q_{Aa|Ii}^+Q_{A|I}^--Q_{Aa|Ii}^-Q_{A|I}^+\,,
\eeqa
where $A$ and $I$ are sets of bosonic and fermionic indices having cardinality $|A|\leq 2$, $|I|\leq 2$, and in the last equation $a\notin A$ and $i\notin I$.
Thanks to the QQ-relations, most Q-functions are not independent, and
only a small subset is needed to reconstruct the full Q-system.

\subsection*{Analytical properties of the Quantum Spectral Curve.} The \adsst\, Quantum Spectral Curve was formulated by prescribing appropriate analytical properties for these Q-functions, which led to the proposal of a $\bP\mu$-system. Both this $\bP\bm{\mu}$-system and the proposed analytical structure of the Q-functions agree with those we derive in this paper. It is therefore useful to collect these analytical properties here. The complete $\bP\bm{\mu}$-system can be found in section \ref{sec:matchQSC0}.

First, in the QSC, the Q-functions $Q_{\emptyset|\emptyset}$ and $Q_{ab|ij}$ are set to 1 in both the dotted and undotted Q-systems. 

The fundamental objects are the $\bP_a$ functions and their dotted counterparts $\dot\bP_a$. These functions have power-like asymptotics at large $u$, which encode the quantum numbers of the states, and two branch points on their main Riemann sheet at $\pm 2$, joined by a short branch cut. Analytic continuation through this cut leads to
different sheets with an infinite ladder of branch points at
$u=\pm2+2in/h$, $n\in\mathbb Z$, which are again joined pairwise by short
cuts. This structure is represented in
figures~\ref{fig:Pfunc-main} and~\ref{fig:Pfunc-continued}.

The continuation of the $\bP$-functions through the cut on the first Riemann sheet is governed by
the gluing equations~\eqref{def:QSCgluing_P}. The functions
$\mu_a{}^{\dot b}$ (and $\mu_{\dot a}{}^{ b}$) on their right-hand side are the entries of the
$\bm{\mu}$ matrix. They are periodic functions on a Riemann section on which
they have long cuts. Finally, equations~\eqref{def:QSCgluing_mu1} determine
the discontinuity of the $\bm{\mu}$ matrix through its cuts. The functions
$\bP_a$, $\dot{\bP}_a$ and $\mu_a{}^{\dot b}$, together with their analytic
properties and equations~\eqref{def:QSCgluing_P}--\eqref{def:QSCgluing_mu1},
form the $\bP\bm{\mu}$-system.

Note that in the QSC the Q-functions $Q_{a|i}$ can be viewed as matrices relating the
$\bP$ to the $\bQ$ functions. In fact, defining
$Q^{a|i}\equiv\epsilon^{ab}\epsilon^{ij}Q_{b|j}$, the normalization
$Q_{\emptyset|\emptyset}=Q_{12|12}=1$ implies
$Q^{a|i}Q_{b|i}=\delta^a_b$. The QQ-relations then give
\beq
\label{def:PQ_rotations}
\begin{gathered}
Q_{a|i}^{\pm}\bQ^i=\bP_a,
\qquad
Q_{a|i}^{\pm}\bP^a=\bQ_i,
\\
(Q^{a|i})^{\pm}\bQ_i=\bP^a,
\qquad
(Q^{a|i})^{\pm}\bP_a=\bQ^i.
\end{gathered}
\eeq

\paragraph{Hodge duality.} The Q-system is invariant under Hodge duality. Denoting by $\bar A$ and
$\bar I$ the multi-indices complementary to $A$ and $I$, respectively, the
Hodge-dual function $Q^{A|I}$ is obtained from $Q_{\bar A|\bar I}$ by
contracting the complementary indices with the corresponding Levi-Civita
tensors. The simultaneous replacement 
\beq
Q_{A|I}\longrightarrow Q^{A|I}
\eeq
maps the QQ-relations into relations of the same form and is therefore a
symmetry of the full Q-system. In particular, it exchanges (up to signs) $\bP_a$ with
$\bP^a$ and $\bQ_i$ with $\bQ^i$.
\subsubsection*{Gauge symmetries.}
The Q-system admits a large class of gauge symmetries, which are transformations that leave the QQ-relations \eqref{def:QQrels} invariant. They are given by:
\beqa
\label{def:gauge_Qsys}
Q_{A|I}&\rightarrow& \frac{g^{[+(|A|-|I|+1)]}}{g^{[-(|A|-|I|+1)]}}Q_{A|I}\,,\\
\label{def:gauge_Qsys2}
Q_{A|I}&\rightarrow& \frac{f^{[+(|A|-|I|)]}}{f^{[-(|A|-|I|)]}}Q_{A|I}\,,
\eeqa
where $g$ and $f$ are non-vanishing functions and we indicate by $| A|$ the cardinality of the set of indices $A$. The normalisation
$Q_{\emptyset|\emptyset}=1$ imposes $g^+=g^-$, in which case the first
transformation acts trivially. The remaining transformation must also be
compatible with the prescribed cuts and large-$u$ asymptotics.

\paragraph{Rescaling symmetry.} A particularly useful residual transformation, which follows from~\eqref{def:gauge_Qsys2} by
choosing $f^+/f^-=c$, is the
rescaling symmetry of the Q-system
\beq
\label{def:constant_Q_rescaling}
Q_{A|I}\longrightarrow c^{|A|-|I|}Q_{A|I},
\qquad c\in\mathbb C.
\eeq
In particular, on the fundamental and Hodge-dual Q-functions it acts as
\beq
\label{def:constant_Q_rescaling_fundamental}
\begin{gathered}
\bP_a\to c\bP_a,
\qquad
\bP^a\to c^{-1}\bP^a,
\qquad
\bQ_i\to c^{-1}\bQ_i,
\\
\qquad
\bQ^i\to c\bQ^i,
\qquad
Q_{a|i}\to Q_{a|i}.
\end{gathered}
\eeq
In principle, the two Q-systems of the \adsst\, model admit independent
constants $c$ and $\dot c$. However if we want to keep invariant the $\bP\bm{\mu}$ system equations \eqref{def:QSCgluing_P}, the two
rescaling constants must be equal, i.e. $\dot c=c$.
\paragraph{Rotations and $H$-rotations.}
The QQ-relations are covariant under independent rotations in the
bosonic and fermionic indices. Defining two constant matrices
$R_{ b},R_{f}\in GL(2)$, these rotations are given by
\beq
\label{def:constant_Q_rotations}
Q_{A|I}\longrightarrow
(R_{ b})_A{}^B(R_{ f})_I{}^JQ_{B|J},
\eeq
where the matrices act in the exterior-power representations associated with the antisymmetric multi-index sets $A$ and $I$. The
Q-function with full indices transforms as
$Q_{12|12}\to\det R_{\rm b}\det R_{\rm f}\,Q_{12|12}$. Hence its unit
normalisation requires
\beq
\label{def:constant_Q_rotations_det}
\det R_{\rm b}\det R_{\rm f}=1.
\eeq
These rotations extend to a functional form known as $H$-rotations. Let
$H_{\rm b}(u)$ and $H_{\rm f}(u)$ be periodic matrices,
\beq
H_{ b}^{[+2]}=H_{ b},
\qquad
H_{ f}^{[+2]}=H_{ f}.
\eeq
Then the H-rotations are defined as
\beq
\label{def:H_rotations}
Q_{A|I}\longrightarrow
\left(H_{ b}^{[|A|-|I|]}\right)_A{}^B
\left(H_{ f}^{[|A|-|I|]}\right)_I{}^JQ_{B|J}
\eeq
Again, the unit normalisation of the Q-function with full indices requires
\beq
\label{def:H_rotations_det}
\det H_{ b}(u)\det H_{ f}(u)=1.
\eeq
In general, an $H$-rotation is allowed only if it also
preserves the branch cut structure and large-$u$ asymptotics. The dotted wing has an analogous
symmetry, and the two rotation matrices are related by the condition that the $\bP\bm{\mu}$ system equations \eqref{def:QSCgluing_P} are left invariant.

\paragraph{$\Lambda$-symmetry.}
The \adsst\, QSC has one further useful residual symmetry. For a constant $\Lambda$ and $x$, the Zhukovski variable, the undotted Q-system transforms as
\beq
\label{def:Lambda_symmetry_undotted}
\bQ_i\longrightarrow x^{+\Lambda}\bQ_i,
\qquad
\bP_a\longrightarrow x^{-\Lambda}\bP_a,
\qquad
Q_{a|i}\longrightarrow Q_{a|i}.
\eeq
The action on the dotted Q-system must be with opposite
powers:
\beq
\label{def:Lambda_symmetry_dotted}
\dot{\bQ}_i\longrightarrow x^{-\Lambda}\dot{\bQ}_i,
\qquad
\dot{\bP}_a\longrightarrow x^{+\Lambda}\dot{\bP}_a,
\qquad
\dot Q_{a|i}\longrightarrow\dot Q_{a|i}.
\eeq
This simultaneous transformation leaves both the QQ-relations and the
$\bP\bm{\mu}$-system invariant. The remaining Q-functions transform as
dictated by the QQ-relations. 

These transformations come from the second gauge symmetry
in~\eqref{def:gauge_Qsys} with
\beq
\frac{f^+}{f^-}=x^{-\Lambda},
\eeq
combined with its inverse in the dotted wing. 

\section{Outline of the derivation of the QSC from the $\bP$\texorpdfstring{$\bm{\mu}$}{m}-system}\label{app:Qsysgen2}

In this appendix, we briefly summarise how, from the $\bP\bm{\mu}$ system, we can recover the full structure of the QSC.  For convenience, we switch to a convention where all shifts are intended to be on a section with short cuts. This means that we introduce a function $\hat{\mu}$ with a short branch cut on the real axis, for $u\in(-2,2)$, and satisfying the following mirror periodicity property:
\beq
\hat\mu^{++}=\hat\mu^\gamma\,.
\eeq
The same property holds for the functions $\hat\lambda,\,\hat{\dot\lambda},\,\hat{\kappa}$ which are the analog with short cuts of $\lambda,\,\dot\lambda,\,\kappa$.

It is then easy to see that the equation~\eqref{def:QSCgluing_mu1} can be rewritten as
\begin{equation}
\label{eq:Pmuwithshortcuts}
(\hat\mu^a_{\; \dot b} )^{++}  = \left( \delta^a_c - \bP^a \bP_c \right) \hat\mu^c_{\dot c} \left( \delta^{\dot c}_{\dot b} + \dot\bP^{\dot c} \dot\bP_{\dot b} \right) .
\end{equation}
Now, we can construct the $2 \times 2$ matrix  $\omega_k^{\;\dot m } \equiv Q_{a|k}^- \mu^a_{\;\dot b} Q^{\dot b| \dot{m} -}$, with $Q_{a|k} Q^{b|k} \equiv \delta_a^b$. As a consequence of the Q-system relations
\begin{equation}
Q_{a|i}^+ = Q_{b|i}^- \left(\delta_a^b + \bP^b \bP_a \right) ,
\end{equation}
and of \eqref{eq:Pmuwithshortcuts}, 
we then find that the $\omega$ matrix is a $2i/h$-periodic function with short cuts, i.e. $\hat{\bm{\omega}}=\hat{\bm{\omega}}^{++}$. The same is true for the dotted counterpart of this matrix. 

Then, since the $\bQ$ functions are related to the $\bP$ functions by
\beq
\bQ_k \equiv Q_{\emptyset|k} = Q_{a|k}^{\pm} \bP^a ,
\eeq
it is immediate to establish, from the $\bP\bm{\mu}$-system identity 
\begin{equation}
\bP_a^{\gamma} = (\hat\mu_a^{\; \dot b} )^{\gamma} \dot\bP_{\dot b} =(\hat\mu_a^{\; \dot b} )^{++} \dot\bP_{\dot b} ,
\end{equation}
that the following identity is true, together with the analogous relation exchanging dotted/undotted indices:
\begin{equation}
\bQ_k^{\gamma} = \omega_k^{\;\dot{m}} \, \dot\bQ_{\dot m}.
\end{equation}
This relation allows us to get to a concise formulation in terms of the \emph{gluing matrix}. In fact, one notices that, starting from the same $\bP$ functions, it is possible to construct two alternative bases for $\bQ$ functions, denoted as $\bQ^{\downarrow}$ and $\bQ^{\uparrow}$, where the first basis is free of singularities in the UHP and the second basis is free of singularities in the LHP. The two bases should be connected by a short cut $2i/h$-periodic $2\times 2$ matrix function $\Omega$. The equations above then imply that
\begin{equation}
\bQ_k^{\downarrow \gamma}(u) = G_k^{\;\dot{m} } \, \dot\bQ_{\dot m}^\uparrow(u),
\end{equation}
where the key point is that the \emph{gluing matrix} $G$ is now constant. It is defined explicitly as
\begin{equation}
G_k^{\;\dot{m}} \equiv \omega_k^{\;\dot n}(u) (\dot{\Omega}^{-1} )_{\dot n}^{\;\dot{m}}(u).
\end{equation}
All these equations have a dotted shadow equation, and in particular there is a second gluing matrix $\dot{G}_{\dot k}^{\; m}$. 

Finally, starting from the two bases $\bQ^{\uparrow}$ and $\bQ^{\downarrow}$, it is easy to define the Q-functions $\bQ^{\text{long}}$ with a single, long branch cut on the real axis:
\beq
\bQ^{\text{long}}_k=\begin{cases}
 \bQ_{k}^{\downarrow},\;\;\,\,\qquad \Im(u)>0\,, \\
   G_k^{\;\dot m}\dot\bQ_{\dot m}^{\uparrow},\quad \Im(u)<0\,,
\end{cases}
\eeq
with a similar equation defining $\dot {\mathbf{Q}}^{\text{long}}$.
\bibliography{bibliography}

\end{document}